\documentclass[aps,prb,10pt,longbibliography,superscriptaddress,eqsecnum,twocolumn,nofootinbib]{revtex4-2}
\usepackage{graphicx}
\usepackage{stmaryrd}
\usepackage{amsmath, amsfonts,amssymb}
\usepackage{mathrsfs}
\usepackage{braket}
\usepackage{color}
\usepackage{xspace}
\usepackage{amscd}
\usepackage{comment}
\usepackage{bm}
\usepackage{bbm}
\definecolor{lightblue}{rgb}{0.13, 0.26, 0.99}

\usepackage{xurl}
\PassOptionsToPackage{hyphens}{url}
\usepackage[
  colorlinks=true,
  urlcolor=blue,
  citecolor=blue,
  linkcolor=blue,
hyperfootnotes=false]{hyperref}

\usepackage{appendix}

\usepackage{mathtools}

\allowdisplaybreaks

\usepackage[normalem]{ulem}

\newcommand{\red}[1]{{\color{red}#1}}

\begin{document}


\title{Magnon harmonic generation in
  antiferromagnets
  : \\ Dynamical symmetry enriched by 
symmetry breaking}

\author{Yuto Jita}
\email{yutojj130@gmail.com}
\affiliation{Department of Physics, Chiba University, Chiba 263-8522, Japan}

\author{Minoru Kanega}
\affiliation{Department of Physics, Chiba University, Chiba 263-8522, Japan}

\author{Takumi Ogawa}
\affiliation{Department of Physics, Ibaraki University, Mito, Ibaraki 310-8512, Japan}

\author{Shunsuke C. Furuya}
\affiliation{Department of Liberal Arts, Saitama Medical University, Moroyama, Saitama 350-0495, Japan}
\affiliation{Institute for Solid State Physics, The University of Tokyo, Kashiwa, 277-8581, Japan}

\author{Masahiro Sato}
\email{sato.phys@chiba-u.jp}
\affiliation{Department of Physics, Chiba University, Chiba 263-8522, Japan}

\date{\today}
\begin{abstract}
  In recent years, techniques of intense THz laser have enabled us to experimentally observe nonlinear spin dynamics in antiferromagnets since the elementary excitations such as magnons reside on a THz to GHz range in antiferromagnets and THz laser thus can directly excite them.
  We numerically and theoretically investigate THz-laser or GHz-wave driven harmonic generations in typical ordered phases of antiferromagnets: N\'eel, canted and weak ferromagnetic phases.
  The radiation waves (harmonic generations) are created by the incident-wave driven magnon dynamics. We point out that magnetic orders and phase transitions can change the spectra of harmonic generations, differently from those of metallic, semiconductor, or atomic-gas systems without (spontaneous) symmetry breakings.
  We consider both the magnon harmonic generation driven by standard single-color laser and that by two-color laser in the antiferromagnets, and find several dynamical symmetries and the corresponding selection rules of the harmonic generations. In particular, a special selection rule emerges in U(1)-spin-rotation symmetric magnetic phases (like ferromagnetic, ferrimagnetic, N\'eel phases and U(1)-symmetric spin-liquid phases) irradiated by $C_{\ell}$-symmetric two-color laser. We also find that the computed spectrum in the weak ferromagnetic phase is consistent with a recent experiment of $\rm HoFeO_3$.
  These results indicate that the magnon harmonic generation spectra provide new information about symmetry or symmetry breaking of antiferromagnets.
\end{abstract}

\maketitle

\section{Introduction}
\label{sec:intro}

\begin{figure}[t!]
  \centering
  \includegraphics[bb = 0 0 653 206, width = \linewidth]{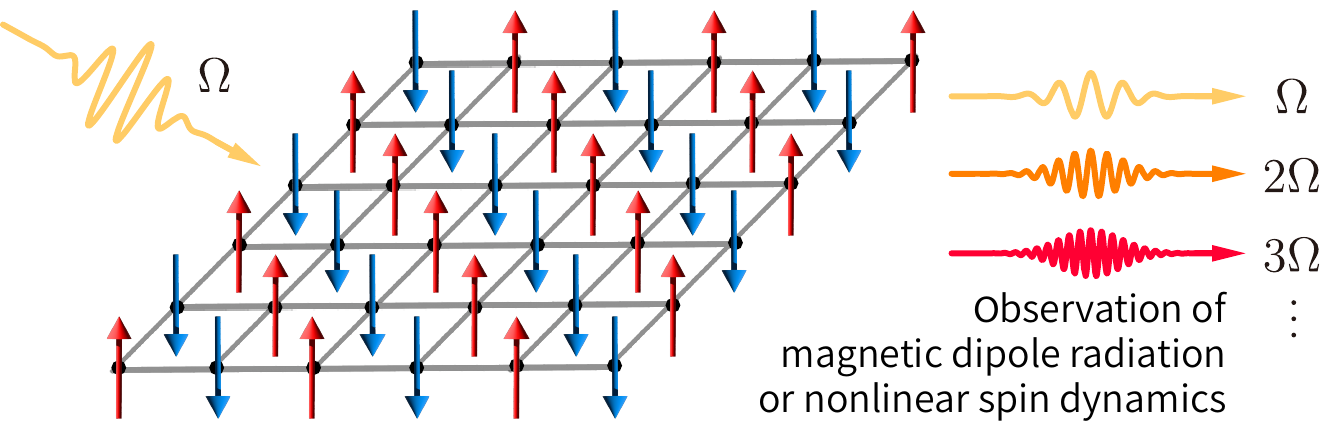}
  \caption{
    Schematic diagram of harmonic generation in antiferromagnetic insulator. When we apply a strong enough THz laser pulse with frequency $\Omega$ to an antiferromagnet, we can observe a radiation wave consisting of mixed waves with frequencies $\Omega$, $2\Omega$, $3\Omega$, $\ldots$ through the spin-light coupling.
  }
  \label{fig:HHG}
\end{figure}

Recent developments of terahertz (THz) laser technologies open up new fields in science and technology.
Nonlinear optical phenomena in \textit{solids}~\cite{Kruchinin2018Colloquium, Yu2019High, Silva2019Topological, Murakami2025Photoinduced} have actively been studied in more recent years.
Among the most active topics is high harmonic generation (HHG)~\cite{Ghimire2018_HHG, Yue2022_HHG, Goulielmakis2022_HHG, Li2023_HHG}, including multi-color-laser-driven HHG~\cite{Vampa2015Linking,Luu2016Highorder,Luu2018Observing,Rana2022Generation,He2022Dynamical,Udono2024_HHG,Kanega2025Twocolor}.
Suppose that we apply an external electromagnetic wave with frequency $\Omega$ to a material, an outgoing wave with multiple frequencies of $\Omega$, $n\Omega$ with $n=1,~2,~3,~\ldots$, is emitted.
This phenomenon is referred to as HHG (see Fig.~\ref{fig:HHG}) and is one of the simplest nonlinear optical responses.

HHG was first observed in atomic gas systems~\cite{McPherson1987_HHG, Ferray1988_HHG}.
Developments in strong-field laser technologies~\cite{Ghimire2014_attopulse} expanded the playing field to solid systems of semiconductors and metals~\cite{Ghimire2011_HHG, Schubert2014_HHG, Hohenleutner2015_HHG, Luu2015_HHG, Ndabashimiye2016_HHG, Vampa2017_HHG, Kaneshima2018_HHG}. For instance, more than twenty-th order peaks of HHG spectra have been observed in a semiconductor GaSe~\cite{Schubert2014_HHG}.
In these experiments, the electric-dipolar radiation is dominant in generating outgoing wave because the coupling between electric-field and charge is the strongest one in the light-matter interaction.
We also witness important scientific and technological advances in a different frequency region, namely, in terahertz (THz) lasers~\cite{Hirori2011_THz, Sato2013_THz, Hafez2016_THz, Dhillon2017_THz, Liu2017_THz, Mukai2016_THz, Kovalev2018_THz, Evangelos2021_THz, Leitenstorfer2023_THz}.
Strong-field THz lasers with magnitude $1-10~\mathrm{MV/cm}$ have been put into practice in the last decades.


THz laser (typically with $0.1-10$ THz frequency) has an extreme importance for its ideal matching with energy scales of elementary excitations in magnetic Mott insulators, especially, antiferromagnets.
THz electromagnetic wave can cause the resonant absorption or generation of magnetic excitations, and strong enough THz laser can even induce nonlinear spin dynamics beyond the linear response.
Hence, intense THz laser is ideal to investigate and control magnetic dynamics in magnetic Mott insulators.
There are numerous experimental studies of THz-laser driven magnetic dynamics in various magnetic Mott insulators~\cite{Kampfrath2011_mag, Kampfrath2013_mag, Vicario2013_mag, Mukai2016_THz, Lu2017_mag, Baierl2016_mag, Bossini2019_mag, Li2020_mag, Walowski2016_mag, Salen2019_mag, Safin2020_mag, Pimenov2006_mag, Takahashi2012_mag, Kubacka2014_mag, Wang2018_mag, Kirilyuk2010_mag, Yamaguchi2010_mag, Baierl2016_2_mag, Sirenko2019_mag, Mashkovich2019_mag, Mashkovich2021_mag, Behovits2023_mag, Kurihara2023_mag, Weiss2023Discovery, Zhang2023_HHG, Zhang2024Terahertz, Huang2024_HHG, Zhang2025_mag}.
Speaking of HHG, several papers reported observations of harmonic generations in antiferromagnetic insulators in the fairly recent past: up to third-order harmonic generations in a pump-probe experiment by an intense THz laser pulse~\cite{Zhang2023_HHG} and up to sixth-order harmonic generations in a double-pulse experiment~\cite{Huang2024_HHG}.
The spin-light coupling is generally much weaker than the charge-light one, but the sufficiently strong THz wave makes it possible to generate third- or sixth-order harmonics.

These experimental studies have stimulated theoretical studies of nonlinear optical responses and harmonic generations in antiferromagnetic insulators~\cite{Takayoshi2014_magthe,Takayoshi2014b_magthe,Sato2016_magthe,sato2014_magthe,
Sato2017_magthe,Mochizuki2010_magthe,Ishizuka2019_magthe,Ikeda2019_HHG,Choi2020Theory,Sato2020Twophoton,Kanega2021_HHG,Sim2023Microscopic,Potts2024Signatures,Yarmohammadi2024Terahertz,Allafi2024Spin,Watanabe2024Exploring,Udono2024_HHG} including multiferroics and quantum spin liquids.
In particular, one can find theoretical studies on HHG in quantum spin systems such as one-dimensional quantum antiferromagnets~\cite{Ikeda2019_HHG}, two-dimensional Kitaev spin-liquids~\cite{Kanega2021_HHG}, and two-dimensional Shastry-Sutherland model in a quantum disordered phase known as the dimer-singlet phase~\cite{Udono2024_HHG}, done in collaboration with our group.
Reference~\cite{Udono2024_HHG} investigated HHG driven by two-color laser pulse~\cite{Eichmann1995_2color, Hache1997_2color, Sun2010_2color, Kitagawa2012Photo, Fleischer2014_2color, Bas2015_2color, Kfir2014_2color, Baykusheva2016_2color, Kfir2016_2color, Kerbstadt2017_2color, Kerbstadt2019_2color, Ikeda2023Photocurrent, Ogawa2024_2color, Mitra2024_2color, Tyulnev2024_2color}.
The research of HHG driven by multi-color laser is growing in addition to that by the usual one-color laser.

Even at this stage, one lacks firm understanding of the relation between HHG and magnetically ordered materials. A key feature of magnetic Mott insulators is that magnetic order and symmetry breaking can be controlled by tuning temperature and external magnetic field.
We are going to show that such an ordering makes the HHG in magnetic Mott insulators richer than those in semiconductor or atomic-gas systems without any symmetry breaking.
This paper undertakes this fundamental question of how basic kinds of magnetic orders and symmetry breakings make an impact on harmonic generations in magnetic Mott insulators.
In particular, we consider the N\'eel order, the canted order, and the weak ferromagnetic (WF) order in antiferromagnets, which are representative magnetically ordered states with magnon excitations~\cite{Holstein1940Field,White2007_spinwave}
in a sub-THz or THz range.




We adopt numerical and theoretical approaches to this question.
On the one hand, we numerically solve the Landau-Lifshitz-Gilbert (LLG) equation of microscopic magnetic moments~\cite{landau1935theory,Brown1963Thermal,Garcia-Palacios1998Langevindynamics,Gilbert2004phenomenological,li2004_LLG,Li2004Thermally,lakshmanan2011_LLG,aron2014_LLG,seki2016_LLG}, faithfully taking magnetic exchange interactions into account.
The LLG equation is well known as a standard semiclassical equation that describes dynamical evolution of magnetic moments, best suited to magnetically ordered phases of our target.
On the other hand, we take advantage of a theoretical concept, dynamical symmetry~\cite{Ikeda2019_HHG, Kanega2021_HHG, Udono2024_HHG,Kanega2024_dyn, Alon1998_dyn, Liu2016_dyn, Morimoto2017_dyn,Neufeld2019_dyn, Chinzei2020_dyn, Ikeda2020_dyn, Qian2024Theory}, to establish basic understanding about the essential role of magnetic orders in magnon harmonic generation spectra.
The numerically obtained spectrum shows a selection rule, systematic disappearance of harmonic generation peaks.
The selection rule possesses rich information about interplay of underlying system and applied laser fields: magnetic orders, direction of ac magnetic field of laser fields, and spatial symmetry of laser fields.
We note that the dynamical symmetry itself has already been applied to several systems~\cite{Ikeda2019_HHG, Kanega2021_HHG, Udono2024_HHG,Kanega2024_dyn, Alon1998_dyn, Liu2016_dyn, Morimoto2017_dyn,Neufeld2019_dyn, Chinzei2020_dyn, Ikeda2020_dyn}. The point is that we investigate the dynamical symmetries and the resulting selection rules in spontaneously symmetry broken magnets. We show how a magnetic phase transition or a symmetry breaking alters the dynamical symmetry, by comparing systematically the N\'eel, canted, WF phases and non-symmetry-breaking magnets. Moreover, we show that two-color laser driven harmonic generation provides us richer information about the systems rather than the one-color laser case. In particular, we show that a special selection rule emerges in two-color laser driven U(1)-spin-rotation symmetric magnetic phases such as the N\'eel phase irrespective of the underlying lattice symmetry. We also adopt the spin-wave theory~\cite{Holstein1940Field,White2007_spinwave} to obtain microscopic insight into harmonic generation spectra in ordered phases in antiferromagnets.
Our numerical and theoretical investigations in this paper point out that the harmonic generation spectrum will provide information on characteristics of magnetic orders and (spontaneous) symmetry breaking.

We organize this paper as follows.
Section~\ref{sec:model} gives two types of the model Hamiltonians for antiferromagnetic Mott insulators.
One model shows the N\'eel and canted phases, depending on the strength of static magnetic field.
These magnetic orders spontaneously break the symmetry of Hamiltonian in different manners.
The other model shows the WF phase, where the magnetic order keeps the full symmetry of Hamiltonian. In Sec.~\ref{sec:laser}, we define the laser pulse we use in the present work. This study focuses on two sorts of laser pulses: A standard one-color laser with linear polarization and a two-color laser whose ac field draws a $C_{\ell+1}$-symmetric (multiple-leaf shape) closed trajectory ($\ell=2,3,\cdots$).
In Sec.~\ref{sec:LLG}, we explain the LLG equation, which generally describes the spin dynamics in magnetically ordered states. We will use this equation of motion to quantitatively investigate the magnon harmonic generation spectrum in antiferromagnets.
Sections~\ref{sec:MagHG} and \ref{sec:2color} are the main parts of this work, in which we comprehensively investigate the magnon harmonic generations driven by THz laser or GHz wave. We explore the harmonic generations driven by one-color laser in Sec.~\ref{sec:MagHG}.
We find the similarity and difference between the N\'eel and canted phases.
These two orders appear in the same Hamiltonian, but we find that these two systems satisfy different dynamical symmetries, and as a result, their harmonic generation spectra follow different selection rules. Namely, we see that the symmetry or symmetry breaking (phase transitions) makes the harmonic generation spectrum richer.
We also compare the spectra of the canted and WF phases. These two states have the same magnetic ordering pattern, but they follow different Hamiltonians. We find the similarity and difference between the spectra of the canted and WF phases.
We verify that the harmonic generation peaks in the WF phase are consistent with a recent experiment of $\rm HoFeO_3$ in Ref.~\cite{Zhang2023_HHG}.
In Sec.~\ref{sec:2color}, we study two-color laser driven harmonic generations and we show that multiple-color laser makes the harmonic spectra richer.
In particular, when we apply $C_\ell$-symmetric two-color laser to the N\'eel phase, we find that non-trivial dynamical symmetries and the resulting selection rules emerge.
We show that these interesting selection rules stem from two properties of the N\'eel state: (i) The N\'eel state possesses the U(1) spin-rotation symmetry and (ii) two-color laser is coupled to magnets through the Zeeman interaction between the total spin and the laser magnetic field.
This selection rule is very special for the U(1)-symmetric magnets and is independent of the underlying lattice structures. In fact, theoretical studies of graphene~\cite{Kanega2025Twocolor} and Shastry-Sutherland magnets~\cite{Udono2024_HHG} have already shown that the harmonic generation spectra in systems irradiated by $C_\ell$-symmetric two-color laser strongly depend on their lattice symmetry.
In Sec.~\ref{sec:NonPerturb}, we discuss a few non-perturbative effects of intense THz laser in the magnon harmonic generation: A power law of the radiation intensity and a red shift of the resonant peak position.
Finally, we summarize our results in Sec.~\ref{sec:summary}. We put several calculations and the detailed properties of the harmonic generation in Appendices.

\section{Model of magnetic Mott insulators}
\label{sec:model}

This section provides details of our models whose laser-driven dynamics is to be investigated in later sections.
We define two model Hamiltonians of magnetic Mott insulator on the square lattice, which show N\'eel, canted, and WF phases.
The Hamiltonians contain spin degree of freedom only.
This is because the charge degree of freedom is frozen in the Mott insulating phases.
We consider the setup that the frequency of applied THz laser and temperature are both much lower than the charge gap. Besides, the temperature needs to be sufficiently low so that the magnetic order develops well. For simplicity, we focus on the harmonic generations in two-dimensional (2D) models of antiferromagnets throughout this paper. This is reasonable because applied THz laser field is usually absorbed by atoms and molecules in the vicinity of the surface of materials.

However, not only atoms on the surface but also atoms moderately far from the surface contribute to the absorption of THz laser field in a quantitative sense.
To enhance the reliability of our results for 2D antiferromagnets, we will briefly explore quasi-2D layered antiferromagnets in Appendix~\ref{app:3d} and show that
similar magnon harmonic generations also occur
in three-dimensional (3D) models of antiferromagnets if the magnetic order in the 3D model is the same type as the 2D model.

\subsection{N\'eel and canted orders}
\label{subsec:NeelCant}

The first model Hamiltonian is
\begin{align}
  \hat{\mathcal H}_{\text{N-C}}
  &= J\sum_{\braket{\bm r,\bm r'}} \hat{\bm S}_{\bm r} \cdot \hat{\bm S}_{\bm r'}-K\sum_{\bm r} (\hat S_{\bm r}^z)^2 \nonumber\\
  &\qquad - B_0 \sum_{\bm r}\hat S_{\bm r}^z,
  \label{H_Neel-Canted_def}
\end{align}
where $\bm r= r_x a_0\bm e_x + r_y a_0\bm e_y$ specifies a site on the square lattice ($a_0$ is the lattice constant) and $\braket{\bm r,\bm r'}$ denotes a nearest-neighbor pair of sites. The coordinates $(r_x,r_y)$ are a pair of integers, and $\bm e_x$ and $\bm e_y$ are unit vectors along the $x$ and $y$ directions, respectively.
$\hat{\bm S}_{\bm r}=(\hat S_{\bm r}^x,~\hat S_{\bm r}^y,~\hat S_{\bm r}^z)^\top$ is the spin operator at the site $\bm r$. Hereafter, we will use the unit of $\hbar=1$, but we occasionally recover $\hbar$, especially when we consider the spin angular momentum.
The periodic boundary condition is imposed on the square lattice.
Parameters $J(>0)$ and $K(>0)$ represent the isotropic Heisenberg antiferromagnetic exchange interaction and the easy-axis single-ion anisotropy along the $z$ axis, respectively, and $B_0$ represents the static magnetic field.
Note that the $g$ factor and the Bohr magneton $\mu_B$ are implicitly included in the parameter $B_0$, allowing us to identify the magnetic field and the Zeeman energy for simplicity.
Throughout this paper, we define the spin operators so that the spins tend to point to the magnetic field $\bm B_0=(0,0,B_0)$ in the energetic sense [see the Zeeman term of Eq.~\eqref{H_Neel-Canted_def}].

\begin{figure}[t!]
  \centering
  \includegraphics[bb = 0 0 888 913, width=0.9\linewidth]{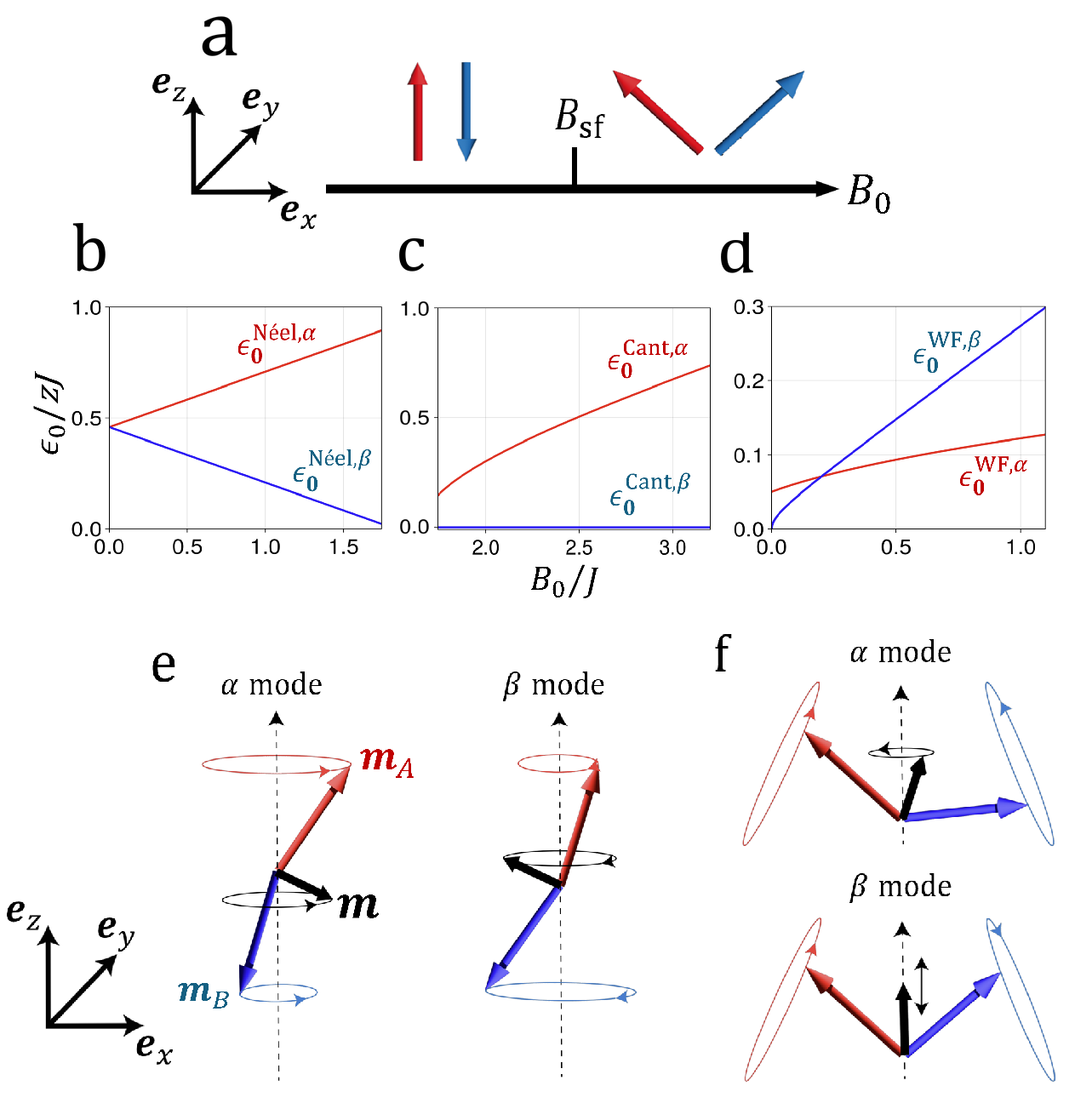}
  \caption{(a) Ground-state phase diagram of model~\eqref{H_Neel-Canted_def}.
    The N\'eel phase exists for weak magnetic field ($B_0<B_{\rm sf}$) and the canted phase appears for strong magnetic field ($B_0>B_{\rm sf}$). The point
    $B_0=B_{\rm sf}$ corresponds to the spin-flop transition, that separates the N\'eel and canted phases.
    Red and blue arrows respectively denote A- and B-sublattice magnetic moments. Vectors ${\bm e}_{x,y,z}$ are the orthonormal basis vectors of the Cartesian coordinate.
    Panels (b), (c), and (d) show the $B_0/J$-dependence of the magnon gap of two magnon bands in the N\'eel phase, the canted phase and the WF phase, respectively.
    Panels (e) and (f) schematically show dynamics of magnon with wavevector $\bm k=\bm 0$ (i.e., spin precession motion) relevant to the harmonic generation.
    The red arrow ($\bm m_A$) and the blue arrow ($\bm m_B$) represent the precessions of A- and B-sublattice spins on the nearest-neighbor sites, respectively.
    The black arrow ($\bm m$) shows the dynamics of uniform magnetization $\bm m = \frac{1}{2}(\bm m_A + \bm m_B)$.
  Panel (e) corresponds to the N\'eel phase and panel (f) corresponds to the canted and WF phases.}
  \label{fig:model}
\end{figure}

Figure~\ref{fig:model}~(a) shows the ground-state phase diagram of the model \eqref{H_Neel-Canted_def}.
The ground state has the N\'eel phase with A and B sublattices for weak magnetic field ($B_0<B_{\rm sf}$).
The N\'eel order spontaneously breaks the translation symmetry of the model \eqref{H_Neel-Canted_def}, leading to the double degeneracy of the ground state and the finite excitation gap.
Reflecting the two-sublattice structure of the N\'eel order along the $z$ axis, there are two kinds of magnons.
Let us call them $\alpha$ mode and $\beta$ mode, in this paper.
The energy bands $\epsilon_{\bm k}^{\text{N\'eel},\alpha}$ and $\epsilon_{\bm k}^{\text{N\'eel},\beta}$ of $\alpha$ and $\beta$ modes have a minimum at $\bm k = \bm 0$ (see Appendix~\ref{app:LSW}).
This magnon gap, an excitation energy gap to create a single magnon, is finite in the presence of an easy-axis magnetic anisotropy $K$.
Figure~\ref{fig:model}~(b) shows the $B_0/J$ dependence of magnon gaps $\epsilon_{\bm k=\bm 0}^{\text{N\'eel},\alpha}$ and $\epsilon_{\bm k=\bm 0}^{\text{N\'eel},\beta}$ [Eqs.~\eqref{magnon_band_Neel_alpha} and \eqref{magnon_band_Neel_beta}].
Figure~\ref{fig:model}~(e) gives a rough sketch of the $\alpha$ and $\beta$ modes in the N\'eel phase.
We can distinguish the $\alpha$ and $\beta$ modes from the viewpoint of the precession of the total magnetic moment.
The precessing directions in these modes are opposite.

The increase of $B_0/J$ reduces the lowest magnon gap $\epsilon_{\bm 0}^{\text{N\'eel},\beta}$.
The $\beta$-mode magnon gets softened at $B_0=B_{\rm sf}$, that is, it has an infinitesimal excitation energy there.
The system undergoes a spin-flop transition at $B_0=B_{\rm sf}$ from the N\'eel phase ($B_0<B_{\rm sf}$) to the canted phase ($B_0>B_{\rm sf}$).
Figure~\ref{fig:model}~(c) shows magnon gaps at $\bm k = \bm 0$ in the canted phases.
The canted phase also hosts the two magnon modes with the same label, $\alpha$ and $\beta$.
The $\beta$ mode in the canted phase remains gapless at $\bm k = \bm 0$, that is, $\epsilon_{\bm 0}^{\text{Cant},\beta}=0$ as shown in Fig.~\ref{fig:model} (c).
This gapless feature comes out of generic Nambu-Goldstone (NG) theorem about spontaneous breaking of continuous symmetry~\cite{Nambu1960QuasiParticles,Goldstone1962Broken,Watanabe2012Unified,Hidaka2013Counting}.
In the canted phase, the U(1) rotation symmetry around the $S^z$ axis (i.e., a continuous symmetry) spontaneously breaks and the gapless $\beta$ mode can be viewed as the NG mode.

Figure~\ref{fig:model}~(f) gives a rough sketch of the $\alpha$ and $\beta$ modes in the canted phase.
The $\alpha$ mode involves the precession of the total magnetic moment.
By contrast, the total magnetic moment in the $\beta$ mode is oscillating along the $z$ axis but not precessing.

Finally, we note the temperature effect in the canted phase. We will consider a low but finite temperature setup of the canted phase to compute the harmonic generation. However, Mermin-Wagner-Hohenberg theorem~\cite{Mermin1966Absence,Hohenberg1967Existence}
tells us that any spontaneous breaking of continuous symmetry (including the $\rm U(1)$ symmetry breaking in the canted phase) cannot occur at finite temperatures in two-dimensional systems. Therefore, strictly speaking, the canted phase in our two-dimensional model disappears once we introduce a finite temperature. We will introduce a very weak easy-plane anisotropy $+D_y\sum_{\bm r}(\hat S_{\bm r}^y)^2$ with $D_y/J=0.01$ so that the canted spin order in the $S^x$-$S^z$ plane survives. We verify in Appendix~\ref{app:anistropy_cant} that the above weak anisotropy does not affect the spectra of the harmonic generation when the laser frequency is close to the gapped ($\alpha$) mode. In real materials, the canted phase is protected by such a weak anisotropy or three dimensionality due to weak inter-plane interactions.

\subsection{Weak-ferromagnetic order}


Here we introduce another square-lattice model that shows the WF phase.
The model has the Hamiltonian,
\begin{align}
  \hat{\mathcal H}_{\rm WF} &= J \sum_{\braket{\bm r,\bm r'}} \hat{\bm S}_{\bm r} \cdot \hat{\bm S}_{\bm r'} - B_0\sum_{\bm r}\hat S_{\bm r}^z \nonumber\\
  &\quad - D\sum_{\bm r} \sum_{\bm \rho=\bm e_x,\bm e_y} (-1)^{(r_x+r_y)} (\hat S_{\bm r}^z \hat S_{\bm r+\bm \rho}^x -\hat S_{\bm r}^x \hat S_{\bm r+\bm \rho}^z),
  \label{H_WF_def}
\end{align}
with the nearest-neighbor Heisenberg antiferromagnetic exchange interaction (first term), the Zeeman energy in the $z$ direction (second term), and the staggered Dzyaloshinskii-Moriya (DM) interaction (third term).
The site $\bm r$ on the square lattice is represented as $\bm r= r_xa_0 \bm e_x+r_ya_0 \bm e_y$, where $r_x$ and $r_y$ are integers.
The DM interaction may be rewritten as $-(-1)^{(r_x+r_y)}\bm D\cdot (\hat {\bm S}_{\bm r}\times\hat {\bm S}_{\bm r+\bm \rho})$ with the DM vector $\bm D=(0,D,0)$.
Since the direction of the DM interaction is alternating along the $x$ and $y$ directions, we call it the staggered DM interaction.

The model \eqref{H_WF_def} has the WF state as the ground state.
The spin configuration in the WF phase is similar to that in the canted phase, and the precession patterns of two magnon modes in the WF phase are also like those of the canted phase, as shown in the panel (f) of Fig.~\ref{fig:model}.
However, the magnon gap shows a different $B_0/J$ dependence.
Figure~\ref{fig:model}~(d) shows the $B_0/J$ dependence of the two magnon gaps, $\epsilon_{\bm 0}^{\mathrm{WF},\alpha}$ and $\epsilon_{\bm 0}^{\mathrm{WF},\beta}$, in the WF phase.
Whereas the canted phase has the gapless mode ($\epsilon_{\bm 0}^{\mathrm{Cant}, \beta}=0$) for $B_0>B_{\rm sf}$, the WF phase has no gapless mode for $B_0/J>0$.
The WF phase of the model \eqref{H_WF_def} has the gapless mode only for $B_0=0$ because the model has the U(1) symmetry around the $y$ axis for $B_0=0$ but not for $B_0>0$.
The magnetic field $B_0$ breaks the U(1) symmetry around the $y$ axis and thus yields the nonzero magnon gap in the WF phase.
This paper compares the harmonic generation in the canted and WF phases to clarify the fact that the harmonic generation spectrum is determined not only by the symmetry of the ground state but also by that of the Hamiltonian.

\section{Laser pulse}
\label{sec:laser}

\begin{figure}[t!]
  \centering
  \includegraphics[bb = 0 0 1636 477, width=0.95\linewidth]{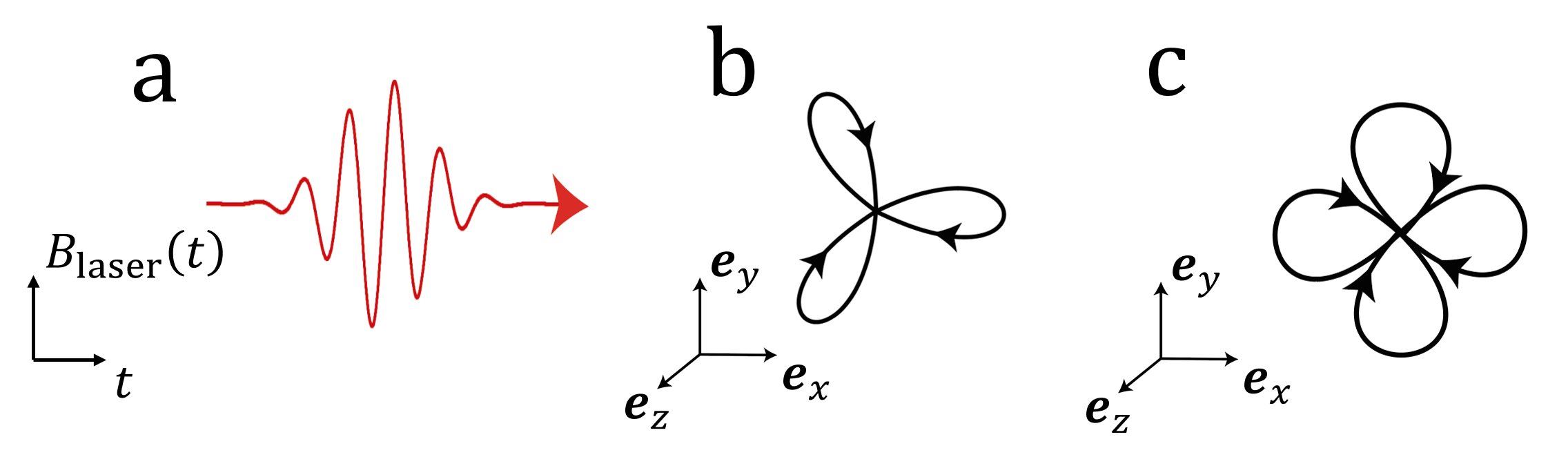}
  \caption{
    (a) Schematic figure of laser pulse field $\bm B_{\rm laser}(t)$ as a function of time $t$.
    Panels (b) and (c) depict the trajectories drawn by the arrowhead of the ac field $[\bm b_{-1}(t)+\bm b_\ell(t)]/\sqrt 2$ in two-color laser pulse $\bm B_{2,\ell}(t)$ [Eq.~\eqref{B_2_def}] with (b) $\ell=2$ and (c) $\ell=3$ for single period $0\le t-t_0 \le 2\pi/\Omega$. We consider the setup that the ac magnetic field $\bm B_{2,\ell}(t)$ is in the $x$-$y$ plane.
    The arrowhead depicts the direction of change in time.
  }
  \label{fig:pulse}
\end{figure}

The previous section defined two models of magnetic Mott insulators.
Here, we take into account the laser pulse and introduce it into our model to discuss the harmonic generation.
As we mentioned, we assume that both the temperature and the laser frequency need to be lower than the Mott gap.
This condition is required to drop the ac electric field of the laser pulse.
Let $\bm B_{\rm ac}(t)$ be the ac magnetic field of the laser pulse.
The magnetic Mott insulator interacts with the temporally oscillating external field $\bm B_{\rm ac}(t)$ through the Zeeman interaction.
Namely, the full Hamiltonian $\hat{\mathcal H}(t)$ incorporates the magnetic interactions and the laser fields as follows.
\begin{align}
  \hat{\mathcal H}(t) &= \hat{\mathcal H}_{\rm mag} +\hat{\mathcal H}_{\rm ext}(t),
  \label{full_Hamiltonian_gen_def} \\
  \hat{\mathcal H}_{\rm ext} (t) &= - \bm B_{\rm ac}(t) \cdot \sum_{\bm r} \hat{\bm S}_{\bm r},
  \label{H_ext_gen_def}
\end{align}
Here, $\hat{\mathcal H}_{\rm mag}$ is the Hamiltonian of the magnetic Mott insulator in the absence of the laser pulse, such as $\hat{\mathcal H}_{\text{N-C}}$ and $\hat{\mathcal H}_{\rm WF}$.
As the ac Zeeman energy \eqref{H_ext_gen_def} shows, we consider the laser pulse as a plane wave coupled to the total spin operator, $\sum_{\bm r} \hat{\bm S}_{\bm r}$.
The electromagnetic wave including the laser pulse with the THz frequency has a wavelength much longer than the typical length scale (lattice spacing) of crystals, and is thus deemed a plane wave when applied to magnetic Mott insulators.

We adopt two kinds of external ac magnetic field due to the laser $\bm B_{\rm ac}(t)$: one- and two-color laser pulses, which we denote $\bm B_1(t)$ and $\bm B_{2,\ell}(t)$, respectively.
Both $\bm B_1(t)$ and $\bm B_{2,\ell}(t)$ are laser-pulse fields with a Gaussian envelope.
\begin{align}
  \bm B_1(t) &= \bm B_{\rm ac} \exp\biggl( - \frac{(t-t_0)^2}{2\sigma^2} \biggr) \cos [\Omega(t-t_0)],
  \label{B_1_def} \\
  \bm B_{2,\ell}(t) &= \frac{\bm b_{-1}(t)+\bm b_\ell(t)}{\sqrt 2}\exp\biggl( -\frac{(t-t_0)^2}{2\sigma^2}\biggr),
  \label{B_2_def} \\
  \bm b_{\ell}(t) &= B_{\rm ac} \Bigl(\cos[\ell \Omega(t-t_0)],~-\sin[\ell \Omega(t-t_0)],~0\Bigr),
  \label{B_ell_def}
\end{align}
where $\Omega$, $\sigma$, and $B_{\rm ac}$ denote the laser frequency, the pulse width, and  the field strength at $t=t_0$ [Fig.~\ref{fig:pulse}~(a)].
The dimensionless parameter $\ell$ takes a positive integer $\ell=1,~2,~3,\cdots$ and defines the spatial symmetry of the two-color laser pulse [Fig.~\ref{fig:pulse}~(b) for $\ell=2$]. In the case of a monochromatic laser, the polarization direction of the laser magnetic field $\bm B_{\rm ac}$ is considered to be in all directions: $x$, $y$, and $z$.
On the other hand, the ac field $\bm B_{2,\ell}(t)$ of two-color laser is chosen to be in the $x$-$y$ plane, i.e., perpendicular to the dc magnetic field $\bm B_0$.

\begin{table}
  \caption{\label{tab:laser}
    Typical THz-laser intensities to observe magnetic harmonic generation~\cite{Ikeda2019_HHG,Sato2021Floquet}.
  In this paper, we will often set the strength of ac magnetic field $B_{\rm ac}=0.1J$, in which $J$ is the strength of exchange interaction of antiferromagnets. Typical range of $J$ is $1\lesssim J/k_{\rm B}\lesssim 100$. As one will see soon, $B_{\rm ac}=0.1J$ is strong enough to observe second, and third magnon harmonic generations. The symbol $c$ is the speed of light.}
  \begin{ruledtabular}
    \begin{tabular}{lll}
      Fields & $J/k_{\rm B}=10$[K] & $J/k_{\rm B}=100$[K]\\
      \hline
      ac magnetic field $B_{\rm ac}=0.1 J$ & $0.744 $[T] & $7.44$ [T] \\
      ac electric field $E_{\rm ac}=c B_{\rm ac}$ & $2.23$ [MV/cm] & $22.3$ [MV/cm]
    \end{tabular}
  \end{ruledtabular}
\end{table}

The pulse width $\sigma$ is fixed to $20\hbar/J$ in the following numerical calculations. In the mathematical sense, the amplitude of Gaussian pulse is always finite in all range of time $-\infty<t-t_0<\infty$.
However, in our numerical calculations, we apply the laser pulse in a finite time window $t_{\rm i}<t-t_0<t_{\rm end}$, in which the radiation time and the ending time are chosen to be $t_{\rm i} = -6\sigma$ and $t_{\rm end} = 9\sigma$. In fact, the laser amplitudes of $\bm B_1(t)$ and $\bm B_{2,\ell}(t)$ are negligibly small outside the above time window. We prepare the thermal equilibrium state of $\hat{\mathcal H}_{\rm mag}$ at $t-t_0=t_{\rm i}$.


While Eq.~\eqref{B_1_def} represents the linearly polarized single-color laser pulse, Eq.~\eqref{B_2_def} represents the two-color laser pulse with a more complex spatial pattern.
Figure~\ref{fig:pulse}~(b) and (c) show trajectories of $[\bm b_{-1}(t)+\bm b_{\ell}(t)]/\sqrt 2$ on the $xy$ plane for $\ell=2$ and $\ell=3$, respectively.
The trajectory for $\ell=2$ has the $C_3$ symmetry, namely, is identical under the $2\pi/3$ rotation around the origin of the $xy$ plane.
Generically, $\bm B_{2,\ell}(t)$ shows the $C_{\ell+1}$ symmetry for $\ell=2,~3,~\cdots$.

Finally, we comment on the intensity of THz laser. If we use current experimental techniques of THz laser~\cite{Hirori2011_THz,Liu2017_THz}, its ac electric field strength can exceed $\sim 1$ MV/cm, which corresponds to the ac magnetic field with $\sim 0.3$ T. On the other hand,
the typical value of the exchange coupling in antiferromagnetic insulators is around $J/k_{\rm B}= 1-100$ K.
Table~\ref{tab:laser} indicates that for standard antiferromagnets, the ac magnetic field with $B_{\rm ac}=0.1 J$ can be created by using the current laser techniques. As one will see soon,
we numerically show that the ac magnetic field with $B_{\rm ac}=0.1 J$ is strong enough to observe lower-order (second, third, etc.) magnon harmonic generations in antiferromagnets. Besides the harmonic generation, we should consider the possibility that the strong THz laser with $B_{\rm ac}=0.1 J$ violates the initial magnetic order at $t=t_{\rm i}$ in antiferromagnets.
If such a violation occurs, the magnetic order would change into a different order (ultrafast spin switching). However, such a switching phenomenon is out of the scope of the present study. In Appendix~\ref{app:amplitude}, we verify that the initial magnetic order is not broken by the THz field, and briefly discuss the other possibility of laser-driven finite-wavevector instability.

\section{Landau-Lifshitz-Gilbert equation}
\label{sec:LLG}

Let us briefly review the LLG equation that our numerical calculations of spin dynamics are based on.
While the spin has a quantum origin, it can be seen as a classical vector with a fixed length when the system belongs to a long-range magnetically ordered phase.
We regard $\hat{\bm S}_{\bm r}$ at each site $\bm r$ as a classical vector $\hbar\bm m_{\bm r}$ with a fixed length, $\hbar m$, and investigate the spin dynamics based on the following stochastic LLG equation.
\begin{align}
  \frac{d\bm m_{\bm r}}{dt} = \frac{1}{\hbar}\bm m_{\bm r} \times \biggl( \frac{\partial \mathcal H(t)}{\partial \bm m_{\bm r}} + \bm \xi_{\bm r}(t) \biggr) + \frac{\alpha}{m} \bm m_{\bm r} \times \frac{d\bm m_{\bm r}}{dt},
  \label{stochastic-LLG_def}
\end{align}
where $\bm \xi_{\bm r}(t)$ is a white-noise random field that represents the thermal fluctuation of $\bm m_{\bm r}$ at time $t$.
The positive parameter $\alpha$ is called Gilbert damping coefficient.
Note that $\mathcal H(t)$ is a classical counterpart of Eq.~\eqref{full_Hamiltonian_gen_def}, where every $\hat{\bm S}_{\bm r}$ is replaced by $\bm m_{\bm r}$.
The random field $\bm \xi_{\bm r}(t)$ satisfies the following relations about the average and correlation.
\begin{align}
  \braket{\bm \xi_{\bm r}(t)} &= 0,
  \label{xi_average} \\
  \braket{\xi_{\bm r}^a(t)\xi_{\bm r'}^b(t')} &= \frac{2\hbar\alpha k_BT_{\bm r}}{m} \delta_{\bm r,\bm r'} \delta_{a,b} \delta(t-t'),
  \label{xi_correlation}
\end{align}
where $T_{\bm r}$ is the temperature at the site $\bm r$ and $a,b = x,y,z$.
We use the following LLG equation, equivalent to Eq.~\eqref{stochastic-LLG_def} but represented with dimensionless operators and parameters, in our numerical simulations
\begin{align}
  \frac{d\bm m_{\bm r}}{d\tau}
  &= \frac{1}{1+\alpha^2}\biggl[
    \bm m_{\bm r} \times \biggl( \frac{\partial \tilde{\mathcal H}(\tau)}{\partial \bm m_{\bm r}} +\tilde{\bm \xi}_{\bm r}(\tau)
    \biggr) \nonumber\\
    &\quad + \frac{\alpha}{m}\bm m_{\bm r} \times \biggl\{\bm m_{\bm r} \times \biggl(\frac{\partial\tilde{\mathcal H}(\tau)}{\partial \bm m_{\bm r}} + \tilde{\bm \xi}_{\bm r}(\tau)
      \biggr)
    \biggr\}
  \biggr],
  \label{stochastic-LLG_dimensionless}
\end{align}
where $\tau = tJ/\hbar$, $\tilde{\mathcal H}={\mathcal H}(t)/J$, $\tilde{\bm \xi}_{\bm r}(\tau) = \bm \xi_{\bm r}(t)/J$.
In our numerical simulations, we set $k_BT/J=10^{-3}$ so that the magnetic orders in the N\'eel, canted, and WF phases survive. The Gilbert damping constant is chosen to be $\alpha=0.01$, which is a typical value of ordered magnets.
Note that we need to take the ensemble average of $N_{\rm av}=1000$ results to reduce the standard deviation due to the random field $\tilde{\bm \xi}_{\bm r}(\tau)$. Throughout the paper, we set the length of magnetic moment $m=1$.

Let us comment on some details on numerical calculations: system size and numerical methods.
We hereafter show numerical results performed on the system with $10\times 10$ sites.
The system size $10\times 10$ turns out to be large enough as a result of comparisons with harmonic spectra in further larger systems (see Appendix~\ref{app:size}).
To obtain harmonic generation spectra, we adopted two numerical techniques: the Heun method~\cite{Heun1900}, implemented in \texttt{DifferentialEquations.jl}~\cite{Rackauckas2017DifferentialEquationsjl}, to solve the LLG equation \eqref{stochastic-LLG_dimensionless}, and the fast Fourier transform (FFT), computed via \texttt{FFTW.jl} (FFTW3)~\cite{FrigoJohnson2005FFTW3}, to turn $\bm m(t)$ into $\tilde{\bm m}(\omega)$.

We prepare the thermal equilibrium state by evolving the system with the LLG equation \eqref{stochastic-LLG_def} with ${\mathcal H}(t) = {\mathcal H}_{\rm mag}$.
The thermal equilibrium state is ready at clock time $t-t_0=t_{\rm i}$.
We start the irradiation of $\bm B_{\rm laser}(t)$ at $t-t_0=t_{\rm i}$ and make the system develop with the LLG equation with ${\mathcal H}(t) = {\mathcal H}_{\rm mag} + {\mathcal H}_{\rm ext}(t)$. As already mentioned, we will take $t_0 = t_{\rm i} + 6\sigma$ so that $|\bm B_{\rm laser}(t_{\rm i})|$ is negligibly small.

\section{Magnon harmonic generation}
\label{sec:MagHG}

Using the normalized stochastic LLG equation \eqref{stochastic-LLG_dimensionless} for the two models \eqref{H_Neel-Canted_def} and \eqref{H_WF_def}, we numerically obtain the real-time dynamics of the magnetic moment $\bm m_{\bm r}$ at each site $\bm r$.
Since the laser-pulse field $\bm B_{\rm ac}(t)$ is coupled to the uniform (total) magnetic moment, as Eq.~\eqref{H_ext_gen_def} shows, we focus on the real-time dynamics of the total magnetic moment,
\begin{align}
  m^a(t) = \frac 1N \sum_{\bm r} m_{\bm r}^a(t), \qquad (a=x,~y,~z),
  \label{ma_tot_def}
\end{align}
and its Fourier transform,
\begin{align}
  \tilde M^a(\omega) = \int_{-\infty}^{\infty} dt \,e^{i\omega t} m^a(t),
  \label{tilde_m_def}
\end{align}
with the total number of sites, $N$.
In practical numerical calculations, we employ the following dimensionless version in a finite time window,
\begin{align}
  \tilde m^a(\omega)  = \frac{J}{\hbar}\int_{-6\sigma}^{9\sigma} dt\, e^{i\omega t} m^a(t),
  \label{tilde_m_practical}
\end{align}
where we have set $t_0=0$.
Using this definition of the Fourier transform with a sufficiently long time window, we can obtain the reliable harmonic generation spectra.

Here, we comment on the relation between magnetic harmonic generation and $\tilde m^a(\omega)$. In THz-wave driven magnetic Mott insulators, the dominant radiation is expected to stem from the magnetic dipolar radiation process~\cite{Jackson2009Classical}. Its intensity $I(\omega)$ is given by
\begin{align}
  I(\omega)\propto |\omega^2 \tilde m^a(\omega)|^2.
  \label{eq:mag_dipole}
\end{align}
On the other hand, several recent experimental studies have indirectly observed the real-time evolution of $m^a(t)$ and its Fourier transform $\tilde m^a(\omega)$, by using magneto-optical phenomena such as Kerr and Faraday effects~\cite{Zhang2023_HHG, Zhang2024Terahertz, Huang2024_HHG, Zhang2025_mag}. Therefore, we will compute the spectra of $|\tilde m^a(\omega)|$ instead of $I(\omega)$ throughout this paper.
To enhance the reliability of our numerical computation, we show the time-step and time-window length dependence of the harmonic generation spectra $|\tilde m^a(\omega)|$ estimated by numerically solving the LLG equation in Appendixes~\ref{app:timestep} and \ref{app:timewindow}.

We note that electric dipolar radiation is usually dominant in HHG of solid electron systems instead of magnetic dipolar radiation. However, motion of electrons is almost frozen in Mott insulators and therefore magnetic dipolar radiation can become the dominant radiation process there. Moreover, if the frequency of applied THz laser is much smaller than Mott gap, the radiation processes driven by electrons' motion are considerably negligible. In such a case, magnetic dipolar radiation is expected to be the leading radiation process.

In what follows, we show numerical results on the spectra $|\tilde m^a(\omega)|$ in the N\'eel, canted, and WF phases, where we see how a symmetry and its breaking determine the dynamical symmetry and, ultimately, the harmonic generation spectra. This section is mainly devoted to our results of one-color laser driven harmonic generations.

From Sec.~\ref{sec:Neel}, we will give numerical results of $m^\alpha(t)$ and ${\tilde m}^\alpha(\omega)$ in different phases at finite temperatures. In figures showing such numerical results, we ignore the symbol of ensemble average $\langle\cdots\rangle$ for the sake of simplicity; for instance, $|{\tilde m}^\alpha(\omega)|$ in the figures stands for $|\langle{\tilde m}^\alpha(\omega)\rangle|$.

\subsection{Dynamical symmetry in general}
\label{sec:DS_in_general}

Before going into specific cases, we describe how the following argument on dynamical symmetry works.
The dynamical symmetry refers to a symmetry that incorporates the time translation and an ordinary symmetry operation such as a global spin rotation.
The dynamical symmetry holds when the externally applied ac field has an approximate or exact periodicity in time.
Suppose that the ac magnetic field $\bm B_{\rm ac}(t)$ is periodic in time, $\bm B_{\rm ac}(t+T_{\rm ac})=\bm B_{\rm ac}(t)$, where $T_{\rm ac}=2\pi/\Omega$ is the laser period.
Then, the Hamiltonian \eqref{full_Hamiltonian_gen_def} trivially satisfies $\hat{\mathcal H}(t+T_{\rm ac}) = \hat{\mathcal H}(t)$.
The Hamiltonian \eqref{full_Hamiltonian_gen_def} can have another relation
\begin{align}
  \hat U\hat{\mathcal H}(t) \hat U^\dag = \hat{\mathcal H}(t+\kappa T_{\rm ac}),
  \label{dyn_sym_gen_def}
\end{align}
with a unitary or antiunitary operator $\hat U$.
The parameter $0<\kappa < 1$ depends on details of $\hat U$ and $\hat{\mathcal H}(t)$. When the Hamiltonian satisfies Eq.~\eqref{dyn_sym_gen_def}, we say that the time-periodic system possesses the dynamical symmetry.
A dynamical symmetry is characterized by the symmetry operator $\hat U$ and the time shift $\kappa T_{\rm ac}$. Therefore, we will label each dynamic symmetry with the following equation
\begin{align}
  (\hat U, \,\,\kappa T_{\rm ac}).
  \label{eq:dynsym_symbol}
\end{align}

In our spin systems irradiated by THz laser, (we will show later) some sorts of dynamical symmetries~\eqref{dyn_sym_gen_def} can emerge since the spin-light coupling term~\eqref{H_ext_gen_def} between the ac field $\bm B_{\rm ac}(t)$ and the spin operator $\hat{\bm S}_{\bm r}$ relates the time translation and the ordinary symmetry operation.
For instance, we find the dynamical symmetry where $\kappa = 1/2$ and $\hat U$ is the $\pi$ spin rotation around the $S^z$ axis in the N\'eel phase.

When we focus on a dynamically symmetric time-periodic system, we can often encounter a selection rule of its harmonic generation spectrum. To find such a selection rule, one needs to consider the transformation property of global observables (such as total magnetization) under the symmetry operation $\hat U$. The transformation is usually described as
\begin{align}
  \hat U\hat{A}(t) \hat U^\dagger = f(\hat{A}(t), \hat{B}(t),\cdots),
  \label{dyn_sym_phys}
\end{align}
where $\hat A(t)$ is an observable of the system and $f(\hat A,\hat B, \cdots)$ is a function of operators $\hat A$, $\hat B$, $\cdots$. Combining Eqs.~\eqref{dyn_sym_gen_def} and \eqref{dyn_sym_phys} leads to a relationship between two expectation values of $\hat A$ at different times $t$ and $t+\kappa T_{\rm ac}$. The relation directly gives a selection rule for the HHG spectrum of $\hat{A}(t)$. In this paper, we will show several concrete examples of dynamical symmetries and selection rules for magnetization $m^\alpha(t)$.

In the above argument around Eqs.~\eqref{dyn_sym_gen_def} and \eqref{dyn_sym_phys}, we assume that the thermal state (including the ground state) of the system before laser application possesses all the symmetries in the static part of the Hamiltonian. However, in the case of magnetic Mott insulators, their static symmetry can often be broken or recovered by tuning temperature and static magnetic field. This nature is quite different from most of semiconductor systems. If such a symmetry breaking occurs, we have to consider not only observables but also the quantum state (the wave function)~\cite{Ikeda2019_HHG,Kanega2021_HHG} to find the dynamical symmetry and the selection rule. We discuss the more detailed analysis and how to derive the selection rule in Appendix~\ref{app:LLG}.

Lastly, we make two comments on dynamical symmetry.
The first comment is associated with time periodicity.
The relation \eqref{dyn_sym_gen_def} stands on the periodicity of the ac field.
The ac magnetic fields \eqref{B_1_def} and \eqref{B_2_def} are oscillating in time but have no complete periodicity because of the Gaussian envelope.
Nevertheless, the above argument of the dynamical symmetry is approximately applicable to our setup.
In the time window $|t-t_0|\lesssim \sigma$, we may approximate $\exp[-(t-t_0)^2/2\sigma^2] \sim 1$ and regard $\bm B_1(t)$ and $\bm B_{2,\ell}(t)$ as periodically oscillating fields with $T_{\rm ac} = 2\pi/\Omega$.
In this sense, the model~\eqref{full_Hamiltonian_gen_def} approximately has the dynamical symmetry.
Laser pulses of $\bm B_1(t)$ and $\bm B_{2,\ell}(t)$ consist of mixed waves with different frequencies around $\Omega$ (i.e., broad-band type), and their width in the frequency space is about $\sim 1/\sigma$. However, an essential point of HHG experiments is whether neighboring peaks ($n$-th and $(n+1)$-th order harmonic peaks) can be distinguished.
Therefore, under the condition of $1/\sigma<\Omega$, the conclusion derived by the dynamical symmetry is expected to be applied to HHG spectra, especially, the peak structures. We show in the remainder of the paper that the approximate dynamical symmetry indeed governs the harmonic generation spectrum. In Appendix~\ref{app:pulselength}, we numerically check the above argument and show the harmonic generation spectrum of the N\'eel phase by changing the laser pulse width.

Secondly, we comment on the time-evolution rule of our  ordered antiferromagnet models. As we mentioned in Sec.~\ref{sec:LLG}, the time evolution of spins (more correctly, the expectation value of spins) can be well described by the LLG equation in the N\'eel, canted, and WF phases instead of Schr\"odinger equation. Namely, spins can be approximated by classical vectors in these ordered phases.
In the above paragraphs, on the other hand, we defined the dynamical symmetry under the assumption that the time evolution of observables follows Schr\"odinger equation (i.e., quantum mechanics). In the following subsections, we will mainly consider several dynamical symmetries based on the LLG equation and the ``classical'' magnetic moment $m^\alpha(t)$ because we will numerically apply the LLG equation to compute the magnon harmonic generation.
In Appendix \ref{app:LLG}, details of dynamical symmetries based on both quantum mechanics and LLG equation are discussed. We find that at least in our models of antiferromagnets, we can derive the same selection rule in both a quantum spin model and its classical version based on the LLG equation.

\subsection{N\'eel phase}
\label{sec:Neel}

\begin{figure*}[t!]
  \centering
  \includegraphics[bb = 0 0 1983 1549, width=0.9\linewidth]{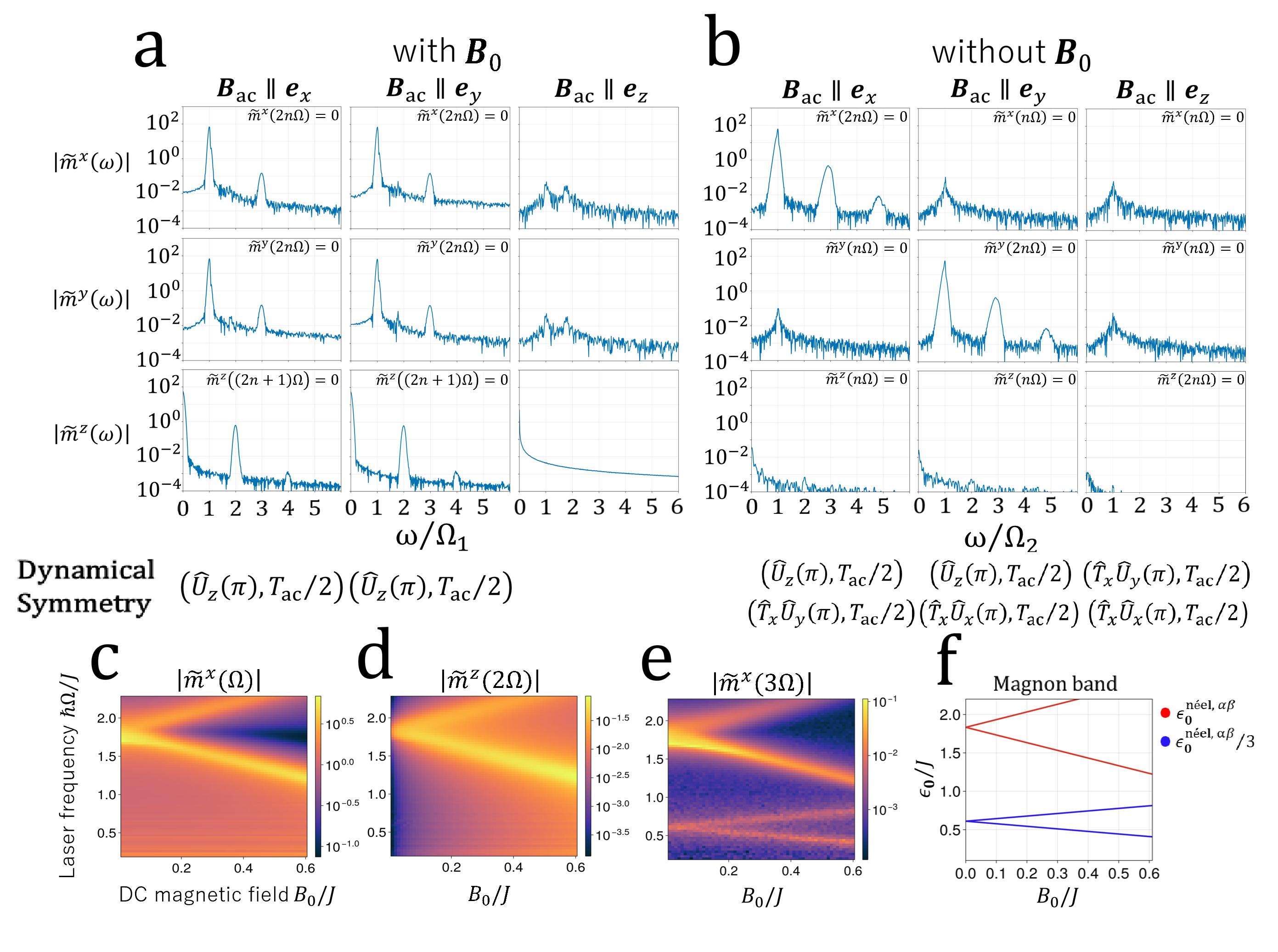}
  \caption{Numerical result of harmonic generation in N\'eel phase at $k_{\rm B}T/J=10^{-3}$.
    Panels (a), (b) show plots of the amplitude of the Fourier transform $\tilde{\bm m}(\omega) = (\tilde m^x(\omega),~\tilde m^y(\omega),~\tilde m^z(\omega))$ of the model \eqref{H_Neel-Canted_def} in the N\'eel phase vs. the normalized frequency $\omega/\Omega$. The static magnetic field $B_0$ is on ($B_0/J=0.5$) in the panel (a) and off ($B_0/J=0.0$) in the panel (b). We set $B_{\rm ac}/J = 0.1$ in both panels. The laser frequency $\Omega=2\pi/T_{\rm ac}$ is set to the resonant frequency of magnon: $\Omega=\Omega_1$ ($\Omega=\Omega_2$) in the presence (absence, respectively) of the magnetic field.
    $\Omega_1$ and $\Omega_2$ are given by $\hbar\Omega_{1} = \epsilon_{\bm{0}}^{\text{N\'eel}, \beta} = 1.33J$ and $\hbar\Omega_{2} = \epsilon_{\bm{0}}^{\text{N\'eel}, \beta} = 1.83J$.
    Each panel consists of $3\times 3$ plots.
    The first, second, and third rows show spectra $|\tilde m^a(\omega)|$ for $a=x,~y,$ and $z$, respectively.
    The first, second, and third columns show spectra when the applied laser pulse is parallel to the $x,~y,$ and $z$ directions, respectively.
    The symmetry operation below each column  [such as $(\hat U_z(\pi),T_{\rm ac}/2)$] denotes the dynamical symmetry in each setup.
    Panels (c), (d), and (e) show the first ($|\tilde m^x(\Omega)|$), second ($|\tilde m^z(2\Omega)|$), and third harmonic generations ($|\tilde m^x(3\Omega)|$), respectively, when $\bm B_{\rm ac}$ is parallel to the $x$ axis.
    Panel (f) shows magnon bands $\epsilon_{\bm{0}}^{\text{N\'eel}, \alpha}$ and $\epsilon_{\bm{0}}^{\text{N\'eel}, \beta}$ (red curves), and their one-third scales (blue curves).
    The latter is plotted for the purpose of comparison with the panel (e). We note that $|{\tilde m}^\alpha(\omega)|$ means $|\langle{\tilde m}^\alpha(\omega)\rangle|$.
  }
  \label{fig:neelhhg}
\end{figure*}

In this subsection, we discuss the numerical results of the magnon harmonic generation driven by one-color laser pulse in the N\'eel phase.
Figure~\ref{fig:neelhhg}~(a) and (b) show the typical spectra of $|\tilde m^a(\omega)|$ in the N\'eel phase.
We show $3\times 3$ patterns of $|\tilde m^a(\omega)|$ with $a=x,~y,$ and $z$ under $\bm B_{\rm ac}\parallel \bm e_x,~\bm e_y,$ and $\bm e_z$.
Figure~\ref{fig:neelhhg}~(a) shows results in the presence of the static magnetic field ($B_0/J=0.5$) and Fig.~\ref{fig:neelhhg}~(b) shows those in the absence of the static magnetic field ($B_0/J=0.0$).
We normalize the frequency $\omega$ with the laser frequency $\Omega$, and it is tuned to the lower magnetic resonance point: $\Omega=\epsilon_{\bm 0}^{\text{N\'eel},\beta}$.

Let us see the first column where the laser field is parallel to the $x$ axis ($\bm B_{\rm ac} \parallel \bm e_x$).
In the presence of the static magnetic field [the first column of Fig.~\ref{fig:neelhhg}~(a)], the spectra $|\tilde m^x(\omega)|$ and $|\tilde m^y(\omega)|$ show clear fundamental and third-harmonic peaks ($\omega/\Omega_1=1$ and $3$) but no second-harmonic peak ($\omega/\Omega_1=2$).
We can generalize this observation to the following statement.
$|\tilde m^x(\omega)|$ and $|\tilde m^y(\omega)|$ show no even-harmonic peaks ($\omega/\Omega_1=2n$ with $n=1,~2,~3,~\cdots$).




\subsubsection{Laser field along \texorpdfstring{$x$}{x} axis}

Let us investigate and compare harmonic generation spectra in the absence or presence of the static magnetic field, closely from the viewpoint of dynamical symmetry.
When the static magnetic field is present and the laser field is parallel to the $x$ axis, the even-order harmonics are absent in the spectra $|\tilde m^x(\omega)|$ and $|\tilde m^y(\omega)|$ because the following dynamical symmetry forbids the even-order harmonics with $\omega/\Omega_1=2n$ ($n\in \mathbb Z$).
This dynamical symmetry is a combination of a time translation by $T_{\rm ac}/2$ and a spin rotation by angle $\pi$ around the $S^z$ axis, $\hat U_z(\pi)$.
The operator $\hat U_z(\pi)$ rotates the spin, $(\hat S_{\bm r}^x,~\hat S_{\bm r}^y,~\hat S_{\bm r}^z) \to (-\hat S_{\bm r}^x,~-\hat S_{\bm r}^y,~\hat S_{\bm r}^z)$.
Namely, the dynamical symmetry is expressed as
\begin{align}
  (\hat U_z(\pi),\,\,T_{\rm ac}/2)
  \label{dyn_sym_w_dc}
\end{align}
Equation~\eqref{dyn_sym_w_dc} holds in
the Hamiltonian \eqref{full_Hamiltonian_gen_def} for $|t-t_0|\lesssim \sigma$.
Detailed calculations around the dynamical symmetry \eqref{dyn_sym_w_dc} are presented in Appendix~\ref{app:dyn_sym_1-color}.

The dynamical symmetry \eqref{dyn_sym_w_dc} gives rise to constraints on the magnetic moment $\bm m(t)$:
\begin{align}
  m^x\biggl(t+ \frac{T_{\rm ac}}{2}\biggr) &= - m^x(t),
  \label{eq:mx} \\
  m^y\biggl(t+ \frac{T_{\rm ac}}{2}\biggr) &= - m^y(t),
  \label{eq:my} \\
  m^z\biggl(t+ \frac{T_{\rm ac}}{2}\biggr) &= m^z(t).
  \label{eq:mz}
\end{align}
The minus sign on the right hand side directly affects the Fourier transform \eqref{tilde_m_def}.
In fact, straightforward calculations lead to
\begin{align}
  \tilde m^a(n\Omega)
  &= \int_0^{T_{\rm ac}/2} dt\, \{1-(-1)^n\} e^{in\Omega t} m^a(t), \quad (a=x,~y), \label{eq:mxmy_Fourier}\\
  \tilde m^z(n\Omega)
  &= \int_0^{T_{\rm ac}/2} dt\, \{ 1+ (-1)^n\} e^{in\Omega t} m^z(t),
  \label{eq:mz_Fourier}
\end{align}
with $t_0=0$.

Note that we take the integration range $0<t<T_{\rm ac}$ instead of $-\infty<t<\infty$ for the following reasons. As we already mentioned in Sec.~\ref{sec:DS_in_general}, the dynamical symmetry holds when $|t-t_0|\lesssim \sigma$.
In this limited time range, we can regard the laser pulse field as the periodically oscillating field with the period $T_{\rm ac}$.
The periodicity allows us to shorten the integration range from $-\infty<t<\infty$ to $0<t<T_{\rm ac}$ with $t_0=0$.
Namely, we can use the Fourier series expansion instead of the Fourier transformation.

The dynamical symmetry thus forbids the even-order harmonic generations with $(-1)^n=1$  to emerge in $|\tilde m^a(n\Omega)|$ for $a=x,~y$ whereas the same symmetry forbids the odd-order harmonic generations with $(-1)^n=-1$ to emerge in $|\tilde m^z(n\Omega)|$, as Fig.~\ref{fig:neelhhg}~(a) shows.
Namely,
\begin{align}
  \tilde m^x(2n\Omega) &= 0,
  \label{mx_zero_w_dc} \\
  \tilde m^y(2n\Omega) &= 0,
  \label{my_zero_w_dc} \\
  \tilde m^z\bigl((2n+1)\Omega) &= 0,
  \label{mz_zero_w_dc}
\end{align}
for $n=0,~1,~2,~\cdots$.

When the static magnetic field $B_0$ is absent, the Hamiltonian \eqref{full_Hamiltonian_gen_def} has a higher symmetry.
The Hamiltonian \eqref{H_Neel-Canted_def} without $B_0$ has the $\mathrm{U(1)} \rtimes \mathbb Z_2$ symmetry due to the U(1) spin rotation around $S^z$ and the time reversal while the same Hamiltonian with nonzero $B_0$ has the U(1) symmetry since the magnetic field breaks the time reversal symmetry.
The larger symmetry of the magnetic Mott insulator leads to a larger dynamical symmetry.
Namely, the full Hamiltonian \eqref{full_Hamiltonian_gen_def} with $B_0=0$ has the following dynamical symmetry
\begin{align}
  (\hat T_x \hat U_y(\pi),\,\,T_{\rm ac}/2)
  \label{dyn_sym_wo_dc}
\end{align}
in addition to Eq.~\eqref{dyn_sym_w_dc}.
Here, $\hat T_x$ is the one-site translation $\bm S_{\bm r} \to \bm S_{\bm r+ a_0\bm e_x}$ along the $x$ axis and $\hat U_y(\pi)$ is the $\pi$ rotation of spin around $S^y$.
Note that $\hat T_x$ is required in Eq.~\eqref{dyn_sym_wo_dc} so that the operator $\hat T_x\hat U_y(\pi)$ keeps the ground state, the N\'eel state, invariant, as we will closely discuss later.

The symmetry under Eq.~\eqref{dyn_sym_wo_dc} yields another selection rule:
\begin{align}
  \tilde m^x(2n\Omega) &= 0,
  \label{mx_zero_wo_dc} \\
  \tilde m^y\bigl((2n+1)\Omega\bigr) &= 0,
  \label{my_zero_wo_dc} \\
  \tilde m^z(2n\Omega) &= 0,
  \label{mz_zero_wo_dc}
\end{align}
for $n=0,~1,~2,~\cdots$.
The presence of two dynamical symmetries [Eqs.~\eqref{dyn_sym_w_dc} and \eqref{dyn_sym_wo_dc}] forbids the even-order harmonic generations of $|\tilde m^x(n\Omega)|$ with $n=0,~2,~4,\cdots$ and every harmonic generation of $|\tilde m^a(n\Omega)|$ for $a=y,~z$ with $n=0,~1,~2,~\cdots$.
In fact, the numerically obtained spectra $|\tilde m^y(\omega)|$ and $|\tilde m^z(\omega)|$ [see the first column of Fig.~\ref{fig:neelhhg}~(b)] have suppressed peak heights at $\omega=n\Omega$, which are roughly $10^{-3}$ times of those in the presence of the static magnetic field.

We find out that the selection rules when the one-color laser-pulse field is along the $x$ or $y$ axis explain the qualitative features of the harmonic generations on the first and second columns of Figs.~\ref{fig:neelhhg}~(a) and (b).

\subsubsection{Laser field along \texorpdfstring{$z$}{z} axis}
\label{subsec:Neel_z}

When the laser field is parallel to the $z$ axis, we find no clear peaks in $|\tilde m^a(\omega)|$ for any $a=x,~y,~z$ regardless of the presence or absence of the static magnetic field.
While the spectra look similar in those cases, the dynamical symmetry is quite different.
In the presence of the static magnetic field [Fig.~\ref{fig:neelhhg}~(a)], the Hamiltonian \eqref{full_Hamiltonian_gen_def} has no dynamical symmetry even approximately.
By contrast, in the absence of the static magnetic field [Fig.~\ref{fig:neelhhg}~(b)], we find that the Hamiltonian \eqref{full_Hamiltonian_gen_def} has two dynamical symmetries, $(\hat T_x\hat U_x(\pi),\,\,T_{\rm ac}/2)$ and $(\hat T_x\hat U_y(\pi),\,\,T_{\rm ac}/2)$.
As a result, we have a selection rule that even-order harmonic peaks in $|\tilde m^z(\omega)|$ are absent in the case of zero field $B_0=0$.
Despite this difference between two cases of zero field and a finite field, the resultant spectra equally have no major harmonic generations.
We can understand that the magnon modes [$\alpha$ and $\beta$ modes in Fig.~\ref{fig:model}~(e)] are precessions around the $z$ axis.
The laser fields oscillating linearly along the $z$ axis do not excite such precessing modes, ending up with the absence of harmonic generations.

This weak response to the ac field along the $z$ direction can also be understood from the spin-wave (magnon) picture. From the Holstein-Primakoff transformation, spin operators in collinear ordered phases including the N\'eel phase are approximated by $\hat S_{\bm r}^z=S-\hat c_{\bm r}^\dagger\hat c_{\bm r}$ and $\hat S^+_{\bm r}\simeq \sqrt{2S}\hat c_{\bm r}$, where $\hat c_{\bm r}$ is the annihilation boson (magnon) operator~\cite{Holstein1940Field,White2007_spinwave}. Therefore, the dynamics of $\hat S^z$ requires a simultaneous two-magnon excitation, while the transverse spin dynamics can occur through one-magnon excitation. This indicates that the longitudinal spin dynamics driven by an ac magnetic field is weak compared with the transverse spin dynamics.

Finally, we briefly comment on the broad peaks in the cases of ${\bm B}_{\rm ac}\parallel {\bm e}_z$. Panels (a) and (b) of Fig.~\ref{fig:neelhhg} show that there seem to be two small peaks in the spectra of $|\tilde m^{x,y}(\omega)|$ under ${\bm B}_{\rm ac}\parallel {\bm e}_z$ and a finite static field $B_0=0.5J$, while there is one peak in $|\tilde m^{x,y}(\omega)|$ under ${\bm B}_{\rm ac}\parallel {\bm e}_z$ and zero field $B_0=0$. In Appendix~\ref{app:neelzstr}, we show that these are not the laser-driven harmonics and rather owing to the effect of thermal fluctuation.

\subsubsection{Magnon picture and angular-momentum conservation}

\begin{figure*}[t]
  \centering
  \includegraphics[bb = 0 0 989 360, width=0.8\linewidth]{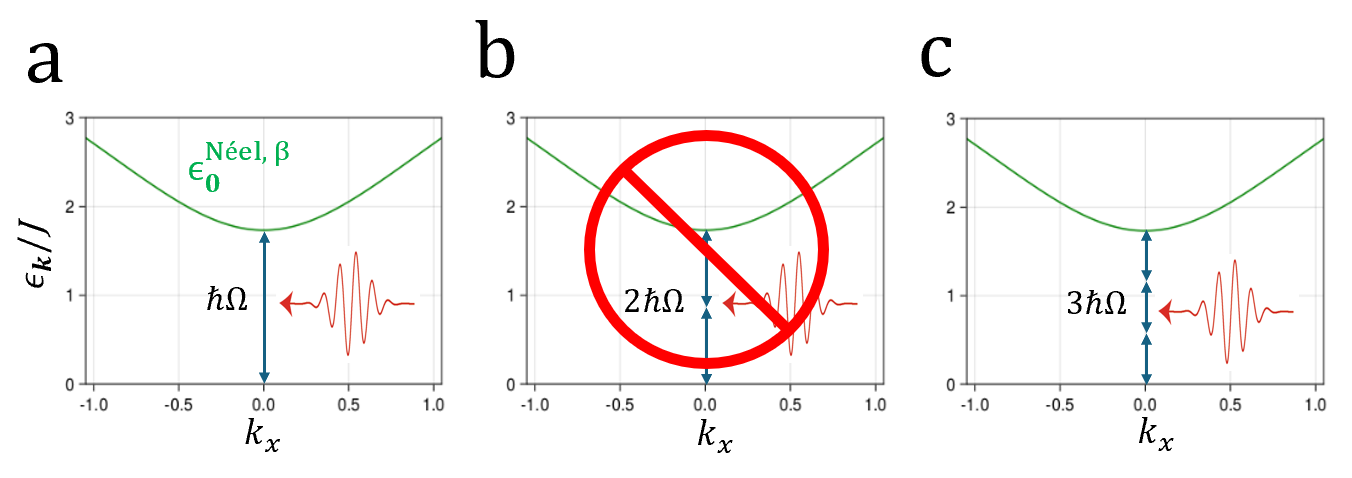}
  \caption{
    Selection rules of magnon harmonic generation in N\'eel phase from viewpoint of angular momentum conservation. We consider the case where a linearly polarized THz laser with ${\bm B}_{\rm ac}\parallel \hat e_x$ is applied [see panels (c)-(e) in Fig.~\ref{fig:neelhhg}].
    Panels (a) and (c) show the resonant absorption of photons.
    A single photon with energy $\hbar \Omega$ creates one magnon when $\hbar\Omega = \epsilon_{\bm 0}^{\text{N\'eel},\beta}$ in panel (a), while three photons, each of which has the energy $\hbar \Omega$, create one magnon when $3\hbar \Omega = \epsilon_{\bm 0}^{\text{N\'eel},\beta}$ in panel (c).
    These resonant absorptions (and the subsequent resonant emissions) of photon(s) are possible since these processes are consistent with the angular momentum conservation.
    Panel (b) shows that two photons, each of which has the energy $\hbar\Omega$, \emph{cannot} create one magnon even though the photon energy meets the resonant condition, $2\hbar\Omega = \epsilon_{\bm 0}^{\text{N\'eel},\beta}$ since this process violates the angular momentum conservation law.
  }
  \label{fig:photonmagnon}
\end{figure*}

The dynamical symmetry enables the qualitative interpretation of the harmonic generation $|\tilde m^a(n\Omega)|$.
Here, we discuss the spectra from the magnon point of view.
Figures~\ref{fig:neelhhg}~(c),~(d), and (e), respectively show the intensity for the first (fundamental), second, and third harmonics.
Though these intensities are mathematically given by $|\tilde m^a(n\Omega)|$ for $n=1,~2,~3$, we practically define them as an average,
\begin{align}
  \int_{(n-\epsilon)\Omega}^{(n+\epsilon)\Omega} d\omega\,|\tilde m^a(\omega)|,
\end{align}
near $\omega=n\Omega$ within a narrow enough width $2\epsilon \Omega$ to
reduce numerical errors. In the panels (c)-(e), we employ $\epsilon = 0.2$ and show these averaged intensities $|\tilde m^a(n\Omega)|$ as a function of laser frequency $\Omega$ and static magnetic field $B_0$.
The high-intensity regions are indicated in these figures as curves with bright color.
By comparing these curves with Fig.~\ref{fig:neelhhg}~(f), we find that the bright curves correspond to the one-magnon energy with $\bm k=\bm 0$, $\epsilon_{\bm 0}^{\text{N\'eel},\mu}$ ($\mu=\alpha,~\beta$) [Figs.~\ref{fig:neelhhg}~(c), (d), and (e)], or $1/3$ of that, $\epsilon_{\bm 0}^{\text{N\'eel},\mu}/3$ ($\mu=\alpha,~\beta$) [Fig.~\ref{fig:neelhhg}~(e)].

We can understand the correspondence of the high-intensity curves to the magnon bands in terms of the spin-wave theory~\cite{Holstein1940Field,White2007_spinwave} and angular momentum conservation.
The fundamental (first) harmonic generation at $\omega=\Omega$ involves the absorption and emission of one photon with energy $\hbar\Omega$.
Such a resonant phenomenon occurs when $\hbar\Omega$ is equal to the magnon gap $\epsilon_{\bm 0}^{\text{N\'eel},\mu}$ ($\mu=\alpha,~\beta$), leading to the high-intensity curve in $|\tilde m^x(\Omega)|$ as well as $|\tilde m^z(2\Omega)|$ and $|\tilde m^x(3\Omega)|$.
We may expect that the magnon bands $\epsilon_{\bm 0}^{\text{N\'eel},\mu}$ with $\mu=\alpha,~\beta$ emerge in $\tilde m^x(n\Omega)$ for every $n=1,~2,~3,~\cdots$ since it corresponds to the resonant absorption of $n$ photons that creates $n$ magnons.
Each photon with energy $\hbar\Omega$ invokes one magnon with energy $\epsilon_{\bm 0}^{\text{N\'eel},\mu}$, leading to the resonance $\hbar\Omega=\epsilon_{\bm 0}^{\text{N\'eel},\mu}$.
For example, the two-photon absorption corresponding to the two-magnon creation is the leading contribution of the bright curve $\hbar\Omega=\epsilon_{\bm 0}^{\text{N\'eel},\mu}$ of $|\tilde m^z(2\Omega)|$ in the panel (d) of Fig.~\ref{fig:neelhhg}. This is because (as we discussed in Sec.~\ref{subsec:Neel_z}) at least a two-magnon creation is necessary to generate an oscillation of $S^z$.
By contrast, the curves corresponding to one third of the magnon band that emerge in $|\tilde m^x(3\Omega)|$ are related to resonant absorption of three photons that creates only one magnon.
Each photon with energy $\hbar\Omega$ has one third of the magnon energy, too small to solely invoke one magnon.
Only after three such photons are assembled do they invoke one magnon, leading to the resonance $\hbar\Omega=\epsilon_{\bm 0}^{\text{N\'eel},\mu}/3$.
Such a channel of resonant absorption is absent in $|\tilde m^z(2\Omega)|$.
Namely, there is no resonant absorption of two photons that would create only one magnon.

To explain this difference between $|\tilde m^z(2\Omega)|$ and $|\tilde m^x(3\Omega)|$, we look at the angular momenta carried by photons and magnons.
Each photon carries the angular momentum $\hbar$ or $-\hbar$ since we apply a linearly polarized THz laser that consists of a right circularly polarized photon with angular momentum $+\hbar$ and a left one with $-\hbar$ (see Table~\ref{tab:angularmomentum}).
Likewise, the angular momentum of each magnon is either $\hbar$ or $-\hbar$ in the N\'eel phase.
The angular momentum of magnon is defined as follows. The Holstein-Primakoff transformation~\cite{Holstein1940Field} relates the $z$ component of the spin to the number of magnons in the N\'eel phase such as $\hat S_{\bm r}^z = S - \hat a_{\bm r}^\dag \hat a_{\bm r}$ or $\hat S_{\bm r}^z = - S + \hat b_{\bm r}^\dag \hat b_{\bm r}$, depending on the sublattice of the square lattice.
Each magnon changes the $z$ component of the spin by $\pm \hbar$, meaning that the magnon has the angular momentum, either $\hbar$ or $-\hbar$. Since the $z$ component of total spin, $\hat S_{\rm tot}^z=\sum_{\bm r}\hat S_{\bm r}^z$, is conserved in the N\'eel phase, the magnon angular momentum is a good quantum number. As we show in Table~\ref{tab:angularmomentum}, in our notation, the angular momentum of the $\alpha$-mode magnon is $-\hbar$, while that of the $\beta$-mode one is $+\hbar$.

When three photons, each of which has the energy $\epsilon_{\bm 0}^{\text{N\'eel},\mu}/3$, invoke one magnon with angular momentum $\hbar$, two photons carry the angular momentum $2\hbar$ and one carries $-\hbar$ so that the total angular momentum of three photons, $\hbar+\hbar + (-\hbar) = \hbar$ coincides with that of the created magnon.
Such angular-momentum conservation cannot be fulfilled when two photons are simultaneously absorbed.
Generally, we may expect the resonance $\hbar \Omega = \epsilon_{\bm 0}^{\text{N\'eel},\mu}/n$ ($\mu=\alpha,~\beta$) is forbidden for even $n=2,~4,~6,~\cdots$.
This argument based on the angular momentum conservation is summarized in Fig.~\ref{fig:photonmagnon}, by using a conceptual sketch.

We note that when an $\alpha$-mode ($\beta$-mode) magnon is excited by applied linearly polarized laser, it means that the antiferromagnet selectively absorbs a circularly polarized wave with angular momentum $-\hbar$ ($+\hbar$). In other words, linearly polarized laser can directly excite an $\alpha$-mode ($\beta$-mode) magnon if the laser frequency is close enough to the resonant value of the $\alpha$ ($\beta$) mode.
This magnon-photon process is possible because (as we mentioned above) linearly polarized light is a linear combination of right and left circularly polarized waves.
The fact that a resonant spin precession can be driven by linearly polarized electromagnetic wave is known well in the field of magnetic resonance such as ESR and NMR~\cite{Slichter1990Principles}.

\begin{table}
  \caption{\label{tab:angularmomentum}
  Angular momenta of magnons in N\'eel phase and photons. In this paper, the sign of spin angular momentum is defined so that the up-spin magnon has the positive angular momentum $+\hbar$.}
  \begin{ruledtabular}
    \begin{tabular}{ll}
      $\alpha$ mode magnon & $-\hbar$ \\
      $\beta$ mode magnon & $+\hbar$ \\
      Right circularly polarized photon &  $+\hbar$\\
      Left circularly polarized photon &  $-\hbar$
    \end{tabular}
  \end{ruledtabular}
\end{table}

\subsection{Canted phase}
\label{sec:Canted}

\begin{figure*}[t!]
  \centering
  \includegraphics[bb = 0 0 1793 1365, width=0.89\linewidth]{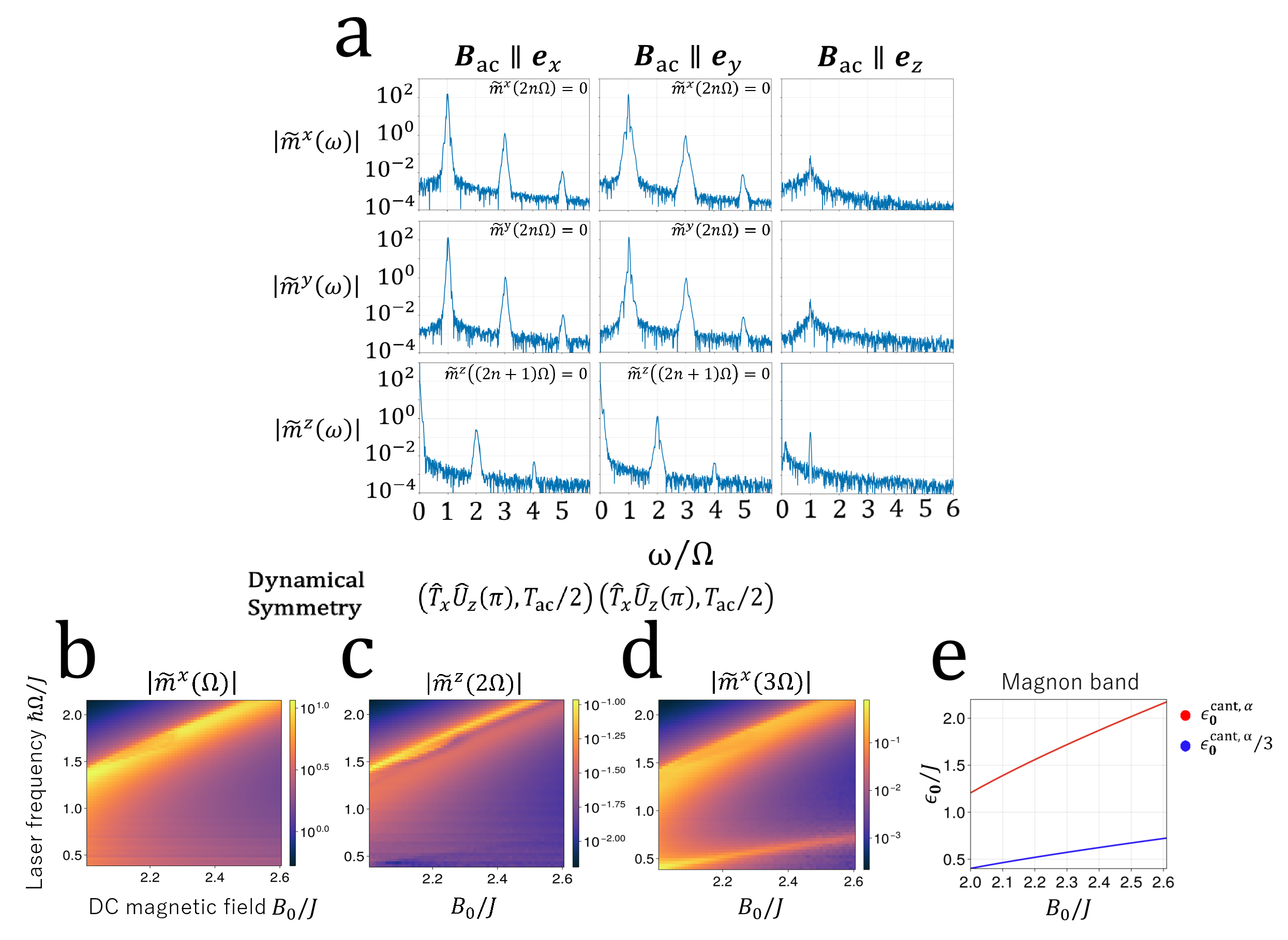}
  \caption{Numerical result of harmonic generation in canted phase at $k_{\rm B}T/J=10^{-3}$.
    (a) Spectra $|\tilde m^a(\omega)|$ of model \eqref{H_Neel-Canted_def} in canted phase, where the spins of the thermal state are chosen to be in $S^x$-$S^z$ plane.
    The first, second, and third rows show spectra $a=x,~y$, and $z$, respectively.
    The first, second, and third columns show spectra when the applied laser pulse is parallel to the $x,~y$, and $z$ axes, respectively. The symmetry operation below each column [such as $(\hat T_x\hat U_z(\pi),T_{\rm ac}/2)$] denotes the dynamical symmetry in each setup.
    We set $B_0/J=3.0$, $B_{\rm ac}/J=0.1$, and the photon energy $\hbar\Omega=2.7J$, that is equivalent to the gap of the $\alpha$-mode magnon, $\epsilon_{\bm{0}}^{\text{Cant}, \alpha}$.
    Panels (b), (c), and (d) show the first ($|\tilde m^x(\Omega)|$), second ($|\tilde m^z(2\Omega)|$), and third ($|\tilde m^x(3\Omega)|$) harmonic generations when $\bm B_{\rm ac}$ is parallel to the $x$ axis.
    Panel (e) shows the magnon band $\epsilon_{\bm{0}}^{\text{Cant}, \alpha}$ (red curve), and its one-third scale (blue curve).
    The latter is plotted for the purpose of comparison with the panel (d). We note that $|{\tilde m}^\alpha(\omega)|$ means $|\langle{\tilde m}^\alpha(\omega)\rangle|$.
  }
  \label{fig:Canthhg}
\end{figure*}

Figure~\ref{fig:Canthhg} shows the harmonic generation spectra in the canted phase, in which the spins are locked in the $S^x$-$S^z$ plane before the laser application.
The model Hamiltonian is the same as that in Sec.~\ref{sec:Neel} but has the canted state as its ground state for large magnetic field, say, $B_0/J=3.0$ in Fig.~\ref{fig:Canthhg}.
Despite the same Hamiltonian, the harmonic generation spectra show different behaviors.
The difference originates from how spontaneous symmetry breaking occurs in the N\'eel and canted phases.
While the N\'eel ordered state has the U(1) spin rotation symmetry around the $S^z$ axis, the canted state has a lower symmetry, the $\mathbb Z_2$ symmetry around the $S^z$ axis.
Namely, the N\'eel state is invariant under the rotation by any continuous angle around the $z$ axis but the canted state is invariant only when the angle is either $0$ or $\pi$.


The laser field with $\bm B_{\rm ac} \parallel \bm e_x$ further lowers the symmetry of the system.
In the N\'eel phase, the symmetry is lowered to the dynamical one~\eqref{dyn_sym_w_dc}.
In the canted phase, the dynamical symmetry is given by
\begin{align}
  (\hat T_x\hat U_z(\pi),\,\,T_{\rm ac}/2)
  \label{dyn_sym_cant}
\end{align}
The difference between two symmetry operations in Eqs.~\eqref{dyn_sym_w_dc} and \eqref{dyn_sym_cant} is the absence or presence of the translation operator $\hat T_x$.
Interestingly, the selection rules in the harmonic generation spectra $|\tilde m^x(\omega)|$ in the N\'eel and canted phases are exactly the same [compare the first columns of Figs.~\ref{fig:neelhhg}~(a) and \ref{fig:Canthhg}~(a)].
The involvement of translation operator (or not) turns out to be irrelevant to the selection rule here since we are focused on the dynamics of the \textit{total} magnetic moment \eqref{ma_tot_def}.

Since there is no dynamical symmetry in the setup of $\bm B_{\rm ac} || \bm e_z$, all peaks of the harmonic generation are expected to appear.
However, we find only the linear response near $\omega=\Omega$ for $\bm B_{\rm ac} || \bm e_z$.
The absence of harmonic generations with $n\ge 2$ is due to the nature of the magnon ($\beta$ mode) created by a longitudinal ac field $\bm B_{\rm ac} || \bm e_z$, which is the gapless NG boson as shown in Fig.~\ref{fig:model} (f).
No photon with finite energy $\hbar\Omega$ can excite the gapless mode with infinitesimal energy. It is inferred that the peak of the linear response is attributed to a ``forced'' excitation of the gapped $\alpha$ mode (precession of total spin in the $S^x$-$S^y$ plane) by the intense laser with $B_{\rm ac}=0.1J$.
We can expect that if the applied laser becomes further strong (although it is hard to prepare such a strong laser in the current laser technique), higher-order harmonic peaks with $n\ge 2$ appear due to the nonlinear interaction processes of the spin-light coupling. In Appendix~\ref{app:canted}, we confirm that this naive expectation is true.

Let us look into the color plots in Figs.~\ref{fig:Canthhg}~(b), (c), and (d).
We find qualitative similarities between these plots and corresponding ones [Figs.~\ref{fig:neelhhg}~(c), (d), and (e)] in the N\'eel phase.
They show the resonant absorption of magnon, resulting in the high intensity at the frequency that equals the magnon band $\epsilon_{\bm 0}^{\text{Cant},\alpha}$.
The third-harmonic generation also shows the resonant absorption at the frequency that equals one third of the magnon band ($\hbar\Omega=\epsilon_{\bm 0}^{\text{Cant},\alpha}/3$) for the same reason as Fig.~\ref{fig:neelhhg}~(e).
The only difference due to the magnetic order (canted vs N\'eel) is quantitative as follows.
The magnon band in the canted phase and its one third scale are plotted in Fig.~\ref{fig:Canthhg}~(e).
As this plot shows, the $B_0/J$ dependence of the magnon band differs quantitatively in the canted and N\'eel phases.
We emphasize that the number of magnon modes is the same in these two phases.
While Fig.~\ref{fig:neelhhg}~(f) shows peaks of both the $\alpha$ and $\beta$ modes, Fig.~\ref{fig:Canthhg}~(e) shows only the peak of the $\alpha$ mode.
The $\beta$ mode does exist in the canted phase but is invisible in Fig.~\ref{fig:Canthhg}~(e) because the plot shows the magnon gap at $\bm k = \bm 0$ and the $\beta$ mode in the canted phase is the NG boson, gapless at this wave vector.

The magnon in the N\'eel phase carries the angular momentum $\pm \hbar$, whose quantization comes out of the collinear nature of the N\'eel order.
The classical N\'eel order shows $\hat S_{\bm r}^z= \pm \hbar S$ and the magnon as quantum fluctuation changes $\hat S_{\bm r}^z$ by $\pm \hbar \times (\text{number of magnons})$.
On the other hand, the canted order is noncollinear as the name implies.
Namely, the canted order breaks the $\rm U(1)$ spin rotation symmetry.
The absence of the $\rm U(1)$ symmetry makes the angular momentum be a  nonconserved quantity.
We thus cannot apply the argument based on the angular momentum (see Fig.~\ref{fig:photonmagnon}) to the canted phase. Nevertheless, we find that the peak structure of the panels (b)-(d) in Fig.~\ref{fig:Canthhg} is similar to that in Fig.~\ref{fig:neelhhg}. Namely, the peak structure in panels (b) and (d) in Fig.~\ref{fig:Canthhg} seems to be interpreted by using the angular momentum transfer between magnons and photons. We might understand this situation as follows. The applied THz laser has a very long wavelength compared with the lattice spacing and the laser field is coupled only to the total spin. This indicates that the laser is ``unaware'' of the detailed spin structure in the unit cell. To the laser, the canted order appears identical to an ordered state with a spatially uniform magnetization (i.e., a ferromagnetic order). Therefore, the argument of the angular momentum 
works well
in explaining the harmonic generation even in the $\rm U(1)$-symmetry breaking canted phase.

\subsection{Weak ferromagnets}
\label{sec:WeakFerro}

\begin{figure*}[t!]
  \centering
  \includegraphics[bb = 0 0 1934 1872, width=0.9\linewidth]{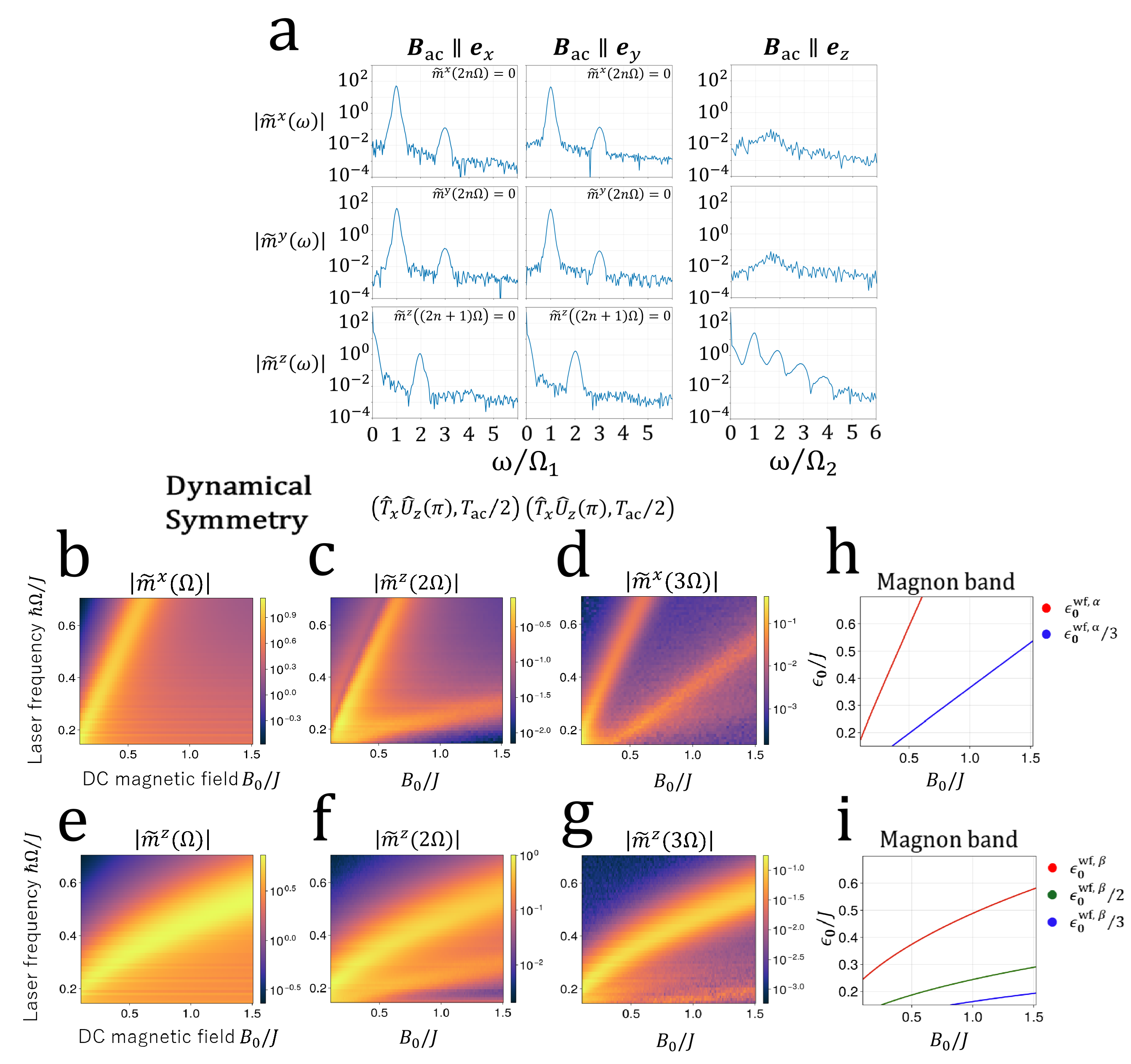}
  \caption{
    Numerical result of harmonic generation in WF phase at $k_{\rm B}T/J=10^{-3}$.
    (a) Spectra $|\tilde m^a(\omega)|$ of model \eqref{H_WF_def} in WF phase.
    The first, second, and third rows show spectra $a=x,~y$, and $z$, respectively.
    The first, second, and third columns show spectra when the applied laser pulse is parallel to the $x,~y$, and $z$ axes, respectively. The symmetry operation below each column [such as $(\hat T_x\hat U_z(\pi),T_{\rm ac}/2)$] denotes the dynamical symmetry in each setup.
    We set $B_0/J=0.5$ and $B_{\rm ac}/J=0.1$. The photon energies $\hbar\Omega_1=0.591J$ and $\hbar\Omega_2=0.374J$ are set to the gaps of the $\alpha$-mode magnon ($\epsilon_{\bm{0}}^{\text{WF}, \alpha}$) and the $\beta$-mode magnon ($\epsilon_{\bm{0}}^{\text{WF}, \beta}$), respectively.
    Panels (b), (c), and (d) show first ($|\tilde m^x(\Omega)|$), second ($|\tilde m^z(2\Omega)|$), and third ($|\tilde m^x(3\Omega)|$) harmonic generations when $\bm B_{\rm ac}$ is parallel to the $x$ axis.
    Panels (e), (f), and (g) show first ($|\tilde m^z(\Omega)|$), second ($|\tilde m^z(2\Omega)|$), and third ($|\tilde m^z(3\Omega)|$) harmonic generations when $\bm B_{\rm ac}$ is parallel to the $z$ axis.
    Panel (h) shows the magnon band $\epsilon_{\bm{0}}^{\text{WF}, \alpha}$ (red curve), and its one-third scale (blue curve).
    The latter is plotted for the purpose of comparison with the panel (d).
    Panel (i) shows the magnon band $\epsilon_{\bm{0}}^{\text{WF}, \beta}$ (red curve), and its half (green curve) and one-third scales (blue curve).
    The latter two are plotted for the purpose of comparison with the panel (g). We note that $|{\tilde m}^\alpha(\omega)|$ means $|\langle{\tilde m}^\alpha(\omega)\rangle|$.
  }
  \label{fig:wfhhg}
\end{figure*}

We move on to the harmonic generation spectra in the WF phase [Fig.~\ref{fig:wfhhg}~(a)]. The spins are set to the $S^x$-$S^z$ plane like the canted phase.
Note that the WF order and the canted order show no clear difference by themselves.
Mechanisms, or Hamiltonians, that yield these magnetic orders are different.
Let us compare the harmonic generation spectra in the WF and canted phases.
For $\bm B_{\rm ac} \parallel \bm e_x$, the spectra in the WF and canted phases resemble each other despite the difference in the Hamiltonians lying behind these orders [Eqs.~\eqref{H_Neel-Canted_def} and \eqref{H_WF_def}].
In particular, the same selection rule \eqref{mx_zero_w_dc}, \eqref{my_zero_w_dc}, and \eqref{mz_zero_w_dc} applies to the spectra in these phases for $\bm B_{\rm ac} \parallel \bm e_x$.
The selection rule is determined by the dynamical symmetry.
A combination of the temporal translation and the symmetry of the ground state results in the selection rule.
Since the canted and WF ground states are qualitatively the same, the resultant dynamical symmetry is similar, although not exactly the same.

It is important to distinguish the symmetry of the ground state from that of the Hamiltonian.
The Hamiltonian $\hat{\mathcal H}_{\text{N-C}}$ has the $\rm U(1)$ global rotation symmetry around the $S^z$ axis, while the Hamiltonian $\hat{\mathcal H}_{\text{WF}}$ has no global rotation symmetry.
The latter still has the $\hat T_x\mathbb Z_2$ symmetry, the combination of the spatial translation and the $\mathbb Z_2$ global rotation around the $S^z$ axis.
The ground state in the canted phase, spontaneously breaking the U(1) rotation symmetry, has the $\hat T_x\mathbb Z_2$ symmetry instead of the U(1) symmetry.
By contrast, the ground state in the WF phase has the full symmetry of the Hamiltonian, the $\hat T_x\mathbb Z_2$ symmetry.
After all, the spectra $|\tilde m^a(\omega)|$ for $\bm B_{\rm ac} \parallel \bm e_x$ show qualitative resemblance originating from the same selection rule in the canted and WF phases.
The absence of the spontaneous symmetry breaking in the WF phase also results in the gapped $\beta$ mode.
As Fig.~\ref{fig:wfhhg}~(i) shows, the $\beta$ mode has the finite gap $\epsilon_{\bm 0}^{\text{WF},\beta}>0$ even for $B_0/J=0$.

We see a clear difference in the spectra $|\tilde m^z(\omega)|$ in the WF and canted phases when $\bm B_{\rm ac} \parallel \bm e_z$.
We find wide but clear harmonic generations $n=1,~2,~\cdots,~5$ in the spectrum $|\tilde m^z(\omega)|$ while that in the canted phase has no such harmonic generations.
This difference in the $z$ direction originates from the nature of the $\beta$ mode.
The $\beta$ mode in the WF phase is gapped but that in the canted phase is gapless, due to the absence or presence of the spontaneous symmetry breaking.
Photons with finite energy can excite gapped modes, leading to harmonic generations in $|\tilde m^z(\omega)|$ for $\bm B_{\rm ac}\parallel \bm e_z$. This magnon harmonic generation spectrum for $\bm B_{\rm ac} \parallel \bm e_z$ in the WF phase is at least qualitatively consistent with a recent observation of the magnon harmonic generation in a WF phase of $\rm HoFeO_3$~\cite{Zhang2023_HHG}, in which the first, second and third harmonic peaks are detected.

Let us see color plots [Figs.~\ref{fig:wfhhg}~(b-g)] for the first, second, and third harmonic generations of $|\tilde m^x(\Omega)|$, $|\tilde m^z(2\Omega)|$ and $|\tilde m^x(3\Omega)|$.
Figures~\ref{fig:wfhhg}~(b), (c), and (d) resemble Figs.~\ref{fig:Canthhg}~(b), (c), and (d), respectively.
The resemblance comes from the similarity of the $\alpha$ mode in the WF and canted phases [compare Figs.~\ref{fig:wfhhg}~(h) and Fig.~\ref{fig:Canthhg}~(e)]. However, we should note that the spectrum of $|\tilde m^z(2\Omega)|$ possesses not only the peak at $\epsilon_{\bm 0}^{\text{WF},\alpha}$ but also that at $\epsilon_{\bm 0}^{\text{WF},\beta}/2$. The latter peak is qualitatively different from the spectrum of $|\tilde m^z(2\Omega)|$ in the canted phase [see Fig.~\ref{fig:Canthhg}(c)].
On the other hand, Fig.~\ref{fig:wfhhg}~(e), (f), and (g), related to the $\beta$ mode [Fig.~\ref{fig:wfhhg}~(i)], are more characteristic to the WF phase. The $\beta$ mode has a finite gap, leading to a resonant absorption of magnons.
The second harmonic $|\tilde m^z(2\Omega)|$ shows the bright curve corresponding to $\epsilon_{\bm 0}^{\text{WF}, \beta}/2$ [compare Figs.~\ref{fig:wfhhg}~(f) and (i)] in addition to the peak at $\epsilon_{\bm 0}^{\text{WF}, \beta}$.
Likewise, $|\tilde m^z(3\Omega)|$ shows two weakly bright curves corresponding to $\epsilon_{\bm 0}^{\text{WF},\beta}/2$ and $\epsilon_{\bm 0}^{\text{WF},\beta}/3$ [compare Figs.~\ref{fig:wfhhg}~(g) and (i)] in addition to $\epsilon_{\bm 0}^{\text{WF}, \beta}$.
The structure of these multiple peaks quite differs from that of the canted phase, and the multiple peaks seem not to be understood from the angular momentum transfer between magnons and photons. In fact,
(as we mentioned) the WF state has no continuous spin rotation symmetry, and the magnons do not have quantized angular momenta $\pm \hbar$.
The simple argument of the angular momentum conservation in the N\'eel phase thus does not apply to the spectra $|\tilde m^a(n\Omega)|$ in the WF phase.

\section{Two-color laser}
\label{sec:2color}

\begin{figure*}[t!]
  \centering
  \includegraphics[bb = 0 0 2273 1036, width=1.0\linewidth]{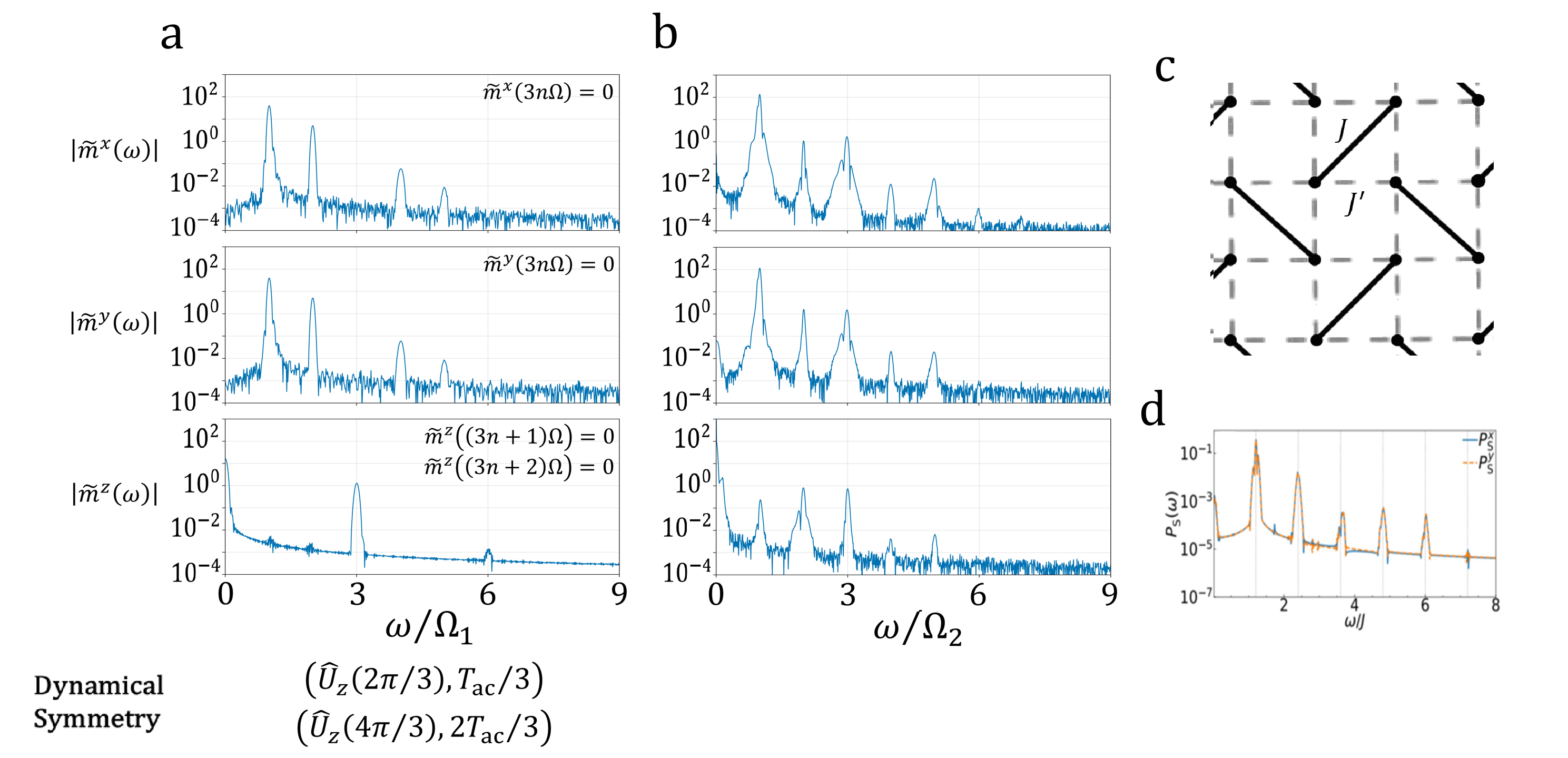}
  \caption{
    (a) Spectra $|\tilde m^a(\omega)|$ for $a=x,~y,~z$ in  N\'eel phase where two-color laser pulse with $C_3$ symmetry around the $z$ axis is applied.
    (b) Spectra $|\tilde m^a(\omega)|$ for $a=x,~y,~z$ in canted phase where two-color laser pulse with $C_3$ symmetry around $z$ axis is applied.
    We set the static magnetic-field strength $B_0$ to (a) $B_0/J=0.1$ and (b) $B_0/J=3.0$.
    We also set $B_{\rm ac}/J=0.1$ in both cases.
    The laser frequencies $\Omega_1$ and $\Omega_2$ are chosen to be the magnetic resonance values: $\hbar\Omega_{1}/J = \epsilon_{\bm{0}}^{\text{N\'eel}, \beta}/J = 1.73$, and $\hbar\Omega_{2}/J = \epsilon_{\bm{0}}^{\text{Cant},\alpha}/J=2.7$. The symmetry operation below the panel (a) denotes the dynamical symmetry.
    (c) Shastry-Sutherland model. The model has antiferromagnetic exchange interactions on nearest-neighbor $J'$ bonds and next-nearest-neighbor $J$ bonds.
  (d) Harmonic generation spectra $|\hat P_s(\omega)|$ of the Shastry-Sutherland model. $\hat P_s(\omega)$ is the electric polarization [Eq.~\eqref{EP_SS_def}]. This panel is the result of Ref.~\cite{Udono2024_HHG}.}
  \label{fig:HHG_2color_C3}
\end{figure*}

So far we have discussed the magnon harmonic generations driven by conventional one-color lasers.
This section is devoted to investigations on how two-color laser fields \eqref{B_2_def} affect the harmonic generation spectrum.
We consider two particular cases: two-color lasers with $C_3$ or $C_4$ (spatial) rotation symmetry.
When $\ell = 2$ ($\ell=3$) in Eq.~\eqref{B_2_def}, the trajectory of the laser field $\bm B_{2,\ell}(t) \propto [\bm b_{-1}(t)+\bm b_{\ell}(t)]$ has the $C_3$ ($C_4$) symmetry, as we explained before.
The spatial symmetry of the laser field is also a key ingredient of the dynamical symmetry.
Application of the two-color laser instead of the one-color one affects the dynamical symmetry and, ultimately, the selection rule of harmonic generations.
In the remainder of this section, we discuss effects of the $C_3$- and $C_4$-symmetric two-color laser fields on the harmonic generation in the N\'eel and canted phases. We also compare the two-color laser driven harmonic generations in these ordered states with that in the dimer-singlet phase in the two-dimensional Shastry-Sutherland model that has been studied in Ref.~\cite{Udono2024_HHG}. The dimer-singlet state is a typical many-spin ground state without spontaneous symmetry breaking and magnetic (dipole) moment, in contrast to ordered phases in antiferromagnets.


\subsection{\texorpdfstring{$C_3$}{C_3}-symmetric laser}
\label{sec:2color_C3}

Let us look into the $\ell=2$ case, where the laser field \eqref{B_2_def} has the $C_3$ spatial rotation symmetry [see Fig.~\ref{fig:pulse}~(b)].
In analogy with the one-color laser, we consider
\begin{align}
  \bm B'_{2,\ell}(t) = \frac{\bm b_{-1}(t)+\bm b_{\ell}(t)}{\sqrt2},
  \label{B_p_def}
\end{align}
instead of $\bm B_{2,\ell}(t)$ of Eq.~\eqref{B_2_def}.
We can approximate $\bm B_{2,\ell}(t) \approx \bm B'_{2,\ell}(t) = ({B'_{2,\ell}}^x(t),~{B'_{2,\ell}}^y(t),~{B'_{2,\ell}}^z(t))^\top$ for $|t-t_0|\ll \sigma$.
Applying the $2\pi/3$ rotation around the $z$ axis to
$\bm B'_{2,\ell}(t)$ with $\ell=2$, we immediately obtain
\begin{align}
  \begin{pmatrix}
    \cos(2\pi/3) & - \sin(2\pi/3) & 0 \\
    \sin(2\pi/3) & \cos (2\pi/3) & 0 \\
    0 & 0 & 1
  \end{pmatrix}
  \bm B'_{2,\ell}(t)
  &= \bm B'_{2,\ell} \biggl( t + \frac{2\pi}{3\Omega} \biggr) \nonumber\\
  &= \bm B'_{2,\ell} \biggl( t + \frac{T_{\rm ac}}{3}\biggr).
\end{align}
This result means that the $C_{3}$ rotation of the $C_3$-symmetric laser field can be compensated by the time translation of the laser field by $T_{\rm ac}/3$.
Since the ac Zeeman interaction is only the time-dependent term in the Hamiltonian,
the time translation of the laser field can be regarded as that of the Hamiltonian. Next, let us focus on the shape of the ac Zeeman interaction, which consists of the inner product of the ac magnetic field and the total spin, $-\bm B_{2,\ell}(t)\cdot\sum_{\bm r}\hat {\bm S}_{\bm r}$. The inner product is unchanged under the simultaneous rotation of the ac field and total spin. Therefore, if we simultaneously perform the above time translation and the $2\pi/3$ spin rotation around the $S^z$ axis to the ac Zeeman interaction [$\hat U_z(2\pi/3)$], the shape of the Zeeman interaction is unchanged. Here, the spin rotation by angle $\theta$ around the $S^z$ axis is defined as follows:
\begin{align}
  \hat U_z(\theta) \hat {\bm S}_{\bm r}\hat U_{z}(\theta)^\dagger =
  \begin{pmatrix}
    \cos\theta & - \sin\theta & 0 \\
    \sin\theta & \cos \theta & 0 \\
    0 & 0 & 1
  \end{pmatrix}\hat {\bm S}_{\bm r}
\end{align}
Both the Hamiltonian \eqref{H_Neel-Canted_def} and the N\'eel-ordered ground state have the U(1) spin rotation symmetry around the $S^z$ axis.
From these facts, we find a dynamical symmetry
\begin{align}
  (\hat U_z(2\pi/3),\,\,T_{\rm ac}/3).
  \label{dyn_sym_1_C3_Neel}
\end{align}
Likewise, by considering the time translation by $2\times T_{\rm ac}/3$ and the spin rotation by $2\times 2\pi/3$, we obtain another dynamical symmetry
\begin{align}
  (\hat U_z(4\pi/3), \,\,2T_{\rm ac}/3).
  \label{dyn_sym_2_C3_Neel}
\end{align}
These dynamical symmetries forbid all the $n$th harmonic generations in $|\tilde m^x(\omega)|$ and $|\tilde m^y(\omega)|$ for $n=3,~6,~9,~\cdots$ [see the top and middle panels of Fig.~\ref{fig:HHG_2color_C3}~(a)].
By contrast, the same dynamical symmetries forbid the $n$th harmonic generations $|\tilde m^z(\omega)|$ for $n\not=3,~6,~9,~\cdots$ [see the bottom panel of Fig.~\ref{fig:HHG_2color_C3}~(a)].

We note that the trajectory of $C_3$-symmetric laser (three-leaf shape) and the symmetry of square lattice (i.e., $C_4$ symmetry) are incompatible in the sense of symmetry. Therefore, one can naively infer the absence of the dynamical symmetry and the selection rule in the square-lattice N\'eel ordered phase. However, as we discussed above, the selection rules indeed exist in the N\'eel phase. This emergence of the selection rules is owing to the fact that the dynamical symmetry operations in Eqs.~\eqref{dyn_sym_1_C3_Neel} and \eqref{dyn_sym_2_C3_Neel} include only global spin rotation and do not include any spatial operation. We will discuss this point in more detail later.

Let us move on to the canted phase. The selection rule disappears when the ground state has the canted order.
Figure~\ref{fig:HHG_2color_C3}~(b) shows that every harmonic generation appears in the canted phase although the symmetry of the Hamiltonian is kept unchanged.
The canted order breaks the $C_3$ spin rotation symmetry and the canted state is not invariant under the spin rotations in Eqs.~\eqref{dyn_sym_1_C3_Neel} and \eqref{dyn_sym_2_C3_Neel}.
Figures~\ref{fig:HHG_2color_C3}~(a) and (b) show that the symmetry (or symmetry breaking) of the magnetic order can change the qualitative properties of the harmonic generation spectra.

\subsection{Importance of lattice structures and Shastry-Sutherland model}
\label{sec:SSmodel}

To discuss the importance of the underlying lattices of magnetic Mott insulators in the harmonic generation spectrum, we compare the $C_3$-symmetric laser driven harmonic generations in the above N\'eel phase and the singlet-dimer phase in the Shastry-Sutherland (SS) model, which has been theoretically studied in Ref.~\cite{Udono2024_HHG}.
Figure~\ref{fig:HHG_2color_C3} (c) shows the lattice of the SS model~\cite{SriramShastry1981Exact}, the square lattice with the diagonal bonds in each plaquette (the minimal square).
These diagonal bonds are alternately aligned so that they do not share any edge points.
The SS model is the spin-$1/2$ quantum spin model whose Hamiltonian is made of two antiferromagnetic exchange interactions,
\begin{align}
  \hat{\mathcal H}_{\rm S}
  &= J' \sum_{\braket{\bm r,\bm r'}} \hat{\bm S}_{\bm r} \cdot \hat{\bm S}_{\bm r'} + J \sum_{\braket{\braket{\bm r,\bm r'}}} \hat{\bm S}_{\bm r} \cdot \hat{\bm S}_{\bm r'},
  \label{H_ShastrySutherland_def}
\end{align}
where $\braket{\bm r,\bm r'}$  denotes the nearest-neighbor bond of the square lattice [dotted bond in Fig.~\ref{fig:HHG_2color_C3}~(c)] and $\braket{\braket{\bm r, \bm r'}}$ denotes the diagonal bond [black bond in Fig.~\ref{fig:HHG_2color_C3}~(c)].
It is established that the Hamiltonian \eqref{H_ShastrySutherland_def} has the dimer-singlet ground state for $0\le J'/J < 0.675$~\cite{Kageyama1999Exact, Miyahara1999Exact, Koga2000Quantum, Corboz2013Tensor}. 
In particular, for $J'=0$, the model \eqref{H_ShastrySutherland_def} consists of mutually isolated antiferromagnetic dimers, whose ground state obviously belongs to the dimer-singlet phase.
As is well known, the singlet state on the $J$ bond
\begin{align}
  \ket{s}= \frac{1}{\sqrt 2} (\ket{\uparrow \downarrow} - \ket{\downarrow\uparrow})
  \label{singlet}
\end{align}
made of two $S=1/2$ spins is invariant under any $\rm SU(2)$ spin rotation.
Therefore, the dimer-singlet ground state has the $\rm SU(2)$ spin rotation symmetry.
We might naively expect that the harmonic generation spectrum of the SS model in the singlet-dimer phase would show a selection rule similar to that of the model \eqref{H_Neel-Canted_def} in the N\'eel phase.
However, Ref.~\cite{Udono2024_HHG} shows no selection rule in the multiferroic SS model irradiated by a $C_3$-symmetric laser. We briefly review this result below.

Let us apply the $C_3$-symmetric two-color laser field to the Shastry-Sutherland model in the following manner.
We regard
\begin{align}
  \hat{\mathcal H}(t)
  &= \hat{\mathcal H}_{\rm S} + \hat{\mathcal H}_{\rm ext}(t), \label{eq:SS_laser1}\\
  \hat{\mathcal H}_{\rm ext}(t) &= - \bm B_{\rm ac}(t) \cdot \sum_{\bm r}\hat{\bm S}_{\bm r} - \bm E_{\rm ac}(t) \cdot \hat{\bm P}_S
  \label{eq:SS_laser}
\end{align}
as the full Hamiltonian including the laser field.
Here, $\hat{\bm P}_S$ is the electric polarization operator.
In addition to the ac magnetic field $\bm B_{\rm ac}(t)$, we take into account the electric field $\bm E_{\rm ac}(t)$ of the two-color laser
since the $\mathrm{SrCu_2(BO_3)_2}$~\cite{Kageyama1999Exact}, a faithful realization of the SS model \eqref{H_ShastrySutherland_def}, shows the magnetoelectric (ME) effect~\cite{Room2004Farinfrared,Cepas2004Theory,Giorgianni2023Ultrafast,Miyahara2023Theory}.
We assume that the ME coupling between the electric polarization and the spin operator~\cite{Pimenov2006_mag,Tokura2014Multiferroics} is the following magneto-striction type,
\begin{align}
  \hat{\bm P_S} = \sum_{\braket{\bm r, \bm r'}} \Bigl(\Pi_{\rm me} \hat{\bm S}_{\bm r}\cdot \hat{\bm S}_{\bm r'}\Bigr) \bm e_{\bm r\bm r'},
  \label{EP_SS_def}
\end{align}
where $\Pi_{\rm me}$ is a coefficient originating from the ME coupling and $\bm e_{\bm r\bm r'}=(\bm r-\bm r')/|\bm r-\bm r'|$ is the unit vector on the $J'$ bond. In magnet $\mathrm{SrCu_2(BO_3)_2}$, this magneto-striction type of the ME coupling is known to be dominant~\cite{Room2004Farinfrared,Cepas2004Theory,Giorgianni2023Ultrafast,Miyahara2023Theory} in the THz regime.

Figure~\ref{fig:HHG_2color_C3}~(d) shows the harmonic generation spectrum $|\hat{\bm P}_S(\omega)|$ in the singlet-dimer phase when the $C_3$-symmetric two-color laser is applied. This is calculated in Ref.~\cite{Udono2024_HHG}.
The spectrum clearly indicates that there is no selection rule and every harmonic generation appears.

Comparing the $C_3$-symmetric laser driven spectra of the N\'eel phase [Fig.~\ref{fig:HHG_2color_C3}~(a)] with that of the SS model [Fig.~\ref{fig:HHG_2color_C3}~(d)], we consider the reason why a higher-symmetric ($\rm SU(2)$-symmetric) dimer-singlet state in the SS model has no selection rule.
As we discussed in Sec.~\ref{sec:2color_C3}, in the N\'eel phase, the change of the ac magnetic field by time translation can be compensated by a global spin rotation around the $S^z$ axis, and as a result, we obtain the dynamical symmetries of Eqs.~\eqref{dyn_sym_1_C3_Neel} and \eqref{dyn_sym_2_C3_Neel}. This logic holds because the spin-light coupling is the Zeeman interaction type and the total spin is coupled to the ac field. However, in the case of the SS model, the spin-light coupling is an ME coupling of Eq.~\eqref{EP_SS_def}, which is invariant under a continuous spin rotation around the $S^z$ axis since the inner product of two spins $\hat{\bm S}_{\bm r}\cdot \hat{\bm S}_{\bm r'}$ in Eq.~\eqref{EP_SS_def} is unchanged by arbitrary spin rotation. This implies that the change of the ac field via a time translation cannot be canceled out by spin rotation, unlike the case of the N\'eel phase. Thus, we cannot find dynamical symmetry in the singlet-dimer state of the SS model with the ME coupling.

An important point is that the ac Zeeman coupling is the interaction between a uniform ac magnetic field and total spin (i.e., a generator of global spin rotation) and thereby it does not include the information about the underlying lattice (spatial) structure of the target system.
However, in usual laser-driven crystal systems, the static part of the Hamiltonian, the ordering pattern, or the light-matter coupling are correlated with the lattice structure. As a result, a spatial symmetry operation ($\pi$ rotation, parity, mirror operations, etc.) is usually necessary to compensate a time shift of the applied two-color laser field. An example of the relation between spatial symmetry of the target and the applied laser trajectory is given by considering the SS model.
The SS model possesses the $C_2$-rotation symmetry but does not have the $C_3$ symmetry. This means that the spatial symmetry of the SS model ($C_2$) is incompatible with the trajectory of the $C_3$-symmetric laser. Therefore, (as we have already discussed) the SS model~\eqref{eq:SS_laser1} has no dynamical symmetry. On the other hand, if we apply a $C_4$-symmetric two-color laser [see Fig.~\ref{fig:pulse}(c)] to the SS model, it is shown in Ref.~\cite{Udono2024_HHG} that a dynamical symmetry and a corresponding selection rule appear since the $C_2$ symmetry of the SS model is  a subgroup of the $C_4$ symmetry of the laser trajectory, that is, $C_2\, \subset\, C_4$.

We can also find a similar relationship between the lattice symmetry and the trajectory of two-color laser field if we consider laser-driven harmonic generation in solid electron systems.
For example, let us consider the harmonic generation in graphene, that is a many-electron system on a honeycomb lattice. A recent theoretical study of Ref.~\cite{Kanega2025Twocolor} shows the following result. The honeycomb lattice on the $x$-$y$ plane possesses the $C_3$ rotation symmetry and thus the lattice is compatible with the trajectory of the $C_3$-symmetric laser. As a result, we obtain the dynamical symmetries associated with rotating operations $\hat U_{\rm rot}(2\pi/3)$ and $\hat U_{\rm rot}(4\pi/3)$. Note that the rotation operator $\hat U_{\rm rot}(\theta)$ means a real-space rotation by $\theta$ around the $z$ axis, unlike the spin rotations in Eqs.~\eqref{dyn_sym_1_C3_Neel} and \eqref{dyn_sym_2_C3_Neel}.
These dynamical symmetries lead to the selection rule that all the $3n$-order harmonic generation peaks ($n$ is an integer) do not appear in graphene irradiated by a $C_3$-symmetric laser. On the other hand, if we apply a different two-color laser, whose trajectory has no rotation symmetry, to graphene, we have no selection rule and all order harmonics emerge.

From these arguments, we can see how exceptional $\rm U(1)$ spin-rotation symmetric systems with ac Zeeman coupling (including two-color laser driven N\'eel phase) are. Namely, (as we emphasized) in U(1)-symmetric spin models, a time shift of two-color laser field can be compensated by a global spin rotation without spatial symmetry operation.
We will extend this argument of two-color laser in the following subsections~\ref{sec:2color_C4} and \ref{sec:generalization}.


\subsection{\texorpdfstring{$C_4$}{C_4}-symmetric laser}
\label{sec:2color_C4}

\begin{figure*}[t!]
  \centering
  \includegraphics[bb = 0 0 1573 1137, width=0.7\linewidth]{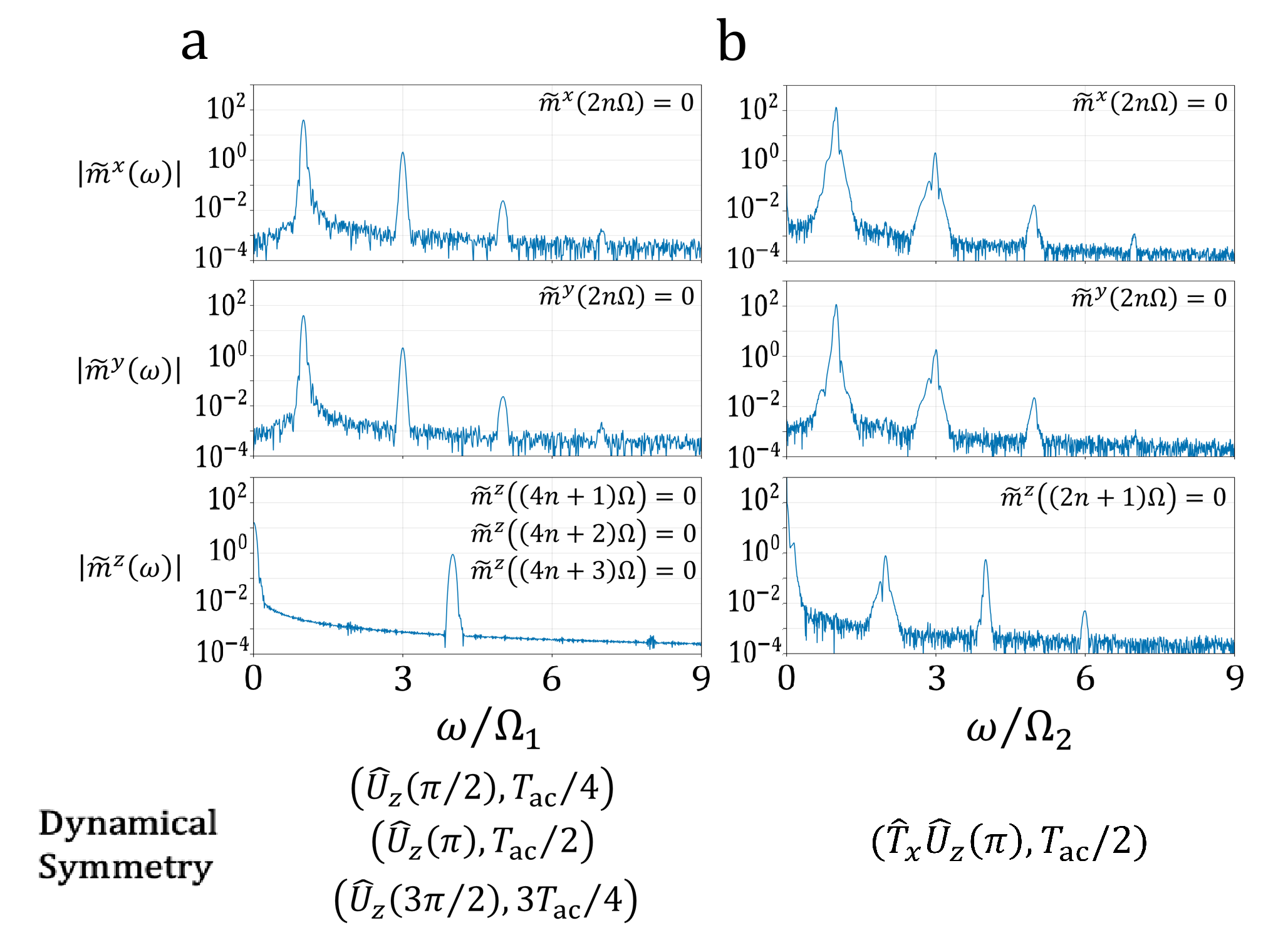}
  \caption{
    (a) Spectra $|\tilde m^a(\omega)|$ for $a=x,~y,~z$ in N\'eel phase where two-color laser pulse with $C_4$ symmetry around $z$ axis is applied.
    (b) Spectra $|\tilde m^a(\omega)|$ for $a=x,~y,~z$ in canted phase where two-color laser pulse with $C_4$ symmetry around $z$ axis is applied.
    We set the magnetic-field strength $B_0$ to (a) $B_0/J=0.1$ and (b) $B_0/J=3.0$.
    We also set $B_{\rm ac}/J = 0.1$ in both cases. The laser frequencies $\Omega_1$ and $\Omega_2$ are respectively given by
  $\hbar\Omega_{1}/J = \epsilon_{\bm{0}}^{\text{N\'eel}, \beta}/J = 1.73$ and $\hbar\Omega_{2}/J = \epsilon_{\bm{0}}^{\text{Cant}, \alpha}/J=2.7$. The symmetry operation below each column denotes the dynamical symmetry in each setup.}
  \label{fig:HHG_2color_C4}
\end{figure*}

This subsection is devoted to the N\'eel and canted phases irradiated by a $C_4$-symmetric laser with $\ell=3$.
Figures~\ref{fig:HHG_2color_C4}~(a) and (b) show the $C_4$-symmetric laser  driven harmonic generation spectra $|\tilde m^x(\omega)|$ in the N\'eel phase and the canted phase, respectively.
In analogy with the $C_3$-symmetric case, we obtain a generic relation between the $C_{\ell+1}$ spatial rotation and the temporal translation by $mT_{\rm ac}/(\ell+1)$ for $\ell=2,~3,~\cdots$ and $m = 1,~2,~\cdots~\ell$:
\begin{align}
  &
  \begin{pmatrix}
    \cos(\tfrac{2m\pi}{\ell+1}) & - \sin(\frac{2m\pi}{\ell+1}) & 0 \\
    \sin(\frac{2m\pi}{\ell+1}) & \cos(\frac{2m\pi}{\ell+1}) & 0 \\
    0 & 0 & 1
  \end{pmatrix}
  \bm B'_{2,\ell}(t)
  \notag \\
  &= \bm B'_{2,\ell} \biggl( t + \frac{mT_{\rm ac}}{\ell+1}\biggr),
  \label{eq:generic_l}
\end{align}
This relation implies that combining the $2m\pi/(\ell+1)$ spin rotation around the $S^z$ axis and the temporal translation by $m T_{\rm ac}/(\ell+1)$, we can transform $\bm B'_{2,\ell}(t)$ to $\bm B'_{2,\ell}(t+m T_{\rm ac}/(\ell+1))$ in the Hamiltonian. In other words, a time shift of the ac magnetic field can be compensated by spin rotation around the $S^z$ axis by a proper angle.
Since the N\'eel-ordered state is symmetric under the U(1) spin rotation around the $S^z$ axis, we find the following dynamical symmetries for $\ell=3$:
\begin{align}
  (\hat U_z(\pi/2),\,\,T_{\rm ac}/4) \label{eq:neel_C4_1}\\
  (\hat U_z(\pi),\,\,T_{\rm ac}/2)
  \label{eq:neel_C4_2}\\
  (\hat U_z(3\pi/2),\,\,3T_{\rm ac}/4).
  \label{eq:neel_C4_3}
\end{align}
Equations~\eqref{eq:neel_C4_1}, \eqref{eq:neel_C4_2}, and \eqref{eq:neel_C4_3} respectively correspond to $m=1$, 2 and 3 of Eq.~\eqref{eq:generic_l}.
These three dynamical symmetries result in the selection rule: $|\tilde m^x(n\Omega)|=|\tilde m^y(n\Omega)|=0$ for even $n=2,~4,~\cdots$ and $|\tilde m^z(n\Omega)|\not=0$ only for $n$ multiple of $4$.

To find the dynamical symmetry of the canted phase driven by a $C_4$-symmetric laser, we have to consider not only the Hamiltonian but also the equilibrium state (ground state) before the laser application. This is because the canted phase spontaneously breaks the $\rm U(1)$ spin-rotation symmetry (although the Hamiltonian is the same as that of the N\'eel phase).
The canted ordered state in the $S^x$-$S^z$ plane is not invariant under a global $\pi/2$ spin rotation $\hat U_z(\pi/2)$, unlike the N\'eel phase. On the other hand, one can find that the state is invariant under the combination of a global $\pi$ spin rotation $\hat U_z(\pi)$ and the one-site translation $\hat T_x$ along the $x$ direction. Therefore, instead of Eqs.~\eqref{eq:neel_C4_1}-\eqref{eq:neel_C4_3}, we obtain the dynamical symmetry
\begin{align}
  (\hat T_x\, \hat U_z(\pi),\,\,T_{\rm ac}/2)
  \label{eq:cant_C4_1}
\end{align}
in the canted phase irradiated by a $C_4$-symmetric laser. As a result, we have the selection rule that the even-order harmonic peaks of $|\tilde m^{x,y}(\omega)|$ are forbidden, while the odd-order peaks of $|\tilde m^{z}(\omega)|$ are forbidden. These two rules are consistent with the numerical result of Fig.~\ref{fig:HHG_2color_C4} (b). We note that the combination of the spin rotation $\hat U_z(\pi)$ and the spatial operation $\hat T_x$ in Eq.~\eqref{eq:cant_C4_1} stems from the underlying lattice structure (square lattice). Namely, the spatial information is important to find the dynamical symmetry in the above canted phase, as in the case of the SS model in Sec.~\ref{sec:SSmodel}.

\subsection{Generalization to U(1)-symmetric states}
\label{sec:generalization}
From the arguments in Secs.~\ref{sec:2color_C3}-\ref{sec:2color_C4}, we can find that the class of the $\rm U(1)$ spin-rotation symmetric spin systems driven by a $C_{\ell+1}$-symmetric two-color laser always possesses the dynamical symmetry and the resulting selection rule if the spin-light coupling is the ac Zeeman interaction type.
The point is that the above statement holds irrespective of details of underlying lattices: Square, honeycomb, triangular, Kagome lattices, etc.
In addition to the N\'eel phase, the above $\rm U(1)$ systems include ferromagnetic phases, ferrimagnetic phases, and $\rm SU(2)$ or $\rm U(1)$ spin-rotation symmetric quantum spin liquid phases. This result is attributed to the fact that the ac magnetic field is coupled to total spin in the ac Zeeman interaction.

The argument around Eq.~\eqref{eq:generic_l} tells us that for a U(1)-symmetric system driven by $C_{\ell+1}$-symmetric two-color laser, one obtains $\ell$ kinds of dynamical symmetries whose symmetry operations are given by
\begin{align}
  \left(\hat U_z\left(2m\pi/(\ell+1)\right),\,\,mT_{\rm ac}/(\ell+1)\right)
  \label{eq:U1_twocolor}
\end{align}
with $m=1,2,\cdots,\ell$. This result implies that the preparation of an ideal two-color laser and the observation of harmonic generations driven by the two-color laser could be a powerful tool to understand the symmetry and the magnetic ordering pattern of target materials.

\section{Non-perturbative natures}
\label{sec:NonPerturb}

In Secs.~\ref{sec:MagHG} and \ref{sec:2color}, we have quantitatively estimated the harmonic generation spectra by numerically solving the LLG equation, without relying on any perturbative arguments.
It is worth comparing our numerical results with perturbative arguments.
This section focuses on the harmonic generations by one-color laser.

\subsection{Laser-strength dependence}

\begin{figure}[t!]
  \centering
  \includegraphics[bb = 0 0 1544 680, width=\linewidth]{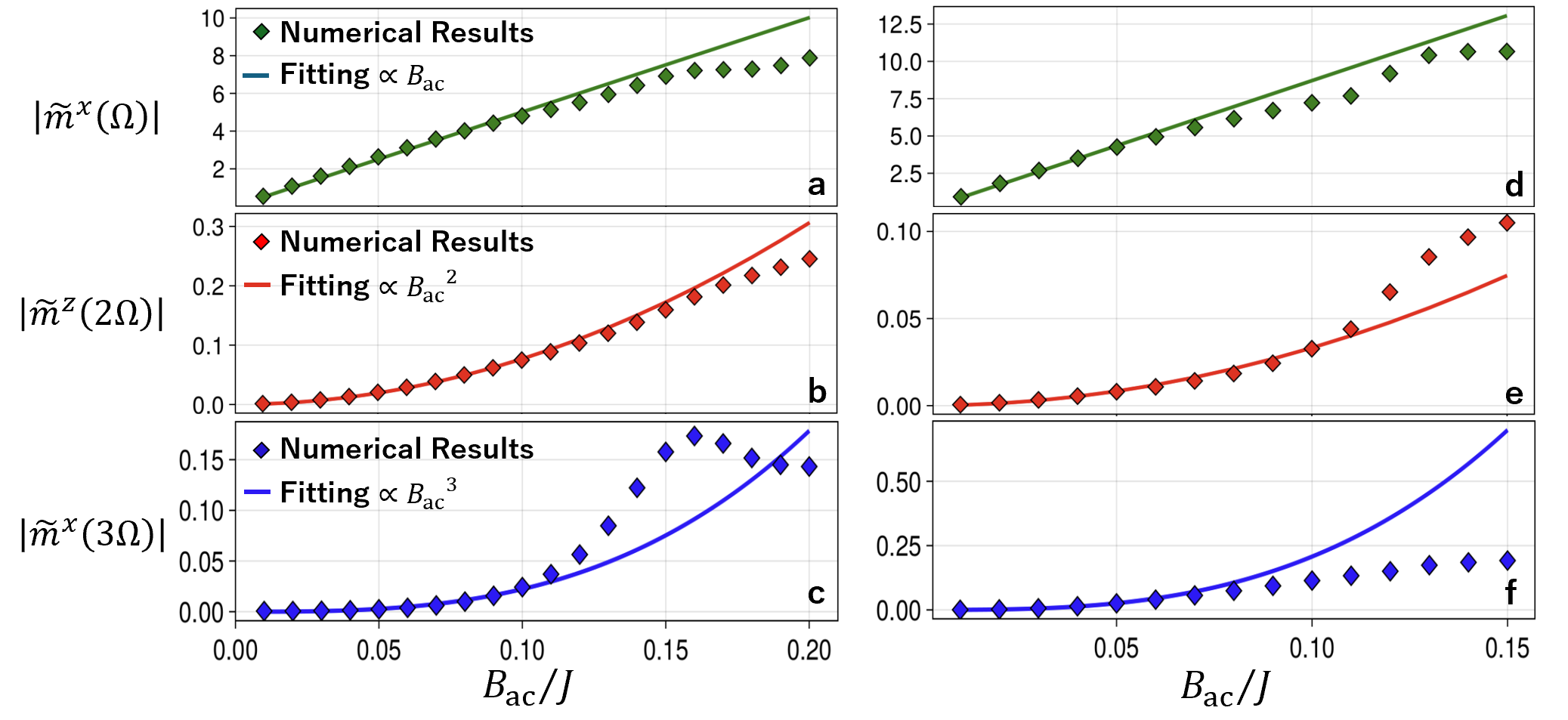}
  \caption{
    Diamond symbols in panels (a), (b), and (c) show $B_{\rm ac}$ dependence of $|\tilde m^x(\Omega)|$, $|\tilde m^z(2\Omega)|$, and $|\tilde m^x(3\Omega)|$, respectively, in the N\'eel phase with static magnetic field $B_0/J=0.5$ when we apply the one-color laser pulse oscillating in the $x$ axis. We set the laser frequency $\Omega$ to the resonant value: $\hbar\Omega=\epsilon_{\bm 0}^{\text{N\'eel},\beta}=1.33J$.
    The solid curves in panels (a), (b), and (c) show fitting curves $\propto B_{\rm ac}^n$ for (a) $n=1$, (b) $n=2$, and (c) $n=3$.
    Diamond symbols in panels (d), (e), and (f) show $B_{\rm ac}$ dependence of $|\tilde m^x(\Omega)|$, $|\tilde m^z(2\Omega)|$, and $|\tilde m^x(3\Omega)|$, respectively, in the canted phase with static magnetic field $B_0/J=3.0$  when we apply the one-color laser pulse oscillating in the $x$ axis. We set the laser frequency $\Omega$ to the resonant value: $\hbar\Omega=\epsilon_{\bm 0}^{\text{Cant},\alpha}=2.7J$.
    The solid curves in panels (d), (e), and (f) show fitting curves $\propto B_{\rm ac}^n$ for (d) $n=1$, (e) $n=2$, and (f) $n=3$.
  }
  \label{fig:str}
\end{figure}

When the Zeeman energy due to the laser field can be seen as a perturbation to the intrinsic Hamiltonian such as Eqs.~\eqref{H_Neel-Canted_def} and \eqref{H_WF_def}, we may expect that the strength of the $n$th harmonic generation, $|\tilde m^a(n\Omega)|$, is proportional to $(B_{\rm ac}/J)^n$, where $n$ photons resonantly turn into $n$ magnons.
As Fig.~\ref{fig:str} shows, this expectation indeed holds for small $B_{\rm ac}/J \lesssim 0.1$.
Figure~\ref{fig:str} also shows that the strength $|\tilde m^a(n\Omega)|$ deviates from the simple power-law curve, which clarifies that our numerical calculations successfully go beyond the perturbative regime, properly including nonperturbative effects.

As we discussed in Table~\ref{tab:laser}, the strength of $B_{\rm ac}=0.1J$ usually corresponds to an ac magnetic field of $\sim 1$ Tesla and an ac electric field of $\sim 1$MV/cm. A recent experimental research has observed the power-law behavior of $|\tilde m^a(n\Omega)|$ with $n=1$, 2 and 3 in the magnon harmonic generation of a WF phase~\cite{Zhang2023_HHG}.
This experiment has utilized ac magnetic fields of Tesla order. The above research has also detected the deviation from the power law when the strength of ac magnetic (electric) field is increased. Therefore, our numerical result in Fig.~\ref{fig:str} is at least qualitatively consistent with the experimental data in Ref.~\cite{Zhang2023_HHG}.

\subsection{Red shift of magnetic resonance}

\begin{figure}[t!]
  \centering
  \includegraphics[bb = 0 0 757 573, width=0.8\linewidth]{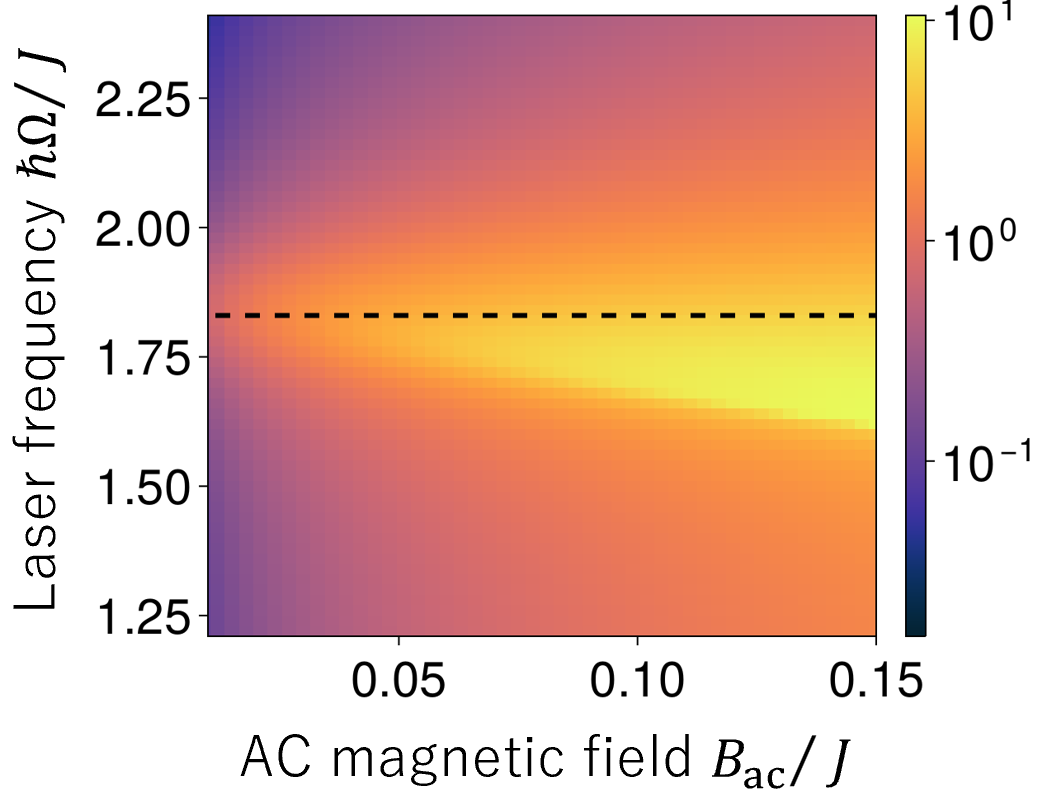}
  \caption{
    Spectrum intensity $|\tilde m^x(\Omega)|$ of first harmonic generation by one-color laser pulse along $x$ axis. We plot $|\tilde m^x(\Omega)|$ as a function of laser frequency $\Omega$ and strength of ac magnetic field $B_{\rm ac}$.
    The dashed line shows the resonant frequency $\epsilon^{\text{N\'eel}, \beta}_{\bm 0}/J = 1.83$, where the static magnetic field is not applied.
  }
  \label{fig:Redshift}
\end{figure}

There is another nonperturbative phenomenon, which we call red shift of magnetic resonance.
As we explained before, the first harmonic generation is related to the resonant absorption and generation of one magnon.
Hence, we may expect that the resonant frequency equals the magnon band at zero wave vector, say, $\epsilon_{\bm 0}^{\text{N\'eel},\beta}$ in the N\'eel phase.
This expectation holds when $B_{\rm ac}/J$ is weak enough but breaks down when it becomes strong.
Figure~\ref{fig:Redshift} shows the spectrum strength $|\tilde m^x(\Omega)|$ in the N\'eel phase under the condition of magnetic resonance.
When $B_{\rm ac}/J\ll 1$, the bright region with large strength is sharp and is located on the horizontal dashed line that represents the magnon gap $\hbar\Omega=\epsilon_{\bm 0}^{\text{N\'eel},\beta}=1.83J$.
As $B_{\rm ac}/J$ increases, the bright region shifts to the low-frequency side, which we call the red shift.
The red shift is the nonperturbative effect.
The larger $B_{\rm ac}/J$, the more the THz laser field excites magnons. Therefore, such a strong THz laser field is expected to lead to the shortening of the spin ``length'' $S$ and ultimately lowering the magnon band $\epsilon_{\bm 0}^{\text{N\'eel},\beta}$.

This red shift has been observed in several experiments~\cite{Mukai2016_THz,Zhang2023_HHG} of THz-laser driven spin dynamics, in which a strong enough THz laser pulse of Tesla order is applied to antiferromagnets.
Our numerical result of Fig.~\ref{fig:Redshift} thereby agrees with these experiments at least qualitatively.

\section{Summary}
\label{sec:summary}

We have investigated the harmonic generation spectrum of antiferromagnetic Mott insulators in ordered phases.
We apply intense THz laser (or GHz wave) pulse fields to these magnetic Mott insulators through the Zeeman interaction between the total spin $\sum_{\bm r} \hat{\bm S}_{\bm r}$ and the oscillating pulse magnetic field $\bm B_{\rm ac}(t)$: $- \sum_{\bm r} \bm B_{\rm ac}(t) \cdot \hat{\bm S}_{\bm r}$.
The real time dynamics of these systems under the laser field is investigated by means of the LLG equation~\eqref{stochastic-LLG_def}.
The LLG equation is known to describe well the dynamics of microscopic magnetic moment $\bm m_{\bm r}$ at each site $\bm r$ if we consider magnetically ordered states in magnets.
The equation contains the phenomenological Gilbert damping effect and the stochastic field $\bm \xi_{\bm r}(t)$ that is introduced to take finite-temperature effects into account.
Solving this stochastic LLG equation numerically, we obtain the spin dynamics of antiferromagnetically ordered phases: The N\'eel, canted and WF phases.

We have considered two model Hamiltonians~\eqref{H_Neel-Canted_def} and \eqref{H_WF_def} to describe magnetic Mott insulators in ordered phases.
The former model shows the N\'eel and canted phases [Fig.~\ref{fig:model}~(a)], where spontaneous symmetry breaking occurs.
The latter model shows the WF phase whose magnetic structure is akin to that in the canted phase.
However, no spontaneous symmetry breaking occurs in the WF phase and the ground state keeps all the symmetries that the Hamiltonian \eqref{H_WF_def} possesses.
We also consider two kinds of the laser field: one-color laser \eqref{B_1_def} and two-color laser \eqref{B_2_def}.
The one-color laser \eqref{B_1_def} is the linearly oscillating field.
The two-color laser \eqref{B_2_def} draws a more complex spatial trajectory whose spatial pattern is determined by the parameter $\ell$ [Fig.~\ref{fig:pulse}~(b) and (c)].
The choices $\ell=2$ and $3$ of the parameter make the two-color laser field exhibit the $C_3$- and $C_4$-symmetric spatial patterns, respectively.

We have obtained plenty of harmonic generation spectra in various situations.
The selection rule of the spectrum critically depends on the dynamical symmetry that stems from the combination of the symmetry of the Hamiltonian, the symmetry of the ground state, and the symmetry of the laser field.
Here, the selection rule is referred to as the existence or absence of a partial series of the harmonic generation.
We first analyze the magnon harmonic generation driven by a standard one-color laser in Sec.~\ref{sec:MagHG}.
When applying a one-color laser field to the model \eqref{H_Neel-Canted_def} in the N\'eel phase, we observe the different selection rules depending on the existence or absence of the static field $B_0$ [see Fig.~\ref{fig:neelhhg} (a) and (b)]. Moreover, we find out that the same even-odd nature of the harmonic generation peaks appears in both N\'eel and canted phases with a finite field of $B_0$ even though these two phases spontaneously break symmetry in different ways (we have different peak structures in the N\'eel and canted phases when we apply a two-color laser as shown in Fig.~\ref{fig:HHG_2color_C3}).
We also detect another contrasting comparison between the canted and WF phases.
The Hamiltonians in these two phases are different: Eq.~\eqref{H_Neel-Canted_def} for the canted phase and Eq.~\eqref{H_WF_def} for the WF phase.
However, the symmetry of their ground states is the same.
Comparing Figs.~\ref{fig:Canthhg} and \ref{fig:wfhhg}, we find different selection rules in the case of ${\bm B}_{\rm ac}\parallel {\bm e}_{z}$ when the Hamiltonians are different but the ground states' symmetry is the same.
We emphasize that the symmetry of the phase and the symmetry of the Hamiltonian can differ in the ordered phases of magnetic Mott insulators, both of which are important for the selection rule in the harmonic generation spectrum. The harmonic generation peak structure of the WF phases for ${\bm B}_{\rm ac}\parallel {\bm e}_{z}$ is consistent with a recent experimental result in Ref.~\cite{Zhang2023_HHG}.
We also note that when the U(1) spin-rotation symmetry is conserved and each magnon has an angular momentum $+\hbar$ or $-\hbar$ like the N\'eel phase, the magnon resonance condition (i.e., the condition for large harmonic generation) can be understood from the angular momentum conservation law (see Fig.~\ref{fig:photonmagnon}).

The application of two-color laser fields makes the situation even richer. Section~\ref{sec:2color} is devoted to the investigation of such two-color laser induced harmonic generations.
We find the characteristic selection rules related to $C_3$ [Fig.~\ref{fig:HHG_2color_C3}~(a)] and $C_4$ symmetry [Fig.~\ref{fig:HHG_2color_C4}~(a)] in the N\'eel phase.
The point is that a time shift of the ac magnetic field can be canceled out by applying a proper spin rotation around the $S^z$ axis if we consider the U(1) spin-rotation symmetric magnets (including the N\'eel state) with an ac Zeeman coupling between the total spin and two-color laser.
We also find that the selection rule is limited in the canted phase since the canted order breaks the spin rotation symmetry around the $S^z$ axis and is incompatible with the $C_3$ symmetry of the two-color laser for $\ell=2$ [Fig.~\ref{fig:HHG_2color_C3}~(b)].
Still, we obtain the limited but nontrivial selection rule in the canted phase for the $C_4$-symmetric laser [Fig.~\ref{fig:HHG_2color_C4}~(b)]. From the results of $C_3$ and $C_4$-symmetric laser induced harmonic generations, we show that a generic selection rule holds in a broad class of two-color laser driven U(1) spin-rotation symmetric magnets with an ac Zeeman interaction. For this class (including the N\'eel phase), one can always have the $\ell$ dynamical symmetries of Eq.~\eqref{eq:U1_twocolor} irrespective of the underlying lattice structure.
We emphasize that these selection rules in two-color laser driven harmonic generation are special for the U(1) symmetric magnets with ac Zeeman interaction. Usually, the spatial pattern of the trajectory of two-color laser (see Fig.~\ref{fig:pulse}) interacts with the lattice structure of laser-irradiated materials, and therefore a matching between the trajectory and the lattice structure is required to obtain a selection rule. This has been shown, for instance, in graphene~\cite{Kanega2025Twocolor} and Shastry-Sutherland magnet~\cite{Udono2024_HHG} irradiated by two-color laser.

In Sec.~\ref{sec:NonPerturb}, we discuss the laser-strength dependence of the magnon harmonic generation spectrum.
By numerically solving the LLG equation, we show that some non-perturbative effects (deviation from the power law in Fig.~\ref{fig:str} and the red shift in Fig.~\ref{fig:Redshift}) emerge when the strength of the ac magnetic field $B_{\rm ac}$ approaches the order of $0.1J$, where $J$ is the strength of the dominant exchange interaction in antiferromagnets.
The intensity $B_{\rm ac}=0.1J$ corresponds to about a Tesla order of the ac magnetic field in real antiferromagnetic materials. These numerical results of the non-perturbative effects are consistent with recent experimental observations~\cite{Mukai2016_THz,Zhang2023_HHG}.

From our comprehensive investigation, we conclude that the harmonic generation spectrum $|\tilde m^a(\omega)|$ of magnetic Mott insulators possesses rich information on magnetic interactions and magnetic orders.
This paper established the fundamental understanding about the harmonic generation spectrum in magnetically ordered phases and the associated dynamical symmetry.

Finally we briefly comment on the difference between the laser-driven quantum magnet following Schr\"odinger (or Heisenberg) equation and the semiclassical magnet following the LLG equation. As we discussed, we have used the LLG equation to numerically analyze the harmonic generation of the THz-laser driven ordered antiferromagnets because spin dynamics of ordered magnets is known to be nicely described by the LLG equation.
However, if we consider a phenomenon deeply associated with the spin commutation relation, there would be a possibility that the result of a quantum ordered magnet largely deviates from that of the corresponding classical magnet.
Finding such a quantum nature of laser-driven ordered magnets might offer an interesting open issue.

\begin{acknowledgments}
  The authors thank Hogara Watanabe for fruitful discussions.
  M.S. is supported by JSPS KAKENHI (Grants No.~JP25K07198, No.~JP25H02112, No.~JP22H05131, No.~JP25H01609 and No.~JP25H01251) and JST, CREST Grant No.~JPMJCR24R5, Japan.
\end{acknowledgments}


\appendix

\section{Linear spin-wave theory}
\label{app:LSW}

This appendix presents the linear spin-wave theory in the ordered phases in antiferromagnets (the N\'eel, canted, and WF phases) for self-containedness of the paper.
The linear spin-wave theory gives us the excitation gap of magnons and shows how the gap depends on parameters including the magnetic field.

As is well known, the spin-wave theory is a semiclassical theory of magnons built on a classical magnetic order.
We derive the spin-wave theory in the three magnetically ordered phases separately but in a parallel way.


\subsection{N\'eel phase}

The square lattice has two sublattices, say A and B.
Any site in the A sublattice is surrounded by four nearest-neighbor sites in the B sublattice and vice versa.
Let $\bm r_{\rm A}$ and $\bm r_{\rm B}$ be sites in the A sublattice and B sublattice, respectively.
We can assume that the classical N\'eel order has $\hat S_{\bm r_{\rm A}}^z = S$  and $\hat S_{\bm r_{\rm B}}^z = -S$.
Let us introduce the Holstein-Primakoff transformation~\cite{Holstein1940Field} as follows.
\begin{align}
  \hat S_{\bm r_{\rm A}}^z &= S- \hat a_{\bm r_{\rm A}}^\dag \hat a_{\bm r_{\rm A}},
  \label{HP_Neel_Sz_A_def} \\
  \hat S_{\bm r_{\rm A}}^+ &= \sqrt{2S- \hat a_{{\bm r}_{\rm A}}^\dag \hat a_{\bm r_{\rm A}}}~\hat{a}_{\bm r_{\rm A}},
  \label{HP_Neel_Splus_A_def} \\
  \hat S_{\bm r_{\rm B}}^z &= -S + \hat b_{\bm r_{\rm B}}^\dag \hat b_{\bm r_{\rm B}},
  \label{HP_Neel_Sz_B_def} \\
  \hat S_{\bm r_{\rm B}}^+ &= \hat{b}_{\bm r_{\rm B}}^\dag~\sqrt{2S-\hat b_{\bm r_{\rm B}}^\dag \hat b_{\bm r_{\rm B}}},
  \label{HP_Neel_Splus_B_def}
\end{align}
where $\hat S_{\bm r}^+ = \hat S_{\bm r}^x+i \hat S_{\bm r}^y$ and $\hat S_{\bm r}^- = (\hat S_{\bm r}^+)^\dag = \hat S_{\bm r}^x -i \hat S_{\bm r}^y$ are the ladder operators of the spin, $\hat a_{\bm r}$ and $\hat b_{\bm r}$ are bosonic annihilation operators at the site $\bm r$.
The annihilation operators $\hat a_{\bm r}$ and $\hat b_{\bm r}$ and their Hermitian conjugate operators, $\hat a_{\bm r}^\dag$ and $\hat b_{\bm r}^\dag$, satisfy the standard bosonic commutation relations such as $[\hat a_{\bm r},~\hat a_{\bm r'}^\dag] =[\hat b_{\bm r},~\hat b_{\bm r'}^\dag] =\delta_{\bm r,\bm r'}$ and $[\hat a_{\bm r},~\hat b_{\bm r'}]=0$.
Here, $\delta_{\bm r,\bm r'}$ is the Kronecker delta.
The Holstein-Primakoff transformation is nonlinear in the boson operators, leading to interactions among magnons.
The linear spin-wave theory drops the third- and higher-order terms about bosonic creation or annihilation operators and thus ignores interactions.
Within the framework of the linear spin-wave theory, the Holstein-Primakoff transformation is approximated as
\begin{align}
  S_{\bm r_{\rm A}}^z &= S-\hat a_{\bm r_{\rm A}}^\dag \hat a_{\bm r_{\rm A}},
  \label{HP_Neel_Sz_A_LSW} \\
  S_{\bm r_{\rm A}}^+ &\approx \sqrt{2S}~\hat a_{\bm r_{\rm A}},
  \label{HP_Neel_Splus_A_LSW} \\
  S_{\bm r_{\rm B}}^z &= - S + \hat b_{\bm r_{\rm B}}^\dag \hat b_{\bm r_{\rm B}},
  \label{HP_Neel_Sz_B_LSW} \\
  S_{\bm r_{\rm B}}^+ &\approx \sqrt{2S}~\hat b_{\bm r_{\rm B}}^\dag.
  \label{HP_Neel_Splus_B_LSW}
\end{align}
Let us plug the approximated Eqs.~\eqref{HP_Neel_Sz_A_LSW}, \eqref{HP_Neel_Splus_A_LSW}, \eqref{HP_Neel_Sz_B_LSW}, \eqref{HP_Neel_Splus_B_LSW} into the Hamiltonian \eqref{H_Neel-Canted_def} and drop the third- and higher-order terms in $\hat a_{\bm r}$ and $\hat a_{\bm r}^\dag$.
\begin{widetext}
  \begin{align}
    \hat{\mathcal H}_{\text{N-C}}
    &\approx - \frac 12 z JNS^2 - K NS^2 + JS \sum_{\bm r_{\rm A}, \bm \rho} (\hat a_{\bm r_{\rm A}} \hat b_{\bm r_{\rm A}+\bm \rho} + \hat a_{\bm r_{\rm A}}^\dag \hat b_{\bm r_{\rm A}+\bm \rho}^\dag + \hat a_{\bm r_{\rm A}}^\dag \hat a_{\bm r_{\rm A}} + \hat b_{\bm r_{\rm A}+\bm \rho}^\dag \hat b_{\bm r_{\rm A}+\bm \rho}) \notag \\
    &\qquad  + (2KS+B_0) \sum_{\bm r_{\rm A}} \hat a_{\bm r_{\rm A}}^\dag \hat a_{\bm r_{\rm A}}+ (2KS-B_0) \sum_{\bm r_{\rm B}} \hat b_{\bm r_{\rm B}}^\dag \hat b_{\bm r_{\rm B}},
  \end{align}
  where $N$ is the site number and $z=4$ is the coordination number.
  The vector $\bm \rho= a_0\bm e_x,~-a_0\bm e_x, ~a_0\bm e_y,~-a_0\bm e_y$ denotes the vector connecting nearest-neighbor sites, and $a_0$ is the lattice spacing.
  Applying the Fourier transform to the bosonic creation and annihilation operators, we obtain
  \begin{align}
    \hat{\mathcal H}_{\text{N-C}}
    &= - \frac 12 zJNS^2 - KNS^2 \nonumber\\
    &\qquad + \sum_{\bm k} \Bigl\{  (zJS+2KS+B_0) \hat a_{\bm k}^\dag \hat a_{\bm k} + (zJS+2KS-B_0) \hat b_{\bm k}^\dag \hat b_{\bm k} +  zJS\gamma_{\bm k}(\hat a_{\bm k} \hat b_{\bm k} + \hat a_{\bm k}^\dag \hat b_{\bm k}^\dag)
    \Bigr\},
    \label{H_Neel_LSW_quadratic}
  \end{align}
  where $\gamma_{\bm k} = \frac 1z \sum_{\bm \rho} e^{i\bm k\cdot \bm \rho} = \frac 12 \{ \cos (k_xa_0) + \cos(k_ya_0) \}$
  and the wavevector $\bm k$ is defined as $\bm k = (k_x,~k_y,~k_z)^\top$.
  We diagonalize the quadratic Hamiltonian \eqref{H_Neel_LSW_quadratic} by performing the Bogoliubov transformation.
  \begin{align}
    \begin{pmatrix}
      \hat \alpha_{\bm k} \\
      \hat \beta_{\bm k}^\dagger
    \end{pmatrix}
    &=
    \begin{pmatrix}
      \cosh\theta_{\bm k} & \sinh \theta_{\bm k} \\
      \sinh\theta_{\bm k} & \cosh \theta_{\bm k}
    \end{pmatrix}
    \begin{pmatrix}
      \hat a_{\bm k} \\
      \hat b_{\bm k}^\dagger
    \end{pmatrix}.
  \end{align}
  The transformed bosonic operators $\hat \alpha_{\bm k}$ and $\hat \beta_{\bm k}$ satisfy the standard commutation relation $[\hat \alpha_{\bm k},~\hat\alpha_{\bm k'}^\dag] = [\hat \beta_{\bm k},~\hat \beta_{\bm k'}^\dag] = \delta_{\bm k,\bm k'}$.
  The other pairs of these bosonic operators are commutative (e.g., $[\hat \alpha_{\bm k},~\hat \beta_{\bm k'}]=0$).
  Choosing the parameter $\theta_{\bm k}$ so that
  \begin{align}
    \tanh (2\theta_{\bm k}) = \frac{zJ\gamma_{\bm k}}{zJ+2K},
  \end{align}
  we can diagonalize the quadratic Hamiltonian \eqref{H_Neel_LSW_quadratic}. The result is
  \begin{align}
    \hat{\mathcal H}_{\text{N-C}}
    &= E_{\text{N\'eel}} + \sum_{\bm k} (\epsilon_{\bm k}^{\text{N\'eel},\alpha} \hat \alpha_{\bm k}^\dag \hat \alpha_{\bm k}
    + \epsilon_{\bm k}^{\text{N\'eel},\beta} \hat \beta_{\bm k}^\dag \hat \beta_{\bm k}),
    \label{H_LSW_diagonalized}
  \end{align}
  with
  \begin{align}
    E_{\text{N\'eel}} &= - \frac 12 zJNS^2 - KNS^2 + \sum_{\bm k} \biggl\{ \sqrt{(zJS+2KS)^2-(zJS\gamma_{\bm k})^2} - (zJS+2KS)
    \biggr\} , \\
    \epsilon_{\bm k}^{\text{N\'eel},\alpha}
    &= \sqrt{(zJS+2KS)^2 - (zJS\gamma_{\bm k})^2} + B_0,
    \label{magnon_band_Neel_alpha} \\
    \epsilon_{\bm k}^{\text{N\'eel},\beta}
    &= \sqrt{(zJS+2KS)^2 - (zJS\gamma_{\bm k})^2} - B_0.
    \label{magnon_band_Neel_beta}
  \end{align}
  We thus obtain the two magnon bands $\epsilon_{\bm k}^{\text{N\'eel},\alpha}$ and $\epsilon_{\bm k}^{\text{N\'eel},\beta}$ in the N\'eel phase.
\end{widetext}

\subsection{Canted phase}
\label{app:LSW_canted}
\newpage
The classical order in the canted phase is sketched in Fig.~\ref{fig:model}~(a), that is,
\begin{align}
  \hat{\bm S}_{\bm r_{\rm A}} =
  \begin{pmatrix}
    -\sin\theta \\
    0 \\
    \cos \theta
  \end{pmatrix}, \qquad
  \hat{\bm S}_{\bm r_{\rm B}} =
  \begin{pmatrix}
    \sin\theta \\
    0 \\
    \cos \theta
  \end{pmatrix},
  \label{classical_canted_config}
\end{align}
where $\theta$ is determined by $B_0/J$ to minimize the energy.
We thus find
\begin{align}
  \theta = \cos^{-1} \biggl( \frac{B_0}{2S(zJ-K)} \biggr).
  \label{canted_angle_classic}
\end{align}
We consider the canted order within the $S^z$-$S^x$ plane without loss of generality since the model \eqref{H_Neel-Canted_def} has the U(1) spin-rotation symmetry around the $S^z$ axis.
The classical canted order is generated by tilting the ferromagnetic order along the $S^z$ axis with angles $\pm \theta$, where the ratio $B_0/J$ uniquely determines $\theta$.
As Fig.~\ref{fig:model}~(a) shows, the spin $\hat{\bm S}_{\bm r_{\rm A}}$ in the A sublattice is tilted by $-\theta$ around the $S^y$ axis and the spin $\hat{\bm S}_{\bm r_{\rm B}}$ in the B sublattice is tilted by $+\theta$ around the $S^y$ axis.
Let us consider another coordinate system $(\hat S_{\bm r}^\xi,~\hat S_{\bm r}^\eta,~\hat S_{\bm r}^\zeta)^\top$ related to the original one $(\hat S_{\bm r}^x,~\hat S_{\bm r}^y,~\hat S_{\bm r}^z)^\top$ through the rotation around the $y$ axis:
\begin{align}
  \begin{pmatrix}
    \hat S_{\bm r_{\rm A}}^x \\
    \hat S_{\bm r_{\rm A}}^y \\
    \hat S_{\bm r_{\rm A}}^z
  \end{pmatrix}
  &=
  \begin{pmatrix}
    \cos\theta & 0 & - \sin \theta \\
    0 & 1 & 0 \\
    \sin\theta & 0 & \cos \theta
  \end{pmatrix}
  \begin{pmatrix}
    \hat S_{\bm r_{\rm A}}^\xi \\
    \hat S_{\bm r_{\rm A}}^\eta \\
    \hat S_{\bm r_{\rm A}}^\zeta
  \end{pmatrix},
  \label{Spin_A_canting_rotation} \\
  \begin{pmatrix}
    \hat S_{\bm r_{\rm B}}^x \\
    \hat S_{\bm r_{\rm B}}^y \\
    \hat S_{\bm r_{\rm B}}^z
  \end{pmatrix}
  &=
  \begin{pmatrix}
    \cos\theta & 0 & \sin \theta \\
    0 & 1 & 0 \\
    - \sin\theta & 0 & \cos \theta
  \end{pmatrix}
  \begin{pmatrix}
    \hat S_{\bm r_{\rm B}}^\xi \\
    \hat S_{\bm r_{\rm B}}^\eta \\
    \hat S_{\bm r_{\rm B}}^\zeta
  \end{pmatrix}.
  \label{Spin_B_canting_rotation}
\end{align}
In the coordinate $(\hat S_{\bm r}^\xi,~\hat S_{\bm r}^\eta,~\hat S_{\bm r}^\zeta)^\top$, the classical canted order \eqref{classical_canted_config} corresponds to the ferromagnetic order,
$(\hat S_{\bm r_{\rm A}}^\xi,~\hat S_{\bm r_{\rm A}}^\eta,~\hat S_{\bm r_{\rm A}}^\zeta)^\top=(\hat S_{\bm r_{\rm B}}^\xi,~\hat S_{\bm r_{\rm B}}^\eta,~\hat S_{\bm r_{\rm B}}^\zeta)^\top= (0,0,1)^\top$.
Hence, we can utilize the spin-wave theory in the ferromagnetic phase.
The spin-wave theory here, however, has two kinds of bosons unlike that in the ferromagnetic phase because of the existence of the two sublattices.
We introduce the following Holstein-Primakoff transformation:
\begin{align}
  \hat S_{\bm r_{\rm A}}^\zeta &= S - \hat a_{\bm r_{\rm A}}^\dag \hat a_{\bm r_{\rm A}}, \\
  \hat S_{\bm r_{\rm A}}^\xi +i \hat S_{\bm r_{\rm A}}^\eta &\approx \sqrt{2S}~\hat a_{\bm r_{\rm A}}, \\
  \hat S_{\bm r_{\rm B}}^\zeta &= S - \hat b_{\bm r_{\rm B}}^\dag \hat b_{\bm r_{\rm B}}, \\
  \hat S_{\bm r_{\rm B}}^\xi+ i \hat S_{\bm r_{\rm B}}^\eta &\approx \sqrt{2S}~\hat b_{\bm r_{\rm B}},
\end{align}
where we have already dropped the higher-order terms.
In the original coordinate system $(\hat S_{\bm r}^x,~\hat S_{\bm r}^y,~\hat S_{\bm r}^z)^\top$, the spin operator is written in terms of these bosonic operators as follows.
\begin{widetext}
  \begin{align}
    \hat S_{\bm r_{\rm A}}^z &= \sqrt{\frac S2} \sin \theta (\hat a_{\bm r_{\rm A}}+ \hat a_{\bm r_{\rm A}}^\dag ) + \cos \theta (S-\hat a_{\bm r_{\rm A}}^\dag \hat a_{\bm r_{\rm A}}), \\
    \hat S_{\bm r_{\rm A}}^+ &\approx \sqrt{\frac S2} ( \cos \theta + 1) \hat a_{\bm r_{\rm A}} + \sqrt{\frac S2} (\cos\theta-1)\hat a_{\bm r_{\rm A}}^\dag - \sin \theta (S-\hat a_{\bm r_{\rm A}}^\dag \hat a_{\bm r_{\rm A}}), \\
    S_{\bm r_{\rm B}}^z &= -\sqrt{\frac S2} \sin \theta (\hat b_{\rm r_{\rm B}}+\hat b_{\bm r_{\rm B}}^\dag) + \cos\theta (S-\hat b_{\bm r_{\rm B}}^\dag \hat b_{\bm r_{\rm B}}), \\
    S_{\bm r_{\rm B}}^+ &\approx \sqrt{\frac S2} (\cos\theta +1) \hat b_{\bm r_{\rm B}} + \sqrt{\frac S2} (\cos\theta - 1) \hat b_{\bm r_{\rm B}}^\dag + \sin \theta (S - \hat b_{\bm r_{\rm B}}^\dag \hat b_{\bm r_{\rm B}}).
  \end{align}
  The rest of the procedure is parallel to that in the N\'eel phase.
  We rewrite the Hamiltonian \eqref{H_Neel-Canted_def} in terms of boson operators.
  \begin{align}
    \hat{\mathcal H}_{\text{N-C}}
    &= \frac 12 zJNS^2 \cos(2\theta) - KNS^2 \cos^2\theta - B_0NS \cos \theta - \frac 12 KNS \sin^2\theta \notag \\
    &\qquad + \sum_{\bm r_{\rm A}}\sqrt{\frac S2} \sin\theta \{2S(zJ-K) \cos\theta-B_0 \} (\hat a_{\bm r_{\rm A}} + \hat a_{\bm r_{\rm A}}^\dag) - \sum_{\bm r_{\rm B}}\sqrt{\frac S2} \sin \theta \{ 2S(zJ-K) \cos\theta - B_0 \} (\hat b_{\bm r_{\rm B}} + \hat b_{\bm r_{\rm B}}^\dag) \notag \\
    &\qquad + \sum_{\bm r_{\rm A},\bm \rho}\biggl\{ \frac 12 JS(\cos(2\theta)+1) (\hat a_{\bm r_{\rm A}} \hat b_{\bm r_{\rm A}+\bm \rho}^\dag + \hat a_{\bm r_{\rm A}}^\dag \hat b_{\bm r_{\rm A}+\bm \rho})
      + \frac 12 JS (\cos(2\theta) - 1) (\hat a_{\bm r_{\rm A}} \hat b_{\bm r_{\rm A}+\bm \rho} + \hat a_{\bm r_{\rm A}}^\dag \hat b_{\bm r_{\rm A}+\bm \rho}^\dag)
    \biggr\} \notag \\
    &\qquad + \sum_{\bm r_{\rm A}} \biggl[- \frac 12 KS \sin^2\theta (\hat a_{\bm r_{\rm A}}\hat a_{\bm r_{\rm A}} + \hat a_{\bm r_{\rm A}}^\dag \hat a_{\bm r_{\rm A}}^\dag) + \{ -zJS\cos(2\theta) + KS(3\cos^2\theta-1) + B_0 \cos\theta \} \hat a_{\bm r_{\rm A}}^\dag \hat a_{\bm r_{\rm A}}
    \biggr] \notag \\
    &\qquad + \sum_{\bm r_{\rm B}} \biggl[
      -\frac 12 KS \sin^2\theta (\hat b_{\bm r_{\rm B}} \hat b_{\bm r_{\rm B}} + \hat b_{\bm r_{\rm B}}^\dag \hat b_{\bm r_{\rm B}}^\dag) + \{-zJS\cos(2\theta) + KS(3\cos^2\theta-1) + B_0 \cos\theta \} \hat b_{\bm r_{\rm B}}^\dag \hat b_{\bm r_{\rm B}}
    \biggr].
  \end{align}
  The linear terms proportional to $(\hat a_{\bm r_{\rm A}}+\hat a_{\bm r_{\rm A}}^\dag)$ or to $(\hat b_{\bm r_{\rm B}}+\hat b_{\bm r_{\rm B}}^\dag)$, which should vanish, indeed vanish because of the relation \eqref{canted_angle_classic} between the canted angle $\theta$ and the ratio $B_0/J$.
  The Fourier transform of this quadratic Hamiltonian is
  \begin{align}
    \hat{\mathcal H}_{\text{N-C}}
    &= \frac 12 zJNS^2\cos(2\theta) - KNS^2\cos^2\theta - B_0NS\cos\theta - \frac 12 KNS \sin^2\theta \notag \\
    &\quad + \sum_{\bm k} \bigl\{ - zJS \cos(2\theta) + KS(3\cos^2\theta-1) +B_0\cos\theta \bigr\} (\hat a_{\bm k}^\dag \hat a_{\bm k} + \hat b_{\bm k}^\dag \hat b_{\bm k}) + \sum_{\bm k} \frac 12 zJS(\cos(2\theta)-1) \gamma_{\bm k} (\hat a_{\bm k} \hat b_{\bm k} + \hat a_{\bm k}^\dag \hat b_{\bm k}^\dag) \notag \\
    &\qquad + \sum_{\bm k} \frac 12 zJS(\cos(2\theta) +1) \gamma_{\bm k} (\hat a_{\bm k} \hat b_{-\bm k}^\dag + \hat a_{\bm k}^\dag \hat b_{-\bm k}) -\sum_{\bm k} \frac 12 KS \sin^2\theta (\hat a_{\bm k} \hat a_{-\bm k} + \hat a_{\bm k}^\dag \hat a_{-\bm k}^\dag + \hat b_{\bm k}\hat b_{-\bm k} + \hat b_{\bm k}^\dag \hat b_{-\bm k}^\dag).
  \end{align}
  To lighten the notation, we introduce the following parameters,
  \begin{align}
    A &= - zJS \cos(2\theta) + KS (3\cos^2\theta-1) + B_0 \cos\theta = zJS -KS\sin^2\theta, \\
    B_{\bm k} &= \frac 12 zJS (\cos(2\theta)-1) \gamma_{\bm k} = - zJS \sin^2\theta \gamma_{\bm k}, \\
    C_{\bm k} &= \frac 12 zJS (\cos(2\theta)+1) \gamma_{\bm k} = zJS \cos^2\theta \gamma_{\bm k}, \\
    F &= - \frac 12 KS \sin^2\theta,
  \end{align}
  and the following operators,
  \begin{align}
    \hat c_{\bm k} &= \frac{1}{\sqrt 2} (\hat a_{\bm k} + \hat b_{-\bm k}), \\
    \hat d_{\bm k} &= \frac{1}{\sqrt 2} (\hat a_{\bm k} - \hat b_{-\bm k}).
  \end{align}
  The Hamiltonian then becomes
  \begin{align}
    \hat{\mathcal H}_{\text{N-C}}
    &= \frac 12 zJNS^2\cos(2\theta) - KNS^2\cos^2\theta-B_0NS\cos\theta - \frac 12 KNS\sin^2\theta \notag \\
    &\qquad + \sum_{\bm k} (A+C_{\bm k}) \hat c_{\bm k}^\dag \hat c_{\bm k} + \sum_{\bm k} (A-C_{\bm k})\hat d_{\bm k}^\dag \hat d_{\bm k} \notag \\
    &\qquad + \sum_{\bm k} \biggl( \frac 12 B_{\bm k} + F\biggr) (\hat c_{\bm k} \hat c_{-\bm k} + \hat c_{\bm k}^\dag \hat c_{-\bm k}^\dag) + \sum_{\bm k} \biggl( - \frac 12 B_{\bm k} + F \biggr) (\hat d_{\bm k} \hat d_{-\bm k} + \hat d_{\bm k}^\dag \hat d_{-\bm k}^\dag).
  \end{align}
  Since $\hat c_{\bm k}$ and $\hat d_{\bm k}$ are mutually decoupled, we are finally ready to apply the following Bogoliubov transformations.
  \begin{align}
    \begin{pmatrix}
      \hat \alpha_{\bm k} \\ \hat \alpha_{-\bm k}^\dag
    \end{pmatrix}
    & =
    \begin{pmatrix}
      \cosh\theta_{\bm k}^\alpha & - \sinh \theta_{\bm k}^\alpha \\
      -\sinh\theta_{\bm k}^\alpha  &  \cosh\theta_{\bm k}^\alpha
    \end{pmatrix}
    \begin{pmatrix}
      \hat c_{\bm k} \\  \hat c_{-\bm k}^\dag
    \end{pmatrix}, \\
    \begin{pmatrix}
      \hat \beta_{\bm k} \\ \hat \beta_{-\bm k}^\dag
    \end{pmatrix}
    & =
    \begin{pmatrix}
      \cosh\theta_{\bm k}^\beta & - \sinh \theta_{\bm k}^\beta \\
      -\sinh\theta_{\bm k}^\beta &  \cosh\theta_{\bm k}^\beta
    \end{pmatrix}
    \begin{pmatrix}
      \hat d_{\bm k} \\  \hat d_{-\bm k}^\dag
    \end{pmatrix}.
  \end{align}
  These transformations from $\hat c_{\bm k}$ and $\hat d_{\bm k}$ to $\hat \alpha_{\bm k}$ and $\hat \beta_{\bm k}$ lead to
  \begin{align}
    \hat{\mathcal H}_{\text{N-C}}
    &= \frac 12 zJNS^2\cos(2\theta) - KNS^2\cos^2\theta-B_0NS\cos\theta - \frac 12 KNS\sin^2\theta \notag \\
    &\qquad +  \sum_{\bm k}\biggl[
      \bigl\{ (A+C_{\bm k}) \cosh(2\theta_{\bm k}^\alpha) + (B_{\bm k} + 2F) \sinh (2\theta_{\bm k}^\alpha) \bigr\}
      + \bigl\{ (A-C_{\bm k}) \cosh(2\theta_{\bm k}^\beta) + (-B_{\bm k} + 2F) \sinh(2\theta_{\bm k}^\beta) \bigr\} -A
    \biggr] \notag \\
    &\qquad + \sum_{\bm k}  \frac 12\bigl\{
    (A+C_{\bm k}) \sinh(2\theta_{\bm k}^\alpha) +  (B_{\bm k}+2F) \cosh(2\theta_{\bm k}^\alpha)\bigr\} (\hat \alpha_{\bm k} \hat \alpha_{-\bm k} + \hat \alpha_{\bm k}^\dag \hat \alpha_{-\bm k}^\dag) \notag \\
    &\qquad + \sum_{\bm k} \frac 12 \bigl\{ (A-C_{\bm k}) \sinh(2\theta_{\bm k}^\beta) + (-B_{\bm k} + 2F) \cosh(2\theta_{\bm k}^\beta) \bigr\} (\hat \beta_{\bm k} \hat \beta_{-\bm k} + \hat \beta_{\bm k}^\dag \hat \beta_{-\bm k}^\dag) \notag \\
    &\qquad + \sum_{\bm k} \bigl\{ (A+C_{\bm k}) \cosh(2\theta_{\bm k}^\alpha) + (B_{\bm k} + 2F) \sinh(2\theta_{\bm k}^\alpha) \bigr\} \hat \alpha_{\bm k}^\dag \hat \alpha_{\bm k} \notag \\
    &\qquad + \sum_{\bm k} \bigl\{ (A-C_{\bm k}) \cosh(2\theta_{\bm k}^\beta) +(-B_{\bm k}+2F) \sinh(2\theta_{\bm k}^\beta) \bigr\} \hat \beta_{\bm k}^\dag \hat \beta_{\bm k}.
  \end{align}
  When $\theta_{\bm k}^\alpha$ and $\theta_{\bm k}^\beta$ satisfy $\tanh (2\theta_{\bm k}^\alpha) = -\frac{B_{\bm k}+2F}{A+C_{\bm k}}$ and $\tanh (2\theta_{\bm k}^\beta) =  \frac{B_{\bm k}-2F}{A-C_{\bm k}}$,
  the off-diagonal terms vanish and the Hamiltonian is diagonalized as
  \begin{align}
    \hat{\mathcal H}_{\text{N-C}}
    &= E_{\rm Cant} + \sum_{\bm k} \epsilon_{\bm k}^{\text{Cant}, \alpha}\hat \alpha_{\bm k}^\dag \hat \alpha_{\bm k} + \sum_{\bm k} \epsilon_{\bm k}^{\text{Cant}, \beta} \hat \beta_{\bm k}^\dag \hat \beta_{\bm k},
  \end{align}
  where
  \begin{align}
    E_{\text{Cant}}
    &= \frac 12 zJNS^2\cos(2\theta) - KNS^2 \cos^2\theta - B_0NS \cos\theta -\frac 12 KNS\sin^2\theta \notag \\
    &\qquad + \sum_{\bm k} \biggl\{
      \frac 12 \sqrt{(A+C_{\bm k})^2-(B_{\bm k}+2F)^2} + \frac 12 \sqrt{(A-C_{\bm k})^2 - (B_{\bm k}-2F)^2} - A
    \biggr\}, \\
    \epsilon_{\bm k}^{\text{Cant}, \alpha} &= \sqrt{(A+C_{\bm k})^2 - (B_{\bm k}+2F)^2},
    \label{magnon_band_canted_alpha} \\
    \epsilon_{\bm k}^{\text{Cant},\beta} &= \sqrt{(A-C_{\bm k})^2 - (B_{\bm k}-2F)^2}.
    \label{magnon_band_canted_beta}
  \end{align}
\end{widetext}
We thus obtain the two magnon bands $\epsilon_{\bm k}^{\text{Cant},\alpha}$ and $\epsilon_{\bm k}^{\text{Cant},\beta}$ in the canted phase.
The magnon band $\epsilon_{\bm k}^{\text{Cant},\beta}$ has a minimum at $\bm k = \bm 0$ and
\begin{align}
  \epsilon_{\bm 0}^{\text{Cant}, \beta} = 0,
\end{align}
independent of $\theta$, that is, of $B_0/J$ [Fig.~\ref{fig:model}~(c)]. This gapless mode corresponds to the NG mode associated with the spontaneous U(1) symmetry breaking in the canted phase.

\subsection{Weak ferromagnets}

As we mentioned in the main text, the ground state in the WF phase is akin to that in the canted phase.
We thus reuse the Holstein-Primakoff transformation from the spin to magnon operators that we developed in the canted phase.
In the linear spin-wave theory regime, we then approximate $\hat{\bm S}_{\bm r_{\rm A}}$ and $\hat{\bm S}_{\bm r_{\rm B}}$ as follows.
\begin{align}
  \hat S_{\bm r_{\rm A}}^x &= \hat S_{\bm r_{\rm A}}^\xi \cos \theta - \hat S_{\bm r_{\rm A}}^\zeta \sin \theta \nonumber\\
  &\approx \sqrt{ \frac S2}\, (\hat a_{\bm r_{\rm A}}+\hat a_{\bm r_{\rm A}}^\dag) \cos \theta - (S-\hat a_{\bm r_{\rm A}}^\dag \hat a_{\bm r_{\rm A}} ) \sin \theta,
  \label{Sx-to-a_WF} \\*
  \hat S_{\bm r_{\rm A}}^y
  &= \hat S_{\bm r_{\rm A}}^\eta 
  \approx -i\sqrt{\frac S2}\, (\hat a_{\bm r_{\rm A}} - \hat a_{\bm r_{\rm A}}^\dag),
  \label{Sy-to-a_WF} \\
  \hat S_{\bm r_{\rm A}}^z
  &= \hat S_{\bm r_{\rm A}}^\xi \sin \theta + \hat S_{\bm r_{\rm A}}^\zeta \cos\theta \nonumber\\
  &\approx \sqrt{\frac S2}\, (\hat a_{\bm r_{\rm A}}+\hat a_{\bm r_{\rm A}}^\dag) \sin \theta + (S-\hat a_{\bm r_{\rm A}}^\dag \hat a_{\bm r_{\rm A}} ) \cos\theta,
  \label{Sz-to-a_WF}
\end{align}
for the A sublattice and
\begin{align}
  \hat S_{\bm r_{\rm B}}^x &= \hat S_{\bm r_{\rm B}}^\xi \cos \theta + \hat S_{\bm r_{\rm B}}^\zeta \sin \theta \nonumber\\
  &\approx \sqrt{ \frac S2}\, (\hat b_{\bm r_{\rm B}}+\hat b_{\bm r_{\rm B}}^\dag) \cos \theta + (S-\hat b_{\bm r_{\rm B}}^\dag \hat b_{\bm r_{\rm B}} ) \sin \theta,
  \label{Sx-to-b_WF} \\
  \hat S_{\bm r_{\rm B}}^y
  &= \hat S_{\bm r_{\rm B}}^\eta 
  \approx -i\sqrt{\frac S2}\, (\hat b_{\bm r_{\rm B}} - \hat b_{\bm r_{\rm B}}^\dag),
  \label{Sy-to-b_WF} \\
  \hat S_{\bm r_{\rm B}}^z
  &= -\hat S_{\bm r_{\rm B}}^\xi \sin \theta + \hat S_{\bm r_{\rm B}}^\zeta \cos\theta \nonumber\\
  &\approx -\sqrt{\frac S2}\, (\hat b_{\bm r_{\rm B}}+\hat b_{\bm r_{\rm B}}^\dag) \sin \theta + (S-\hat b_{\bm r_{\rm B}}^\dag \hat b_{\bm r_{\rm B}} ) \cos\theta,
  \label{Sz-to-b_WF}
\end{align}
for the B sublattice.
Using these bosonic operators, we rewrite the Hamiltonian \eqref{H_WF_def} as
\begin{widetext}
  \begin{align}
    \hat{\mathcal H}_{\rm WF}
    &= \frac 12 zJ NS^2 \cos(2\theta) - \frac 12 zDNS^2\sin (2\theta) - B_0NS \cos \theta \nonumber\\
    &\qquad +\sum_{\bm r_{\rm A}}{ \bigl\{ zS(-J\cos(2\theta) + D\sin(2\theta))+B_0\cos\theta \bigr\}} \hat a_{\bm r_{\rm A}}^\dag \hat a_{\bm r_{\rm A}} \nonumber\\
    &\qquad + \sum_{\bm r_{\rm B}} { \bigl\{ zS(-J\cos(2\theta) +D \sin (2\theta))+ B_0 \cos\theta
    \bigr\}}\hat b_{\bm r_{\rm B}}^\dag \hat b_{\bm r_{\rm B}} \nonumber\\
    &\qquad +\sum_{\braket{\bm r_{\rm A},\bm r_{\rm B}}} \frac S2 (J\cos(2\theta) - D\sin (2\theta) - J) (\hat a_{\bm r_{\rm A}}\hat b_{\bm r_{\rm B}} +\hat a_{\bm r_{\rm A}}^\dag \hat b_{\bm r_{\rm B}}^\dag) \nonumber\\
    &\qquad + \sum_{\braket{\bm r_{\rm A},\bm r_{\rm B}}} \frac S2 (J \cos(2\theta) - D\sin(2\theta) + J) (\hat a_{\bm r_{\rm A}} \hat b_{\bm r_{\rm B}}^\dag + \hat a_{\bm r_{\rm A}}^\dag \hat b_{\bm r_{\rm B}}),
  \end{align}
  where we have dropped third- or higher-order terms about bosonic operators.
  The linear terms about $\hat a_{\bm r}$ or $\hat b_{\bm r}$ vanish in analogy with the canted phase.
  The Fourier transformation allows us to rewrite the Hamiltonian as
  \begin{align}
    \hat{\mathcal H}_{\rm WF}
    &= \frac 12 zJ NS^2 \cos(2\theta) - \frac 12 zDNS^2\sin (2\theta) - B_0NS \cos \theta \nonumber\\
    &\qquad +\sum_{\bm k} A_{\bm k} (\hat a_{\bm k}^\dag \hat a_{\bm k} + \hat b_{\bm k}^\dag \hat b_{\bm k})+\sum_{\bm k} B_{\bm k}(\hat a_{\bm k} \hat b_{\bm k} + \hat a_{\bm k}^\dag \hat b_{\bm k}^\dag) + \sum_{\bm k} C_{\bm k} (\hat a_{\bm k} \hat b_{-\bm k}^\dag + \hat a_{\bm k}^\dag \hat b_{-\bm k}),
    \label{H_WF_FT}
  \end{align}
  where $A_{\bm k}$, $B_{\bm k}$, and $C_{\bm k}$ are
  \begin{align}
    A_{\bm k} &= zS (-J\cos(2\theta) + D\sin(2\theta)) + B_0 \cos\theta, \\
    B_{\bm k} &= \frac 12 zS \gamma_{\bm k} (J \cos (2\theta) -D \sin(2\theta) - J), \\
    C_{\bm k} &= \frac 12 zS \gamma_{\bm k} (J \cos(2\theta) -D \sin(2\theta) + J).
  \end{align}
  Diagonalizing the Hamiltonian in analogy with the canted phase, we obtain
  \begin{align}
    \hat{\mathcal H}_{\rm WF}
    &= E_{\rm WF} + \sum_{\bm k} (\epsilon_{\bm k}^{\text{WF}, \alpha}\hat\alpha_{\bm k}^\dag \hat\alpha_{\bm k} + \epsilon_{\bm k}^{\text{WF},\beta}\hat\beta_{\bm k}^\dag \hat\beta_{\bm k}),
  \end{align}
  with the ground-state energy $E_{\rm WF}$ and the magnon bands,
  \begin{align}
    \epsilon_{\bm k}^{\text{WF},\alpha}
    &= \sqrt{(A_{\bm k}+C_{\bm k})^2 - B_{\bm k}^2}, \\
    \epsilon_{\bm k}^{\text{WF},\beta}
    &= \sqrt{(A_{\bm k}-C_{\bm k})^2-B_{\bm k}^2}.
  \end{align}
\end{widetext}

\section{Dynamical symmetries}
\label{app:LLG}

Models of magnetic Mott insulators under an ac magnetic field often possess a dynamical symmetry related to the period $T_{\rm ac}$ of the ac magnetic field when the ac field can be viewed as an ideal continuous wave (not a finite-length pulse).
In this section, we derive the dynamical symmetry of magnetic Mott insulators in a general form and also consider dynamical symmetries in the presence of spontaneous magnetic orders dealt with in the main text. To simplify the setup, we assume that the equation of motion has no dissipation effect, that is, we consider simple closed (isolated) magnetic systems irradiated by an ac field.

\subsection{Dynamical symmetry in quantum systems}
\label{app:dyn_sym_quantum}

Let us consider a quantum spin model $\hat{\mathcal H}_{\rm mag}$ subject to a spin-light coupling $\hat{\mathcal H}_{\rm ext}(t)$, whose temporal period is $T_{\rm ac}$: $\hat{\mathcal H}_{\rm ext}(t)=\hat{\mathcal H}_{\rm ext}(t+T_{\rm ac})$. In this study, we mainly consider the ac Zeeman interaction as the spin-light coupling. To discuss the dynamical symmetry, we focus on the ideal setup that the applied laser is a continuous wave (not a finite-length pulse) and the temporal periodicity $\hat{\mathcal H}_{\rm ext}(t)=\hat{\mathcal H}_{\rm ext}(t+T_{\rm ac})$ holds from the initial time.

Suppose that the static system of $\hat{\mathcal H}_{\rm mag}$ has a symmetry generated by a unitary operator $\hat U$. Namely, $\hat U \hat{\mathcal H}_{\rm mag} \hat U^\dag = \hat{\mathcal H}_{\rm mag}$. (More generally speaking, the operator $\hat U$ is a unitary or an antiunitary operator.)
In the presence of the periodic field, $\hat U$ does not keep the total Hamiltonian
$\hat{\mathcal H}(t) = \hat{\mathcal H}_{\rm mag} + \hat{\mathcal H}_{\rm ext}(t)$ invariant in general: $\hat U\hat{\mathcal H}(t)\hat U^\dagger=\hat{\mathcal H}(t)$ does not hold in general.
Instead, let us consider the case that a time translation of the time-dependent Hamiltonian $\hat{\mathcal H}(t)$ by $\kappa T_{\rm ac}$ is equivalent to its symmetry operation $\hat U$:
\begin{align}
  \hat U \hat{\mathcal H}(t) \hat U^\dag = \hat{\mathcal H}(t+\kappa T_{\rm ac}),
  \label{UHU_shift}
\end{align}
where we assume that the constant $\kappa$ satisfies $0 < \kappa <1$ without loss of generality thanks to the time periodicity of the ac field. The case of $\kappa=1$ corresponds to the trivial equation: $\hat{\mathcal H}(t+\kappa T_{\rm ac})=\hat{\mathcal H}(t+T_{\rm ac})=\hat{\mathcal H}(t)$.
If a time periodic system satisfies the relation~\eqref{UHU_shift}, we say that the system has dynamical symmetry. Namely, the dynamical symmetry is a sort of the symmetry including a time shift. We here note that in Eq.~\eqref{UHU_shift}, the Hamiltonian is defined in the Schr\"odinger picture and it has the time dependence only through the external ac field. Since each dynamical symmetry is characterized by the symmetry operator $\hat U$ and the time shift $\kappa T_{\rm ac}$, (as we used in the main text) we represent each dynamical symmetry using the symbol $(\hat U,\kappa T_{\rm ac})$.

As we briefly discussed in Sec.~\ref{sec:DS_in_general}, the relation of Eq.~\eqref{UHU_shift} is not enough to derive the selection rule for the harmonic generation spectrum. In addition to Eq.~\eqref{UHU_shift}, we should focus on the transformation laws for symmetry operations on physical quantities. In particular, in the present target of laser-driven magnets, the main observable is the spin operator. Therefore, we should consider the following symmetry relation,
\begin{align}
  \hat U\hat {\bm S}_{\bm r}\hat U^\dagger =f(\{\hat{\bm S}_{\bm r'}\}),
  \label{eq:sym_S}
\end{align}
where $f(\{\hat{\bm S}_{\bm r'}\})$ is a function of $\{\hat{\bm S}_{\bm r'}\}$. For example, when $\hat U$ is the $\pi$ spin-rotation operator around the $S^z$ axis ($\hat U=\hat U_z(\pi)$), the above equation is given by $\hat U_z(\pi)\hat S^{x,y}_{\bm r}\hat U_z^\dagger(\pi)=-\hat S^{x,y}_{\bm r}$.

Combining Eqs.~\eqref{UHU_shift} and \eqref{eq:sym_S}, we can often derive a selection rule of the harmonic generation spectrum. To see the logic from these two equations to the selection rule, we consider the time-evolution operator
\begin{align}
  \hat U_{\rm time}(s,s') = \mathcal T\left[\exp\left({\frac{1}{i\hbar}\int_{s'}^{s}dt^{\prime}\hat{\mathcal H}(t^{\prime})}\right)\right],
  \label{eq:timeev}
\end{align}
where $\mathcal T$ stands for the time-ordering operator and the Hamiltonian $\hat{\mathcal H}(t)$ is in the Schr\"odinger picture. Here we assume that the initial time $s'$, which is equivalent to the beginning of the laser application, is long enough ago ($s'\to-\infty$) and the initial state $|\psi_0\rangle$ at $s'$ is given by the ground state or the thermal equilibrium state of $\hat{\mathcal H}_{\rm mag}$.
Under this assumption, we can re-express the time-evolution operator as
\begin{align}
  \hat U_{\rm time}(s)=\hat U_{\rm time}(s,-\infty)
  \label{eq:timeev2}
\end{align}
By taking the time derivative of $\hat U_{\rm time}(t)$, we find the time-evolution operator satisfies the following equation of motion
\begin{align}
  \frac{d}{dt}\hat U_{\rm time}(t) &= \mathcal T\left[\frac{1}{i\hbar}\hat{\mathcal H}(t)\exp\left({\frac{1}{i\hbar}\int_{-\infty}^{t}dt^{\prime}\hat{\mathcal H}(t^{\prime})}\right)\right]\nonumber\\
  & = \frac{1}{i\hbar}\hat{\mathcal H}(t)\hat U_{\rm time}(t).
  \label{eq:timeev_EOM}
\end{align}
By sandwiching this equation between $\hat U$ and $\hat U^\dagger$, we obtain the following equation
\begin{align}
  \frac{d}{dt}\hat U\hat U_{\rm time}(t)\hat U^\dagger
  & = \frac{1}{i\hbar}\hat U\hat{\mathcal H}(t)\hat U^\dagger\hat U\hat U_{\rm time}(t)\hat U^\dagger\nonumber\\
  & = \frac{1}{i\hbar}\hat{\mathcal H}(t+\kappa T_{\rm ac}) \hat U\hat U_{\rm time}(t)\hat U^\dagger,
  \label{eq:timeev_EOM2}
\end{align}
where we have used Eq.~\eqref{UHU_shift} and $\hat U\hat U^\dagger=\hat 1$. Comparing the original equation of motion \eqref{eq:timeev_EOM} with Eq.~\eqref{eq:timeev_EOM2}, we may assume that the relation $\hat U \hat U_{\rm time}(t) \hat U^\dag \propto \hat U_{\rm time}(t+\kappa T_{\rm ac})$ holds, namely,
\begin{align}
  \hat U \hat U_{\rm time}(t) \hat U^\dag &= e^{i\phi}\hat U_{\rm time}(t+\kappa T_{\rm ac}),
  \label{UUtU_shift}
\end{align}
where the phase $\phi$ is a real number.


Using Eqs.~\eqref{UUtU_shift}, \eqref{UHU_shift} and \eqref{eq:sym_S}, we can compute the expectation value of spin operators at time $t$ as follows:
\begin{align}
  \bm M_{\bm r}(t)&\equiv \langle\psi(t)|\hat{\bm S}_{\bm r} |\psi(t)\rangle
  \nonumber\\
  &=\langle \psi_0|\hat U_{\rm time}(t)^\dagger \hat{\bm S}_{\bm r} \hat U_{\rm time}(t)|\psi_0\rangle \nonumber\\
  &=\langle \psi_0| \hat U^\dagger \hat U  \hat U_{\rm time}(t)^\dagger \hat U^\dagger \hat U \hat{\bm S}_{\bm r} \hat U^\dagger \hat U \hat U_{\rm time}(t) \hat U^\dagger \hat U |\psi_0\rangle  \nonumber\\
  &=\langle \psi_0| \hat U^\dagger \hat U_{\rm time}(t+\kappa T_{\rm ac})^\dagger f(\{\hat{\bm S}_{\bm r'}\}) \hat U_{\rm time}(t+\kappa T_{\rm ac}) \hat U |\psi_0\rangle  \nonumber\\
  &=\langle \psi_0|\hat U_{\rm time}(t+\kappa T_{\rm ac})^\dagger f(\{\hat{\bm S}_{\bm r'}\}) \hat U_{\rm time}(t+\kappa T_{\rm ac})|\psi_0\rangle  \nonumber\\
  &= \langle \psi(t+\kappa T_{\rm ac})| f(\{\hat{\bm S}_{\bm r'}\}) |\psi(t+\kappa T_{\rm ac})\rangle  \nonumber\\
  &= f(\{ {\bm M}_{\bm r'}(t+\kappa T_{\rm ac})\}),
  \label{eq:expvalue_t}
\end{align}
where $|\psi(t)\rangle =\hat U_{\rm time}(t)|\psi_0\rangle$ is the wave function at time $t$.
In this equation, we have used $\hat U^\dagger \hat U=\hat 1$ on the third line, and used Eqs.~\eqref{eq:sym_S} and \eqref{UUtU_shift} on the fourth line. On the fifth line,  we have assumed that the initial state $|\psi_0\rangle$ is invariant under the symmetry operation $\hat U$: $\hat U|\psi_0\rangle=e^{i\phi_1}|\psi_0\rangle$ with the phase $\phi_1$ being a real number.
Equation~\eqref{eq:expvalue_t} provides a relation between the expectation values of spins at two different times, $\bm M_{\bm r}(t)$ and $\bm M_{\bm r}(t+\kappa T_{\rm ac})$.
The harmonic generation spectrum is given by the Fourier transform of $\bm M_{\bm r}(t)$ along the time direction, and thereby Eq.~\eqref{eq:expvalue_t} generally restricts the form (frequency dependence) of the Fourier transform, namely, Eq.~\eqref{eq:expvalue_t} leads to a selection rule. We will discuss several concrete examples of such selection rules in the following subsections.

So far we have considered only the case where the initial state (ground state or equilibrium state of $\hat{\mathcal H}_{\rm mag}$) is invariant under the symmetry operation $\hat U$. This argument holds if we discuss dynamical symmetry in a magnetic system where spontaneous symmetry breaking (SSB) does not occur. On the other hand, if the symmetry associated with the operator $\hat U$ is spontaneously broken in the initial state $\ket{\psi_0}$ before the laser application, the state $\ket{\psi_0}$ is generally changed by the action of $\hat U$: $\hat U\ket{\psi_0}\neq \ket{\psi_0}$. In such a case, the system is no longer dynamically symmetric and Eq.~\eqref{eq:expvalue_t} does not hold [although the Hamiltonian still satisfies Eq.~\eqref{UHU_shift}]. That is, for the case of SSB, one has to consider symmetry relations of both the operator and the wave function simultaneously.
Even in such an SSB case, if one finds a symmetry operator $\hat U$ that satisfies Eq.~\eqref{UHU_shift} and $\hat U\ket{\psi_0}=e^{i\phi_1}\ket{\psi_0}$, there is a possibility that one can find a dynamical symmetry associated with $\hat U$.

\subsection{Dynamical symmetry in LL equation}
\label{app:dyn_sym_LL}

In the main text, we have adopted the LLG equation instead of the quantum Heisenberg equation because the system belongs to the magnetically ordered phase.
The long-range magnetic order justifies the application of the semiclassical equation of motion.
However, there is a priori no guarantee that the dynamical symmetry sustains after the semiclassical approximation.
We thus show that the dynamical symmetry also exists when the system evolves in time in accordance with the semiclassical equation of motion instead of the Heisenberg equation.

We start from the Landau-Lifshitz (LL) equation,
\begin{align}
  \frac{d\bm m_{\bm r}(t)}{dt} = \bm m_{\bm r}(t) \times \frac{\partial \mathcal H(t)}{\partial \bm m_{\bm r}(t)}.
  \label{LL_eq_m}
\end{align}
Here, $\bm m_{\bm r}(t)$ is the magnetic moment at the site $\bm r$ and the time $t$, the semiclassical counterpart of the quantum spin operator $\hat{\bm S}_{\bm r}(t)$.
The semiclassical counterpart of the Heisenberg equation is the LL equation \eqref{LL_eq_m} rather than the LLG equation \eqref{stochastic-LLG_def}.
The latter contains the damping term arising out of various effects, which cannot be described by the microscopic spin Hamiltonian. As in the case of the LLG equation in Sec.~\ref{sec:LLG}, the classical Hamiltonian $\mathcal H(t)={\mathcal H}_{\rm mag} + {\mathcal H}_{\rm ext}(t)$ is the classical version of Eq.~\eqref{full_Hamiltonian_gen_def}, where every $\hat{\bm S}_{\bm r}$ is replaced by $\bm m_{\bm r}(t)$. We note that in these semiclassical systems following the LL equation, the set of the local magnetization $\{\bm m_{\bm r}\}$ stands for both the many-spin state and the spin expectation value (one no longer distinguishes the spin operator and the wave function).

Suppose a unitary operator $\hat U$ that satisfies
\begin{align}
  \hat U \mathcal H(t) \hat U^\dag = \mathcal H(t+\kappa T_{\rm ac})
  \label{UHU_shift_LL}
\end{align}
in analogy with Eq.~\eqref{UHU_shift}.
Equation~\eqref{UHU_shift_LL} is defined as the dynamical symmetry in classical magnets following the LL equation. We note that like the quantum situation, the time dependence of the Hamiltonian in Eq.~\eqref{UHU_shift_LL} is included only through the external ac field ${\bm B}_{\rm ac}(t)$.

Similarly to the quantum case, the dynamical symmetry of Eq.~\eqref{UHU_shift_LL} is not enough to derive the selection rule of harmonic generation spectrum. We should consider the symmetry operation for the magnetization $\bm m_{\bm r}$. Let us define $\bm m'_{\bm r'}(t)$ as
\begin{align}
  \bm m'_{\bm r'}(t) = \hat U \bm m_{\bm r}(t) \hat U^\dag.
  \label{m'_def_by_UmU}
\end{align}
The sites $\bm r'$ and $\bm r$ in Eq.~\eqref{m'_def_by_UmU} are not necessarily identical since $\hat U$ can include a spatial operation such as a rotation, one-site translation along a certain axis, etc. For instance, if $\hat U$ is the one-site translation along the $x$ direction, $\hat T_x$, $\bm m'_{\bm r'}={\bm m}_{\bm r+a_0\bm e_x}$.

Using Eqs.~\eqref{LL_eq_m}, \eqref{UHU_shift_LL}, and \eqref{m'_def_by_UmU}, we try to find a relation between $\bm m'_{\bm r'}(t)$ and $\bm m_{\bm r}(t)$. By sandwiching the LL equation of Eq.~\eqref{LL_eq_m} between $\hat U$ and $\hat U^\dagger$, we have
\begin{align}
  \frac{d\bm m'_{\bm r'}(t)}{dt} & = \frac{d \hat U \bm m_{\bm r}(t)\hat U^\dagger}{dt}\nonumber\\
  &=\hat U\bm m_{\bm r}(t)\hat U^\dagger \times\hat U \frac{\partial \mathcal H}{\partial \bm m_{\bm r}} (\{\bm m_{\bm r}\},{\bm B}_{\rm ac}(t))\hat U^\dagger.\nonumber\\
  &= \bm m'_{\bm r'}(t)\times \frac{\partial \mathcal H}{\partial \bm m'_{\bm r'}} (\{\bm m'_{\bm r'}\},{\bm B}_{\rm ac}(t+\kappa T_{\rm ac})),
  \label{LL_eq_m2}
\end{align}
where we have used the time independent nature of $\hat U$  and $\hat U\hat U^\dagger=\hat 1$ on the second line and used Eqs.~\eqref{UHU_shift_LL} and \eqref{m'_def_by_UmU} on the third line. Since the torque term $\frac{\partial \mathcal H}{\partial \bm m_{\bm r}}$ is a function of the set of local magnetizations $\{\bm m_{\bm r}\}$ and the laser field $\bm B_{\rm ac}(t)$, we have added the arguments $(\{\bm m_{\bm r}\},{\bm B}_{\rm ac}(t))$ to emphasize their dependence. Equation~\eqref{LL_eq_m2} indicates that the "magnetization" $\bm m'_{\bm r'}(t)$ is also a solution of the LL equation at time $t+\kappa T_{\rm ac}$. The uniqueness of solution of the LL equation thus leads us to the conclusion that
\begin{align}
  \bm m'_{\bm r'}(t)=\bm m_{\bm r'}(t+\kappa T_{\rm ac}).
  \label{m'-to-m}
\end{align}
This equation gives the relationship between two local magnetizations at different times $t$ and $t+\kappa T_{\rm ac}$ like Eq.~\eqref{eq:expvalue_t}. Therefore, Eq.~\eqref{m'-to-m} can lead to a selection rule of the harmonic generation spectrum.

In deriving Eq.~\eqref{m'-to-m}, we have assumed that the initial spin configuration
$\mathcal{M}_0 =\{ {\bm m}_{\bm r}^0\}$ at $t \to-\infty$ is invariant under $\hat U$, namely,
\begin{align}
  \hat U \mathcal{M}_0 =\mathcal{M}_0
  \label{eq:initialspin}
\end{align}
Here, $\hat U \mathcal{M}_0$ denotes the configuration obtained by applying the transformation in Eq.~\eqref{m'_def_by_UmU}, including its action on the site labels. This invariance is the initial-condition assumption used in applying the uniqueness of solutions of the LL equation to obtain Eq.~\eqref{m'-to-m}.
If the initial state spontaneously breaks this symmetry, Eq.~\eqref{eq:initialspin} is not satisfied, and Eq.~\eqref{m'-to-m} and the corresponding selection rule cannot, in general, be inferred from Eq.~\eqref{UHU_shift_LL} alone.

\subsection{Selection rules in high-harmonic spectra: one-color laser case}
\label{app:dyn_sym_1-color}

Let us derive the selection rule of the spectrum related to the dynamical symmetry in the case where the one-color laser field with the frequency $\Omega$ along the $x$ axis is applied to the magnetic Mott insulator \eqref{H_Neel-Canted_def} with the static magnetic field.
The total Hamiltonian is given by
\begin{align}
  \hat{\mathcal H}(t) &= \hat{\mathcal H}_{\text{N-C}} - B_{\rm ac} \cos(\Omega t) \sum_{\bm r} \hat S_{\bm r}^x.
\end{align}
This Hamiltonian has a dynamical symmetry defined by the following combination
\begin{align}
  (\hat U_z(\pi),T_{\rm ac}/2),
  \label{U_N-C_one-color_def}
\end{align}
which is made of the time translation by $T_{\rm ac}/2$ and the $\pi$ rotation of the total spin around $S^z$.
The Hamiltonian $\hat{\mathcal H}_{\text{N-C}}$ of Eq.~\eqref{H_Neel-Canted_def} with the uniaxial symmetry is obviously invariant under the spin rotation $\hat U_z(\pi)$.
The ac Zeeman energy $\hat{\mathcal H}_{\rm ext}(t)=-B_{\rm ac}\cos(\Omega t) \sum_{\bm r} \hat S_{\bm r}^x$ is changed by the time translation and the spin rotation as follows:
\begin{widetext}
  \begin{align}
    \label{eq:Hext_timeshift}
    \hat{\mathcal H}_{\rm ext}(t+T_{\rm ac}/2)&=- B_{\rm ac} \cos (\Omega (t+T_{\rm ac}/2)) \sum_{\bm r} \hat S_{\bm r}^x
    =+ B_{\rm ac} \cos(\Omega t) \sum_{\bm r}\hat S_{\bm r}^x=-\hat{\mathcal H}_{\rm ext}(t), \\
    \hat U_z(\pi)\hat{\mathcal H}_{\rm ext}(t)\hat U_{z}^\dag(\pi)&=\hat U_z(\pi) \Bigl( -B_{\rm ac} \cos (\Omega t) \sum_{\bm r} \hat S_{\bm r}^x \Bigr) \hat U_{z}^\dag(\pi)
    =+ B_{\rm ac} \cos(\Omega t) \sum_{\bm r}\hat S_{\bm r}^x=-\hat{\mathcal H}_{\rm ext}(t).
    \label{eq:Hext_U1}
  \end{align}
\end{widetext}
Namely, both the time shift by $T_{\rm ac}/2$ and the spin rotation $\hat U_z(\pi)$ turn out to change the sign of the ac field $B_{\rm ac} \to - B_{\rm ac}$. The relation similar to Eqs.~\eqref{eq:Hext_timeshift} and \eqref{eq:Hext_U1} also holds in the classical Hamiltonian, where $\hat {\bm S}_{\bm r}$ is replaced with $\hat {\bm m}_{\bm r}$.
Next, we should focus on the symmetry relation in terms of spin operators. One can easily find the relation of $\hat U_z(\pi)\hat S^{x,y}_{\bm r}\hat U_z^\dagger(\pi)=-\hat S^{x,y}_{\bm r}$ and $\hat U_z(\pi)\hat S^{z}_{\bm r}\hat U_z^\dagger(\pi)=\hat S^{z}_{\bm r}$. The same relation holds for the classical magnetic moment $\hat {\bm m}_{\bm r}$ if the U(1) spin-rotation symmetry around the $S^z$ axis is not spontaneously broken.
From these symmetry relations for the Hamiltonian and spins,
we find that the dynamical symmetry of Eq.~\eqref{U_N-C_one-color_def} yields the following constraint on the uniform (total) magnetization $\bm m(t)$:
\begin{align}
  m^x(t+T_{\rm ac}/2) &= - m^x(t), \\
  m^y(t+T_{\rm ac}/2) &= - m^y(t), \\
  m^z(t+T_{\rm ac}/2) &= m^z(t).
\end{align}
These equations correspond to Eqs.~\eqref{eq:expvalue_t} or \eqref{m'-to-m}.
These constraints manifest themselves as the selection rule of the spectrum.
Indeed, the Fourier transforms $\tilde m^a(\omega)$ are rewritten as
\begin{align}
  \tilde m^a(\omega)
  &= \int_0^{T_{\rm ac}} dt \, e^{i\omega t} m^a(t) \nonumber\\
  &= \int_0^{T_{\rm ac}/2} dt\, \Bigl(e^{i\omega t} m^a(t) +  e^{i\omega(t+T_{\rm ac}/2)} m^a(t+T_{\rm ac}/2)
  \Bigr) \nonumber\\
  &= \int_0^{T_{\rm ac}/2} dt\, \bigl\{ 1+ e^{i\omega T_{\rm ac}/2} \sigma_a\bigr\}e^{i\omega t} m^a(t),
\end{align}
where $\sigma_a = -1$ for $a=x,~y$ and $\sigma_z=1$.
Since $T_{\rm ac}=2\pi/\Omega$, we obtain
\begin{align}
  \tilde m^x(n\Omega)
  &= \int_0^{T_{\rm ac}/2} dt\, \{1-(-1)^n\} e^{in\Omega t}m^x(t), \\
  \tilde m^y(n\Omega)
  &= \int_0^{T_{\rm ac}/2} dt\, \{1-(-1)^n\} e^{in\Omega t}m^y(t), \\
  \tilde m^z(n\Omega)
  &= \int_0^{T_{\rm ac}/2} dt\, \{1+(-1)^n\} e^{in\Omega t}m^z(t).
\end{align}
Therefore, we find the selection rule,
\begin{align}
  \tilde m^x(2n\Omega) &= 0, \\
  \tilde m^y(2n\Omega) &= 0, \\
  \tilde m^z\bigl((2n+1)\Omega\bigr) &= 0,
\end{align}
for $n=0,~1,~2,~3,~\cdots$.

At the end of this section, we briefly comment on the spontaneous symmetry breaking of the system $\hat{\mathcal H}_{\text{N-C}}$. If we consider the N\'eel phase, the above selection rule indeed holds. However, the U(1) spin rotation symmetry is spontaneously broken in the canted phase and the relation of $\hat U_z(\pi)\hat m^{x,y}_{\bm r}\hat U_z^\dagger(\pi)=-\hat m^{x,y}_{\bm r}$ is violated. Therefore, the dynamical symmetry of Eq.~\eqref{U_N-C_one-color_def} is broken in the canted phase. Instead, we can find the other dynamical symmetry of Eq.~\eqref{dyn_sym_cant} there (see Sec.~\ref{sec:Canted}).

\subsection{Selection rules in high-harmonic spectra: two-color laser case}
\label{app:dyn_sym_2-color}
In this section, we consider the dynamical symmetries in antiferromagnets irradiated by two-color laser, and derive the selection rules of their harmonic generation spectra.

\subsubsection{$C_3$-symmetric case}

In this subsection, we derive the selection rules from the dynamical symmetries of $(\hat U_z(2\pi/3),\,\,T_{\rm ac}/3)$ [Eq.~\eqref{dyn_sym_1_C3_Neel}] and $(\hat U_z(4\pi/3),\,\,2T_{\rm ac}/3)$ [Eq.~\eqref{dyn_sym_2_C3_Neel}]. Namely, we consider the magnet driven by $C_3$-symmetric laser.
Under the $2\pi/3$ spin rotation $\hat U_z(2\pi/3)$, the uniform magnetization $\bm m(t)$ is transformed
\begin{align}
  \hat U_z(2\pi/3) m^x(t)\hat U_z^\dagger(2\pi/3) &= - \frac 12 m^x(t) + \frac{\sqrt 3}{2}m^y(t),\nonumber\\
  \hat U_z(2\pi/3) m^y(t)\hat U_z^\dagger(2\pi/3) &= -\frac{\sqrt 3}2 m^x(t)  -\frac{1}{2}m^y(t)
\end{align}
A similar equation can also be obtained for the $4\pi/3$ spin rotation $\hat U_z(4\pi/3)$ by using the relation $\hat U_z^\dag (2\pi/3) = \hat U_z(4\pi/3)$.
Therefore, we find
\begin{align}
  m^x(t+T_{\rm ac}/3) &= - \frac 12 m^x(t) + \frac{\sqrt 3}{2}m^y(t), \\
  m^y(t+T_{\rm ac}/3)& = -\frac{\sqrt 3}2 m^x(t)  -\frac{1}{2}m^y(t), \\
  m^z(t+T_{\rm ac}/3) &= m^z(t).
\end{align}
from the dynamical symmetry $(\hat U_z(2\pi/3),\,\,T_{\rm ac}/3)$.
Similarly, the other dynamical symmetry \eqref{dyn_sym_2_C3_Neel} leads to
\begin{align}
  m^x(t+2T_{\rm ac}/3) &= - \frac 12 m^x(t) - \frac{\sqrt 3}{2}m^y(t), \\
  m^y(t+2T_{\rm ac}/3) &= \frac{\sqrt 3}2 m^x(t)  -\frac{1}{2}m^y(t), \\
  m^z(t+2T_{\rm ac}/3) &= m^z(t).
\end{align}
These relations affect the Fourier transform $\tilde m^a(n\Omega)$ relevant to the $n$th harmonic generation as follows.
\begin{widetext}
  \begin{align}
    \tilde m^a(n\Omega)
    &= \int_0^{T_{\rm ac}} dt \, e^{in\Omega t} m^a(t) \nonumber\\
    &= \int_0^{T_{\rm ac}/3} dt \, e^{in\Omega t}\Bigl(
      m^a(t) + e^{i2\pi n/3} m^a(t+T_{\rm ac}/3) + e^{i4\pi n/3} m^a(t+2T_{\rm ac}/3)
    \Bigr).
  \end{align}
  Using these relations between $m^a(t+T_{\rm ac}/3)$ [$m^a(t+2T_{\rm ac}/3)$] and $m^a(t)$, we obtain
  \begin{align}
    \tilde m^x(n\Omega)
    &= \int_0^{T_{\rm ac}/3} dt\, e^{in \Omega t} \Bigl[
      \bigl\{ 1- \cos (2\pi n/3) \bigr\}m^x(t) + i\sqrt 3 \sin (2\pi n/3) m^y(t)
    \Bigr], \\
    \tilde m^y(n\Omega)
    &= \int_0^{T_{\rm ac}/3} dt\, e^{in \Omega t} \Bigl[ -i\sqrt 3 \sin (2\pi n/3) m^x(t)+
      \bigl\{ 1- \cos (2\pi n/3) \bigr\}m^y(t)
    \Bigr], \\
    \tilde m^z(n\Omega)
    &= \int_0^{T_{\rm ac}/3} dt \, e^{in\Omega t} \bigl[1+ 2\cos(2\pi n/3)
    \bigr] m^z(t).
  \end{align}
  Therefore, we reach the selection rule that $|\tilde m^x(n\Omega)|=|\tilde m^y(n\Omega)| = 0$ for $n=0 \mod 3$ and $|\tilde m^z(n\Omega)| = 0$ for $n=\pm 1 \mod 3$.
\end{widetext}

\subsubsection{$C_4$-symmetric case}

In analogy with the $C_3$-symmetric case, we derive the selection rule for the $C_4$-symmetric case. First, we consider the dynamical symmetry of $(\hat U_z(\pi/2),\,\,T_{\rm ac}/4)$ [Eq.~\eqref{eq:neel_C4_1}].
If we apply the $\pi/2$ spin rotation $\hat U_z(\pi/2)$ to the total magnetic moment $\bm m(t)$, it changes as follows:
\begin{align}
  \hat U_z(\pi/2) m^x(t)\hat U_z^\dagger(\pi/2) &= m^y(t),\nonumber\\
  \hat U_z(\pi/2) m^y(t)\hat U_z^\dagger(\pi/2)  &= -m^x(t).
\end{align}
This leads to the relation between $\bm m(t)$ and $\bm m(t+T_{\rm ac}/4)$
\begin{align}
  m^x(t+T_{\rm ac}/4) &= m^y (t), \\
  m^y(t+T_{\rm ac}/4) &= - m^x(t), \\
  m^z(t+T_{\rm ac}/4) &= m^z(t).
\end{align}
Likewise, the other two dynamical symmetries [Eqs.~\eqref{eq:neel_C4_2} and \eqref{eq:neel_C4_3}] lead to
\begin{align}
  m^x(t+T_{\rm ac}/2) &= -m^x (t), \\
  m^y(t+T_{\rm ac}/2) &= - m^y(t), \\
  m^z(t+T_{\rm ac}/2) &= m^z(t),
\end{align}
and
\begin{align}
  m^x(t+3T_{\rm ac}/4) &= -m^y (t), \\
  m^y(t+3T_{\rm ac}/4) &= m^x(t), \\
  m^z(t+3T_{\rm ac}/4) &= m^z(t).
\end{align}
These relations result in
\begin{widetext}
  \begin{align}
    \tilde m^x(n\Omega)
    &= \int_0^{T_{\rm ac}/4} dt \, e^{in\Omega t} \Bigl[
      \bigl\{ 1- (-1)^n \bigr\} m^x(t) +2i \sin (\pi n/2)m^y(t)
    \Bigr], \\
    \tilde m^y(n\Omega)
    &= \int_0^{T_{\rm ac}/4} dt \, e^{in\Omega t} \Bigl[-2i \sin (\pi n/2)m^x(t) +\bigl\{ 1- (-1)^n \bigr\} m^y(t) \Bigr], \\
    \tilde m^z(n\Omega)
    &= \int_0^{T_{\rm ac}/4} dt \, e^{in\Omega t} \bigl\{1+(-1)^n + 2\cos(\pi n/2) \bigr\} m^z(t).
  \end{align}
\end{widetext}
Hence, we obtain the selection rules: $\tilde m^x(n\Omega) = \tilde m^y(n\Omega) = 0$ for $n=0 \mod 2$ and $\tilde m^z(n\Omega) = 0$ for $n=1,~2,~3 \mod 4$.


\section{Laser pulse width}
\label{app:pulselength}
In this section, we briefly discuss the effect of varying the laser pulse length on the shape of harmonic generation peaks. 
The following content is easier to understand after reading Sec.~\ref{sec:Neel}.

\begin{figure}
  \centering
  \includegraphics[width=0.9\linewidth]{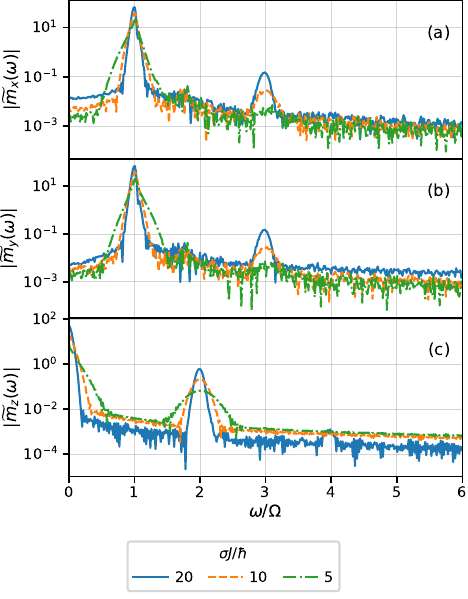}
  \caption{Laser-pulse width $\sigma$ dependence of harmonic generation spectra in the N\'eel phase irradiated by linearly polarized THz laser with frequency $\Omega=\Omega_1$ and ${\bm B}_{\rm ac}\parallel {\bm e}_{x}$. Blue, orange and green curves are the results of $\sigma J/\hbar=20, 10$ and $5$, respectively. Panels (a), (b), and (c) respectively correspond to the results of $|\tilde m^x(\omega)|$, $|\tilde m^y(\omega)|$, and $|\tilde m^z(\omega)|$. We set $k_{\rm B}T/J=10^{-3}$, $B_{\rm ac}/J=0.1$, and $B_0/J=0.5$.
    The results of blue curve with $\sigma J/\hbar=20$ are equivalent to the first column of Fig.~\ref{fig:neelhhg} (a).
  }
  \label{fig:Neel-sigma-dep}
\end{figure}

As we argued at the end of Sec.~\ref{sec:DS_in_general}, if the condition $1/\sigma<\Omega$ is satisfied in laser-pulse driven systems ($\sigma$ is the laser pulse width along the time direction and $\Omega$ is the laser frequency), the peaks of the HHG spectrum are expected to be observed and to follow the selection rule derived from the dynamical symmetry. To quantitatively verify this argument, we numerically compute the harmonic generation spectra of the N\'eel phase irradiated by THz laser pulse, by changing the pulse width.
The result computed by the LLG equation is summarized in Fig.~\ref{fig:Neel-sigma-dep}.
As expected, the peaks become wider with decreasing pulse width. However, one can verify that the harmonic peaks up to the second or third order can be observed and the peak structure agrees with the selection rule derived by the dynamical symmetry even in the short pulse width $\sigma J/\hbar=5$. In the setup of Fig.~\ref{fig:Neel-sigma-dep}, $1/\sigma<\Omega$ (i.e., $\hbar/(J\sigma)<\hbar\Omega/J$) is satisfied due to $\hbar\Omega=1.33J$.
These results indicate that even under application of a finite-width laser pulse, the selection rule derived by a dynamical symmetry holds in the harmonic generation spectrum if the width is sufficiently large.


\section{Effect of small anisotropy $D_y$ in the canted phase}
\label{app:anistropy_cant}
In this section, we briefly consider the effects of small magnetic anisotropy $D_y$ on the harmonic generation spectrum in the canted phase at finite temperatures.
The following content is easier to understand after reading Sec.~\ref{sec:Canted}.

\begin{figure}
  \centering
  \includegraphics[width=0.9\linewidth]{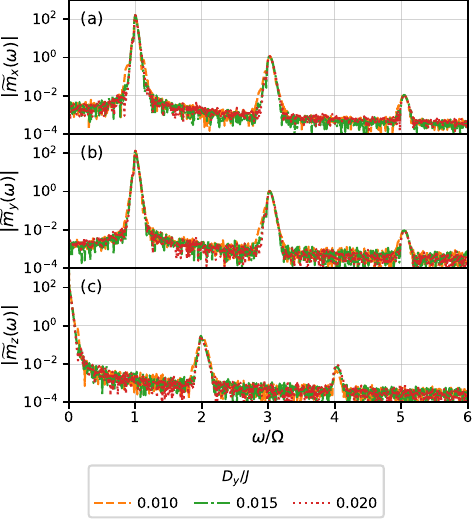}
  \caption{Harmonic generation spectra in canted phase irradiated by linearly polarized THz laser with frequency $\Omega=\Omega_1$ and ${\bm B}_{\rm ac}\parallel {\bm e}_{x}$. Orange, green and red curves are the results of $D_y/ J=0.01, 0.015$ and $0.02$, respectively. Panels (a), (b), and (c) respectively correspond to the results of $|\tilde m^x(\omega)|$, $|\tilde m^y(\omega)|$, and $|\tilde m^z(\omega)|$. We set $k_{\rm B}T/J=10^{-3}$, $B_{\rm ac}/J=0.1$, $B_0/J=3.0$ and $\hbar\Omega=2.7J=\epsilon_{\bm{0}}^{\text{Cant}, \alpha}$.This laser frequency (the magnon gap of the $\alpha$ mode) would be much larger than the excitation gap induced by small $D_y$.
    The result of orange curve with $D_y/J=0.01$ is equivalent to the first column of Fig.~\ref{fig:Canthhg} (a).
  }
  \label{fig:Canted-Dy-dep}
\end{figure}

In Sec.~\ref{subsec:NeelCant}, this anisotropy term
$D_y\sum_{\bm r}(\hat S_{\bm r}^y)^2$ was introduced to incorporate the effects of weak magnetic anisotropy or three-dimensionality, which stabilize the canted phase in actual magnetic materials. Here, we numerically verify that a sufficiently small anisotropy $D_y$ is irrelevant for the harmonic generation spectrum in the canted phase if we focus on a higher laser frequency regime rather than the small magnon gap induced by $D_y$. The numerical result of the $D_y$ dependence of the spectra is given in Fig.~\ref{fig:Canted-Dy-dep}.
One can see from the figure that the shapes of the harmonic generation spectra are almost invariant under changing the values of small $D_y$. The quantitative computation of the magnon gap induced by $D_y$ is beyond the scope of the present work, but we can expect that the gap is much smaller than the laser frequency $\hbar\Omega=2.7J$ used in Fig.~\ref{fig:Canted-Dy-dep} since we set $D_y$ to sufficiently smaller values than $J$.
From these results, we conclude that the numerically computed spectra for the canted phase with a small $D_y/J=0.01$ in Fig.~\ref{fig:Canthhg} are reliable.


\section{Amplitude of laser-driven spin precession}
\label{app:amplitude}
In this section, we briefly discuss how large the intense THz-laser driven spin precession is. The content below is easier to understand after reading Sec.~\ref{sec:WeakFerro}.

The harmonic generation in magnets is a typical nonlinear optical phenomenon, and thereby a sufficiently strong THz laser is necessary to observe high-order harmonics.
On the other hand, if the THz laser is too strong, the magnetic order at the initial state before laser application would be broken by the effect of the laser field.
Let us verify that the THz magnetic field used in this study ($B_{\rm ac}/J=0.1$) does not violate the magnetic order.
\begin{figure}
  \centering
  \includegraphics[width=\linewidth]{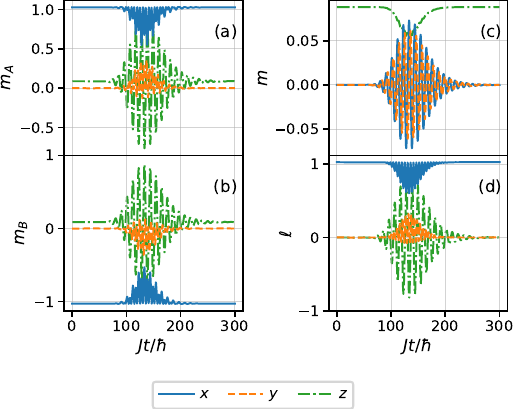}
  \caption{Real time evolution of uniform, staggered, and sublattice magnetizations in THz-laser pulse driven WF phase, where $\bm B_{\rm ac}\parallel {\bm e}_x$. (a) $m_A^\alpha=N_A^{-1}\sum_{{\bm r}\in A}m_{\bm r}^\alpha$ and (b) $m_B^\alpha=N_B^{-1}\sum_{{\bm r}\in B}m_{\bm r}^\alpha$ are respectively the $\alpha$ component of A- and B-sublattice magnetizations per one site, where $N_A=N_B$ is the total site number in the A (or B) sublattice. (c) $m^\alpha=\frac{1}{2}(m_A^\alpha+m_B^\alpha)$ and (d) $\ell^\alpha=\frac{1}{2}(m_A^\alpha-m_B^\alpha)$ are respectively the $\alpha$ component of the uniform and staggered magnetizations. Blue, orange, and green curves correspond to the $x$, $y$ and $z$ components of each magnetization, respectively. We set $k_{\rm B}T/J=10^{-3}$, $B_{\rm ac}/J=0.1$, $B_0/J=0.5$ and the laser frequency $\hbar\Omega=0.591J=\epsilon_{\bm{0}}^{\text{WF}, \alpha}$. These results correspond to the first column of Fig.~\ref{fig:wfhhg} (a).
  }
  \label{fig:WF-time-evolution}
\end{figure}
Figure~\ref{fig:WF-time-evolution} shows the numerically computed time evolution of uniform, staggered, and sublattice magnetizations in the WF phase during THz-laser application.
Panels (a), (b) and (d) indicate that the precession motions of sublattice and staggered magnetizations are extremely large and in particular, the $z$ components approach the maximum values $|\ell^z|=|m_{A,B}^z|=1$, while these magnetizations finally return to the initial values after laser application. Panel (c) shows that the precession of the uniform magnetization is small compared with those of $\bm \ell$ and ${\bm m}_{A,B}$.
From these results, we conclude that in our setup, one observes the harmonic generation without breakdown of the magnetic order. Similarly, there is no breakdown of the initial magnetic order in the case of N\'eel and canted phases.

Strictly speaking, the time evolution of uniform and staggered magnetization (Fig.~\ref{fig:WF-time-evolution}) alone does not rule out the possibility of generating magnetic order at arbitrary wavevectors via THz-laser application. In magnetic materials, for instance, magnons at finite wavevectors can be resonantly generated through parametric resonance~\cite{Suhl1957theory, Lvov1994Wave, Bracher2017Parallel, Zheng2023Tutorial}. However, in experiments of harmonic generation in magnetic materials, the observed quantity is typically limited to uniform magnetization (i.e., magnetization at zero wavevector $\bm k=\bm 0$), while observing magnetization or excitations at finite wave vectors on the ultrafast (THz) timescale remains challenging. Consequently, this paper focuses on the harmonic generation spectra, while leaving the discussion of instabilities at finite wavevectors as an intriguing topic for future research.

\section{Finite-size effects}
\label{app:size}


\begin{figure}[t!]
  \centering
  \includegraphics[bb = 0 0 1536 1200, width=\linewidth]{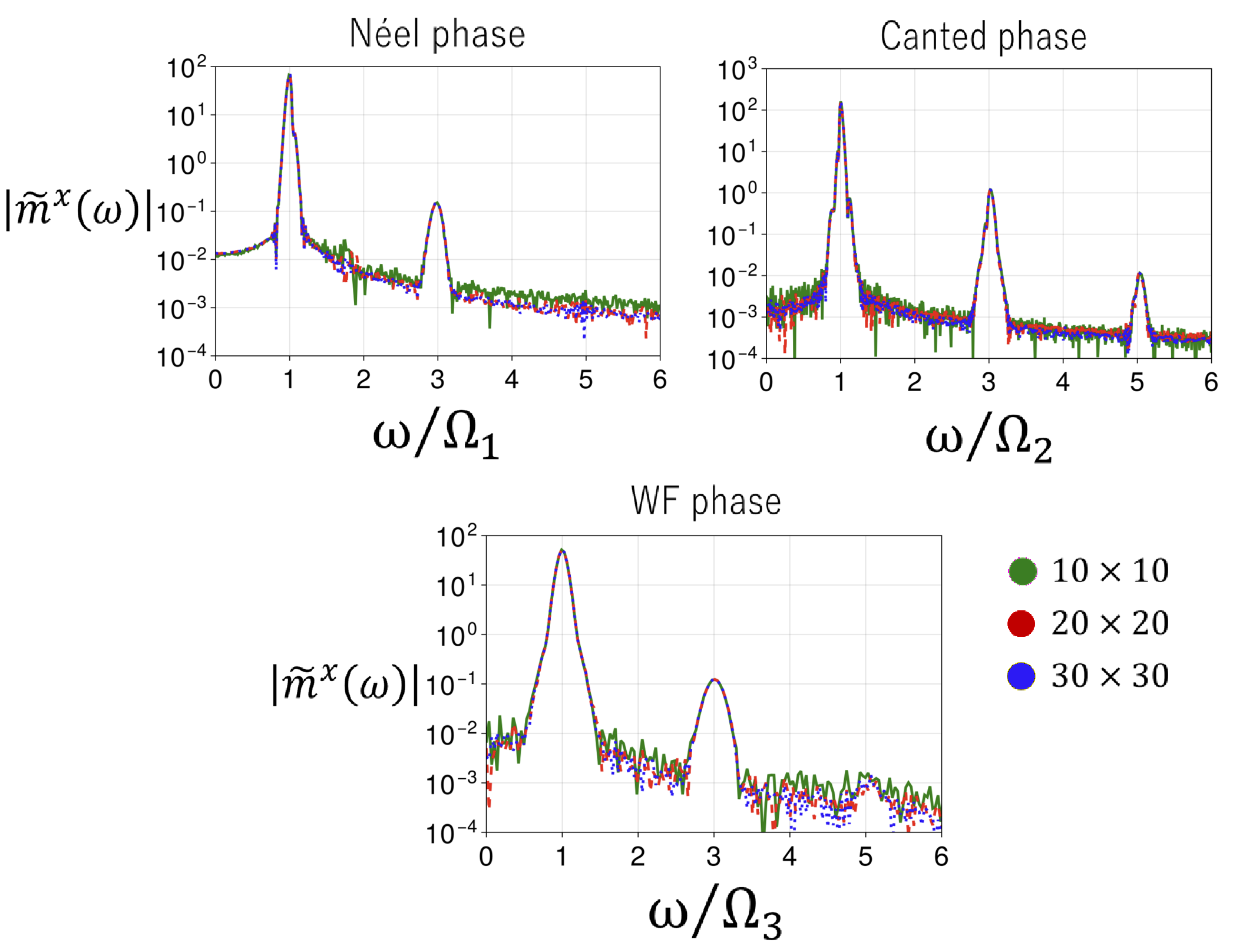}
  \caption{
    Comparison of harmonic generation spectra with system sizes, $10\times 10$, $20\times 20$, and $30\times 30$.
    Panels (a), (b), and (c) show the spectrum $|\tilde m^x(\omega)|$ in the N\'eel, canted, and WF phases, respectively.
    We set the static magnetic field strength to $B_0/J=$ (a) $0.5$, (b) $3.0$, and (c) $0.5$.
    We apply a laser pulse field parallel to the $x$ axis to the system.
    The horizontal axes are plotted in units of $\Omega_n$ ($n=1,~2,~3$).
    The unit $\Omega_n$ denotes the resonant frequency in each phase given by $\hbar\Omega_{1}/J = \epsilon_{\bm{0}}^{\text{N\'eel}, \beta}/J = 1.33, \ \hbar\Omega_{2}/J = \epsilon_{\bm{0}}^{\text{Cant}, \alpha}/J = 2.7, \ \hbar\Omega_{3}/J = \epsilon_{\bm{0}}^{\text{WF}, \alpha}/J = 0.591$.
  }
  \label{fig:sizedep}
\end{figure}

Let us examine the system-size dependence of the harmonic generation spectrum and confirm that the $10\times 10$ size is large enough to guarantee the convergence of our numerical results to the infinite-size limit.
Figure~\ref{fig:sizedep} gives the numerically estimated harmonic generation spectra in the N\'eel, canted and WF phases with different system sizes: Antiferromagnetic models on square lattices consisting of $10\times 10$, $20\times 20$ and $30\times 30$ sites. The result
clearly shows that the $10\times 10$ system size is large enough.
The lineshape of every harmonic generation peak shown there exhibits almost no size dependence.
We can find subtle system-size dependence only in the background noise.


\section{Time step in numerical calculation of LLG equation}
\label{app:timestep}
\begin{figure}
  \centering
  \includegraphics[width=\linewidth]{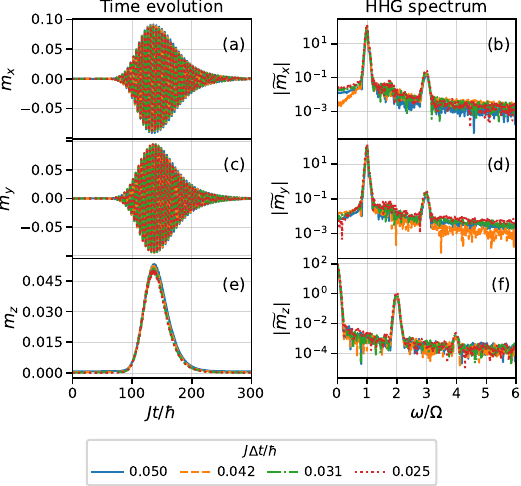}
  \caption{Time-step dependence of time evolution of magnetization $m^\alpha(t)$ and absolute value of its Fourier component $|{\tilde m}^\alpha(\omega)|$ in N\'eel phase irradiated by THz laser pulse with $\bm B_{\rm ac}\parallel {\bm e}_x$. Blue, orange, green and red curves are respectively the numerical results of $J\Delta t/\hbar=0.05$, $0.042$, $0.031$, and $0.025$. We set $k_{\rm B}T/J=10^{-3}$, $B_{\rm ac}/J=0.1$, $B_0/J=0.5$, and $Jt_{\rm i}/\hbar=0$.
    The results of blue curve with $J\Delta t/\hbar=0.05$ are equivalent to the first column of Fig.~\ref{fig:neelhhg} (a).
  }
  \label{fig:Neel-dt-dep}
\end{figure}

In this section, we examine the time-step dependence of the harmonic generation spectrum that we compute by numerically solving the LLG equation. The dimensionless time step $J\Delta t/\hbar$ is defined as the shortest time width when we numerically solve the LLG equation. To check the time-step dependence, let us focus on the harmonic generation in the N\'eel phase. Figure~\ref{fig:Neel-dt-dep} gives the numerically computed time evolution of total magnetization $m^\alpha(t)$ and its harmonic generation spectrum $|{\tilde m}^\alpha(\omega)|$, by changing the value of the time steps $J\Delta t/\hbar=0.05$-$0.025$.
From the figure, one sees that the lineshape of every harmonic generation peak exhibits almost no time-step dependence (although the noise part in the no-peak regime shows a moderate time-step dependence). Therefore, we conclude that the time step $J\Delta t/\hbar=0.05$, which is used in the present study, is sufficiently short to obtain the reliable lineshape of the harmonic generation peaks.

\section{Time window in numerical calculation of LLG equation}
\label{app:timewindow}
Similarly to Appendix~\ref{app:timestep}, we here discuss the
time-window length dependence of the harmonic generation spectrum. As we explained in Sec.~\ref{sec:laser}, we have taken $t_{\rm i} = -6\sigma$ and $t_{\rm end} = 9\sigma$ in the numerical calculation of the present study.
However, there is the possibility that if the end point of the observation $t_{\rm end}$ is increased, the lineshape of the harmonic spectrum moderately changes. To verify that $t_{\rm end} = 9\sigma$ is long enough to obtain the reliable spectrum, we numerically compute the harmonic generation spectra in the WF phase, by varying the value of $t_{\rm end}$ while fixing $t_{\rm i} = -6\sigma$.
\begin{figure}
  \centering
  \includegraphics[width=0.9\linewidth]{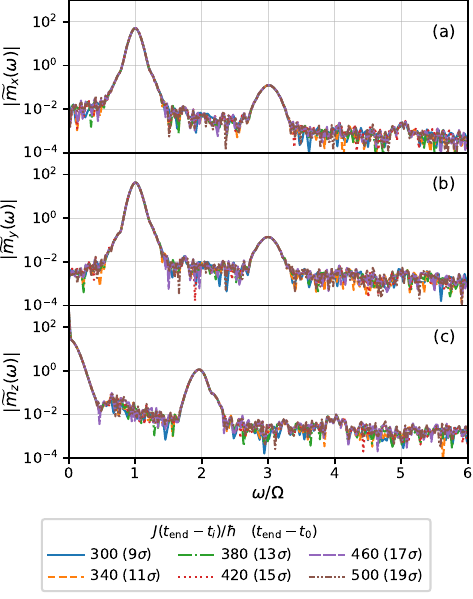}
  \caption{Time-window length ($t_{\rm end}-t_{0}$) dependence of harmonic generation spectrum in WF phase irradiated by THz laser pulse with $\bm B_{\rm ac}\parallel {\bm e}_x$. Panels (a), (b), and (c) respectively show $|{\tilde m}^x(\omega)|$, $|{\tilde m}^y(\omega)|$, and $|{\tilde m}^z(\omega)|$, by varying $t_{\rm end}$ from $9\sigma$ to $19\sigma$ while fixing $t_{\rm i}-t_0=-6\sigma$.
    We set $k_{\rm B}T/J=10^{-3}$, $B_{\rm ac}/J=0.1$, $B_0/J=0.5$ and the laser frequency $\hbar\Omega=0.591J=\epsilon_{\bm{0}}^{\text{WF}, \alpha}$. The result of $t_{\rm end}-t_{\rm i}=300$ ($t_{\rm end}=9\sigma$) corresponds to the first column of Fig.~\ref{fig:wfhhg} (a).
  }
  \label{fig:WF-FFT-time-window-dep}
\end{figure}
Figure~\ref{fig:WF-FFT-time-window-dep} shows the time-window length (i.e., $t_{\rm end}$) dependence of the harmonic generation spectra in the WF phase. Panels (a)-(c) clearly indicate that $t_{\rm end} = 9\sigma$ is sufficiently long and there is almost no $t_{\rm end}$ dependence of the spectra for $t_{\rm end} \geq 9\sigma$.
Thus, our setup of $t_{\rm end} = 9\sigma$ is justified.

\section{Classical picture of magnetic resonance}
\label{app:resonance}

This section provides a short review of the magnetic resonances in antiferromagnets from the classical spin viewpoint.
Namely, we discuss the magnetic resonance by regarding the spin operator $\hat{\bm S}_{\bm r}$ as the classical vector $\bm m_{\bm r}$ with a fixed length. The classical picture of magnetic resonance tells us that the precession motion of the uniform magnetic moment draws an ellipse in the canted and WF phases, while it draws a circle in the N\'eel phase [see Fig.~\ref{fig:model} (e) and (f)]. As a result, the intensity of harmonic generation peaks exhibits the dependence of the polarization of the ac magnetic field in the canted and WF phases.

\subsection{N\'eel phase}

Classically, magnetic resonance in magnetic Mott insulators occurs when the external ac magnetic field induces resonant precession of uniform (total) magnetic moment $\bm m =N^{-1}\sum_{\bm r} \bm m_{\bm r}$.
The magnetic resonance involves the uniform magnetic moment since the external ac magnetic field (electromagnetic wave) can be deemed the plane wave, as we mentioned already in the main text.

The mean-field description is sufficient to obtain the classical picture of magnetic resonance in magnetically ordered phases such as the N\'eel phase.
In the N\'eel phase, the uniform magnetic moment $\bm m$ per one site is made of two sublattice moments, $\bm m_{\rm A}$ and $\bm m_{\rm B}$.  The relation $\bm m = \frac{1}{2}(\bm m_{\rm A} + \bm m_{\rm B})$ holds. Suppose that these modes satisfy the following mean-field LL equation,
\begin{align}
  \hbar \frac{d\bm m_{\rm A}}{dt} &= \bm m_{\rm A} \times \frac{\partial \mathcal H_{\text{N-C;MF}}}{\partial \bm m_{\rm A}},
  \label{LL_Neel_A} \\
  \hbar \frac{d\bm m_{\rm B}}{dt} &= \bm m_{\rm B} \times \frac{\partial \mathcal H_{\text{N-C;MF}}}{\partial \bm m_{\rm B}}.
  \label{LL_Neel_B}
\end{align}
Here, $\mathcal H_{\text{N-C;MF}}$ is the mean-field Hamiltonian corresponding to Eq.~\eqref{H_Neel-Canted_def}:
\begin{align}
  \mathcal H_{\text{N-C;MF}} &= zJ \bm m_{\rm A} \cdot \bm m_{\rm B} - B_0 (m_{\rm A}^z+m_{\rm B}^z) \notag \\
  &\qquad -K\{(m_{\rm A}^z)^2+(m_{\rm B}^z)^2\}.
\end{align}
Magnetic resonance in the N\'eel phase has the property that the spin precessions in the A and B sublattices both draw circles, but they have different radii.

The radius $M_\mu$ ($\mu=\mathrm{A,~B}$) is defined as follows.
We may rewrite the precessing mode $\bm m_\mu = (m_\mu^x,~m_\mu^y,~m_\mu^z)$ as
\begin{align}
  m_{\rm A}^x &= M_{\rm A} \cos \omega t, \\
  m_{\rm A}^y &= M_{\rm A} \sin \omega t, \\
  m_{\rm A}^z &= \sqrt{S^2-M_{\rm A}^2},
\end{align}
for $\mu = \mathrm A$ and
\begin{align}
  m_{\rm B}^x
  &= -M_{\rm B} \cos \omega t, \\
  m_{\rm B}^y
  &= -M_{\rm B} \sin \omega t, \\
  m_{\rm B}^z
  &= -\sqrt{S^2-M_{\rm B}^2}
\end{align}
for $\mu=\mathrm B$.
Substituting these $\bm m_{\rm A}$ and $\bm m_{\rm B}$ into the LL equations \eqref{LL_Neel_A} and \eqref{LL_Neel_B}, we obtain a relation between the radii $M_{\rm A}$ and $M_{\rm B}$.
The $x$ and $y$ components of Eq.~\eqref{LL_Neel_A} lead to an equation for $m_{\rm A}^+ = m_{\rm A}^x + i m_{\rm A}^y$,
\begin{align}
  \hbar \frac{dm_{\rm A}^+}{dt} &= -im_{\rm A}^+(zJm_{\rm B}^z-B_0-2K m_{\rm A}^z) + i zJ m_{\rm A}^z m_{\rm B}^+
  \label{LL_Neel_A_mplus}
\end{align}
Let us drop second- or higher-order terms about $M_{\mu}$ by approximating $m_{\rm A}^z\approx S$ and $m_{\rm B}^z \approx -S$.
Since $m_{\rm A}^+ = M_{\rm A}e^{i\omega t}$ and $m_{\rm B}^+ = - M_{\rm B} e^{i\omega t}$, we can rewrite Eq.~\eqref{LL_Neel_A_mplus} as
\begin{align}
  i\hbar \omega M_{\rm A}
  &\approx -iM_{\rm A}(-zJS-B_0-2KS)-izJSM_{\rm B}
  \label{amp_A_Neel}
\end{align}
Likewise, $m_{\rm B}^+=m_{\rm B}^x + i m_{\rm B}^y$ satisfies
\begin{align}
  \hbar \frac{dm_{\rm B}^+}{dt} &= -im_{\rm B}^+(zJm_{\rm A}^z-B_0-2K m_{\rm B}^z) +izJ m_{\rm B}^z m_{\rm A}^+
  \label{LL_Neel_B_mplus}
\end{align}
We thus obtain
\begin{align}
  -i\hbar \omega M_{\rm B}
  &\approx iM_{\rm B} (zJS-B_0+2KS)-izJSM_{\rm A}.
  \label{amp_B_Neel}
\end{align}
Equations~\eqref{amp_A_Neel} and \eqref{amp_B_Neel} give the following equation for a ratio
\begin{align}
  X = \frac{M_{\rm A}}{M_{\rm B}}
  \label{ratio_AB}
\end{align}
about the radii $M_{\rm A}$ and $M_{\rm B}$:
\begin{align}
  -X = \frac{X(-zJS-B_0-2KS)+zJS}{-(zJS-B_0+2KS)+zJSX}.
\end{align}
This is rewritten as
\begin{align}
  X^2 - 2\biggl( 1+\frac{2K}{zJ}\biggr) X +1 = 0.
\end{align}
Obviously, this equation has two solutions $X=X_\pm$, that is,
\begin{align}
  X_\pm
  &= 1+ \frac{2K}{zJ} \pm \sqrt{\biggl( 1+\frac{2K}{zJ}\biggr)^2 - 1}.
\end{align}
The ratio \eqref{ratio_AB} equals $1$ only when $K=0$. The easy-axis single-ion anisotropy $K>0$ makes the ratio deviate from $1$.
The solutions $X_+$ and $X_-$ correspond to the $\alpha$ mode and the $\beta$ mode with $\bm k=\bm 0$, respectively [see Fig.~\ref{fig:model}~(e)].
The two precession modes draw the circle trajectory on the $S^x$-$S^y$ plane, whose radius $M_{\rm \mu}$ depends on the index $\mu=\mathrm{A,~B}$.

\subsection{Canted phase}

The precession in the canted phase draws elliptical orbits instead of the circular one, as we derive below.
In the $\xi\eta\zeta$ coordinate introduced in Appendix~\ref{app:LSW_canted}, the magnetic moment in the $\mu=\mathrm{A,B}$ sublattice is given by
\begin{align}
  m_{\mu}^\xi &= M^\xi \cos \omega t, \\
  m_{\mu}^\eta &= M^\eta \sin \omega t, \\
  m_{\mu}^\zeta &= \sqrt{S^2-(m^\xi_\mu)^2-(m^\eta_\mu)^2}.
\end{align}
In this equation, we assume that $\bm m_{\rm A}$ and $\bm m_{\rm B}$ follow the precession motion with the same phase. This means that we consider the $\alpha$ mode [see Fig.~\ref{fig:model} (f)].
We can translate the $\xi\eta\zeta$ coordinate into the $xyz$ coordinate by the following rotations:
\begin{align}
  m_{\rm A}^x &= m_{\rm A}^\xi \cos \theta  - m_{\rm A}^\zeta \sin \theta, \\
  m_{\rm A}^y &= m_{\rm A}^\eta, \\
  m_{\rm A}^z &= m_{\rm A}^\xi \sin \theta + m_{\rm A}^\zeta \cos \theta,
\end{align}
for the A sublattice and
\begin{align}
  m_{\rm B}^x &= m_{\rm B}^\xi \cos \theta +m_{\rm B}^\zeta \sin \theta, \\
  m_{\rm B}^y &= m_{\rm B}^\eta, \\
  m_{\rm B}^z &= -m_{\rm B}^\xi \sin \theta+ m_{\rm B}^\zeta \cos\theta,
\end{align}
for the B sublattice.
Here, $\theta$ is determined by $B_0$ and $J$ [see Eq.~\eqref{canted_angle_classic}].
Substituting these into the LL equations,
\begin{align}
  \hbar \frac{dm_{\rm A}^x}{dt}
  &= \left({\bm m}_{\rm A}\times \frac{\partial \mathcal H_{\text{N-C;MF}}}{\partial {\bm m}_{\rm A}}\right)^x
  \notag \\
  &= m_{\rm A}^y(zJm_{\rm B}^z-B_0-2Km_{\rm A}^z)-zJm_{\rm A}^z m_{\rm B}^y, \\
  \hbar \frac{dm_{\rm A}^y}{dt}
  &= \left({\bm m}_{\rm A}\times \frac{\partial \mathcal H_{\text{N-C;MF}}}{\partial {\bm m}_{\rm A}}\right)^y
  \notag \\
  &=-m_{\rm A}^x(zJm_{\rm B}^z-B_0-2Km_{\rm A}^z) + zJ m_{\rm A}^z m_{\rm B}^x,
\end{align}
and dropping second- and higher-order terms about $M^\xi$ and $M^\eta$, we obtain
\begin{widetext}
  \begin{align}
    \hbar M^\xi (-\omega \sin \omega t) \cos \theta
    &\approx  -M^\eta \sin \omega t(B_0+ 2KS\cos\theta),
    \label{amp_xi_canted}
    \\
    \hbar M^\eta \omega \cos \omega t
    &\approx M^\xi \cos \omega t (B_0\cos \theta + 2KS \cos 2\theta) + S \sin \theta \cos \theta \bigl[2S(zJ-K) \cos \theta -B_0] \nonumber\\
    &=  M^\xi \cos \omega t (B_0\cos \theta + 2KS \cos 2\theta).
    \label{amp_eta_canted}
  \end{align}
  In the last line, we used Eq.~\eqref{canted_angle_classic} to eliminate $O((M^\xi)^0)$ terms
\end{widetext}
Let us define another ratio,
\begin{align}
  Y = \frac{M^\xi}{M^\eta},
  \label{ratio_xi_eta_def}
\end{align}
which defines the anisotropy of the ellipse.
The precessing mode $(m_{\rm A}^\xi,~m_{\rm A}^\eta,~m_{\rm A}^\zeta)$ draws the elliptic trajectory on the $\xi\eta$ plane for $Y\not=1$ and the circle trajectory for $Y=1$.
Dividing Eq.~\eqref{amp_xi_canted} by Eq.~\eqref{amp_eta_canted}, we find an equation about $Y$,
\begin{align}
  -Y \cos\theta &\approx \frac{B_0+2KS\cos\theta}{-Y (B_0\cos\theta + 2KS \cos2\theta)},
\end{align}
whose solution is
\begin{align}
  Y = \sqrt{\frac{B_0+2KS\cos\theta}{\cos\theta (B_0\cos\theta + 2KS \cos2\theta)}}.
  \label{ratio_xi_eta_result}
\end{align}

\begin{figure}[h]
  \centering
  \includegraphics[bb = 0 0 850 509, width = \linewidth]{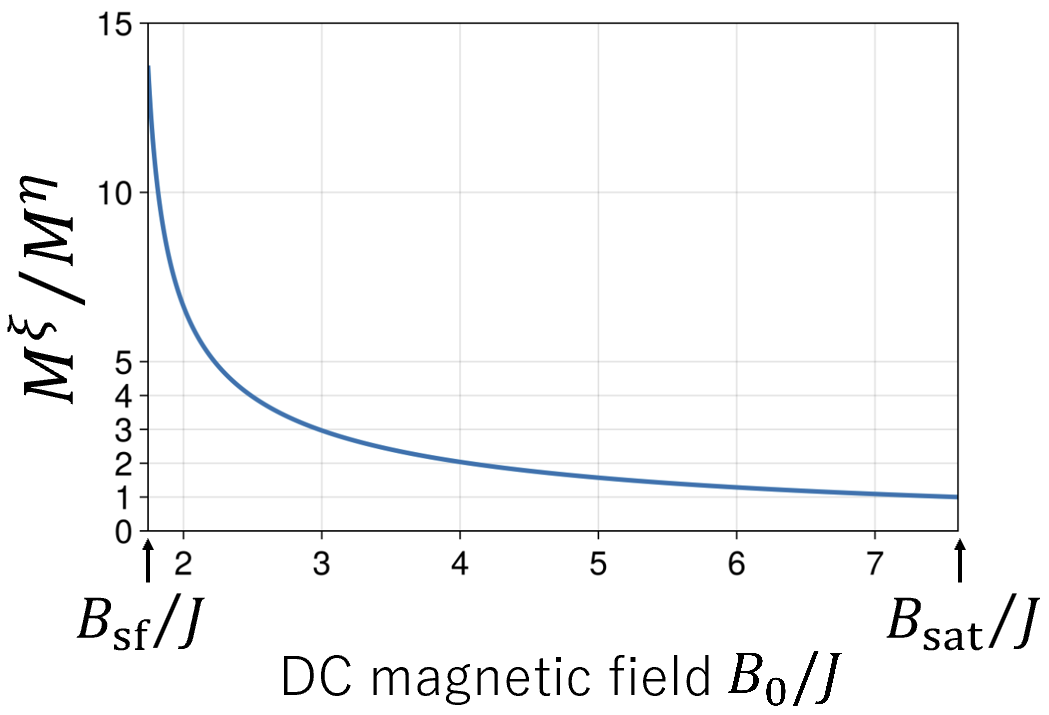}
  \caption{
    $B_0/J$ dependence of ratio $Y$ in canted phase.
    We set $K/J=0.2$.
  }
  \label{fig:cantdaen}
\end{figure}

Figure~\ref{fig:cantdaen} shows the ratio \eqref{ratio_xi_eta_result} as a function of $B_0/J$.
The ellipse approaches the circle as $B_0$ approaches the saturation transition point, $B_0\to B_{\rm sat}-0$.
By contrast, the ellipse is more distorted as $B_0$ approaches the spin-flop transition point, $B_0\to B_{\rm sf}+0$.

From the above result, the uniform magnetization $\bm m=\frac{1}{2}(\bm m_A+\bm m_B)$ per unit cell obeys a precession motion in the $S^x$-$S^y$ plane and its orbit is an ellipse. The $S^x$ axis corresponds to the major axis and the ratio between major and minor radii is given by $Y\cos\theta$.

\subsection{Weak-ferromagnetic phase}

The magnetic structure in the WF phase is akin to that in the canted phase.
The precession in the WF phase as well as the canted phase draws the elliptic trajectory in the $\xi\eta$ plane.
The difference between these two phases arises from the Hamiltonian.

We apply the mean-field approximation to the Hamiltonian \eqref{H_WF_def} and consider
\begin{align}
  \mathcal H_{\text{WF;MF}}
  &= zJ \bm m_{\rm A} \cdot \bm m_{\rm B} - zD (m_{\rm A}^z m_{\rm B}^x- m_{\rm A}^x m_{\rm B}^z) \nonumber\\
  &\qquad  - B_0 (m_{\rm A}^z+m_{\rm B}^z).
\end{align}
We then obtain the LL equation,
\begin{align}
  \hbar \frac{d\bm m_{\mu}}{dt} = \bm m_{\mu} \times \frac{\partial \mathcal H_{\rm WF;MF}}{\partial \bm m_\mu} \qquad (\mu=\mathrm{A,~B}).
  \label{LL_WF_mu}
\end{align}
This equation is rewritten as
\begin{align}
  \frac{\partial \mathcal H_{\rm WF;MF}}{\partial m_\mu^x}
  &= zJ m_{\mu'}^x +\sigma zD m_{\mu'}^z, \\
  \frac{\partial \mathcal H_{\rm WF;MF}}{\partial m_\mu^y}
  &= zJ m_{\mu'}^y, \\
  \frac{\partial \mathcal H_{\rm WF;MF}}{\partial m_\mu^z}
  &= zJ m_{\mu'}^z- \sigma zDm_{\mu'}^x - B_0,
\end{align}
where $\mu'=\mathrm{A}$ for $\mu=\mathrm{B}$ and $\mu'=\mathrm{B}$ for $\mu=\mathrm{A}$,
and we take $\sigma = 1$ for $\mu=\mathrm{A}$ and $\sigma=-1$ for $\mu=\mathrm{B}$.
Note that the sign $\sigma$ in $\mathcal H_{\rm WF;MF}$ disappears in the effective field $\partial \mathcal H_{\rm WF;MF}/\partial \bm m_\mu$ that $\bm m_{\mu}$ feels since the antisymmetry of the Dzyaloshinskii-Moriya interaction cancels it.
We consider the following precessing motion in the $\xi\eta\zeta$ coordinate system,
\begin{align}
  m_\mu^\xi &= M^\xi \cos \omega t, \\
  m_\mu^\eta &= M^\eta \sin \omega t, \\
  m_\mu^\zeta &= \sqrt{S^2-(m_\mu^\xi)^2-(m_\mu^\eta)^2},
\end{align}
and rotate it by angles $\pm \theta$ around $\eta$ axis to return to the $xyz$ coordinate system in analogy with the canted phase.
In this equation, we assume that $\bm m_{\rm A}$ and $\bm m_{\rm B}$ follow the precession motion with the same phase. This means
that we consider the $\alpha$ mode [see Fig.~\ref{fig:model} (f)].

\begin{figure}[t!]
  \centering
  \includegraphics[bb = 0 0 650 409, width = \linewidth]{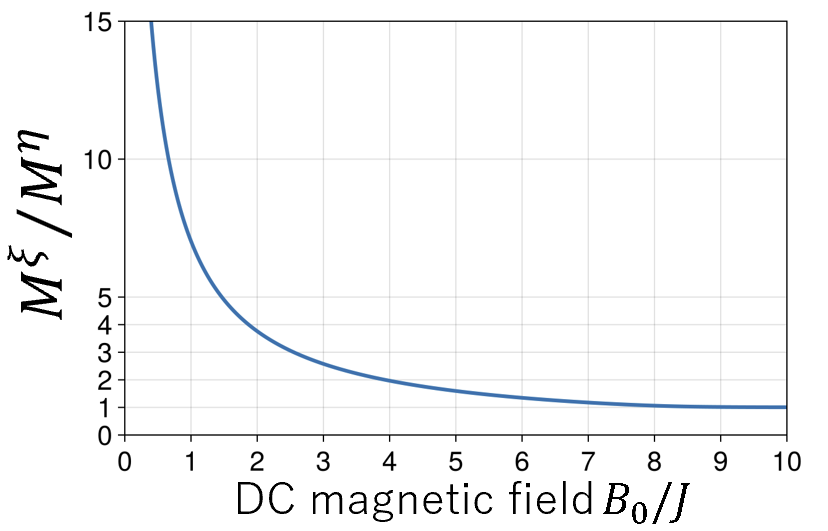}
  \caption{
    $B_0/J$ dependence of ratio $Y$ in WF phase. We set $D/J=0.05$.
  }
  \label{fig:wfdaen}
\end{figure}

The LL equations for $m_{\rm A}^x$ and $m_{\rm A}^y$ become
\begin{widetext}
  \begin{align}
    \hbar M^\xi (-\omega \sin \omega t) \cos \theta
    &\approx M^\eta\sin \omega t(zJS\cos\theta -zDS \sin \theta - B_0) - zJSM^\eta \sin \omega t\cos\theta \nonumber\\
    &\approx -M^\eta\sin \omega t(B_0+zDS \sin \theta), \\
    \hbar M^\eta \omega \cos \omega t
    &\approx  (M^\xi \cos\omega t \sin \theta +S \cos\theta)\{zJ(M^\xi \cos \omega t \cos \theta +S \sin \theta) +zD(-M^\xi \cos \omega t \sin \theta + S \cos \theta) \} \nonumber\\
    &\qquad
    - (M^\xi \cos \omega t \cos \theta - S \sin \theta) \bigl\{zJ(-M^\xi \cos \omega t \sin \theta +S \cos \theta) 
    -zD(M^\xi \cos \omega t \cos \theta + S \sin \theta) -B_0 \bigr\} \nonumber\\
    &\approx B_0M^\xi \cos \omega t \cos \theta + zJS^2 \sin 2\theta + zDS^2 \cos 2\theta - B_0S \sin \theta \nonumber\\
    &= B_0 M^\xi \cos \omega t \cos \theta.
  \end{align}
\end{widetext}
In the last line, we used a relation about the ground-state energy density,
\begin{align}
  \frac{\partial e_{\rm WF}^0}{\partial \theta} = -zJS^2 \sin 2\theta- zDS^2 \cos 2\theta + B_0S\sin \theta =0,
\end{align}
where $e_{\rm WF}^0$ is the ground-state energy density [see Eq.~\eqref{H_WF_FT}], namely,
\begin{align}
  e_{\rm WF}^0 = \frac 12 zJS^2 \cos 2\theta - \frac 12 zDS^2 \sin 2\theta - B_0S \cos \theta.
\end{align}
The stability of the ground state is rephrased as $\partial e_{\rm WF}^0/\partial \theta = 0$.
We thus obtain the following equation about the ratio $Y=M^\xi/M^\eta$.
\begin{align}
  -Y^2 \cos \theta \approx -\frac{B_0+zDS \sin \theta}{B_0\cos \theta}
\end{align}
Since $Y>0$, this equation has the unique solution,
\begin{align}
  Y = \sqrt{\frac{B_0+zDS \sin \theta}{B_0\cos^2\theta}}.
\end{align}
As Fig.~\ref{fig:wfdaen} shows, the ratio $Y$ diverges as $B_0/J \to +0$ and converges to $1$ as $B_0/J \to + \infty$.

Similarly to the canted phase, the uniform magnetization $\bm m=\frac{1}{2}(\bm m_A+\bm m_B)$ follows an elliptic precession in the $S^x$-$S^y$ plane. The $S^x$ axis is the major axis and the ratio between the major and minor radii is $Y\cos\theta$.

\begin{figure}[t!]
  \centering
  \includegraphics[bb = 120 0 1176 1227, width = 0.9\linewidth]{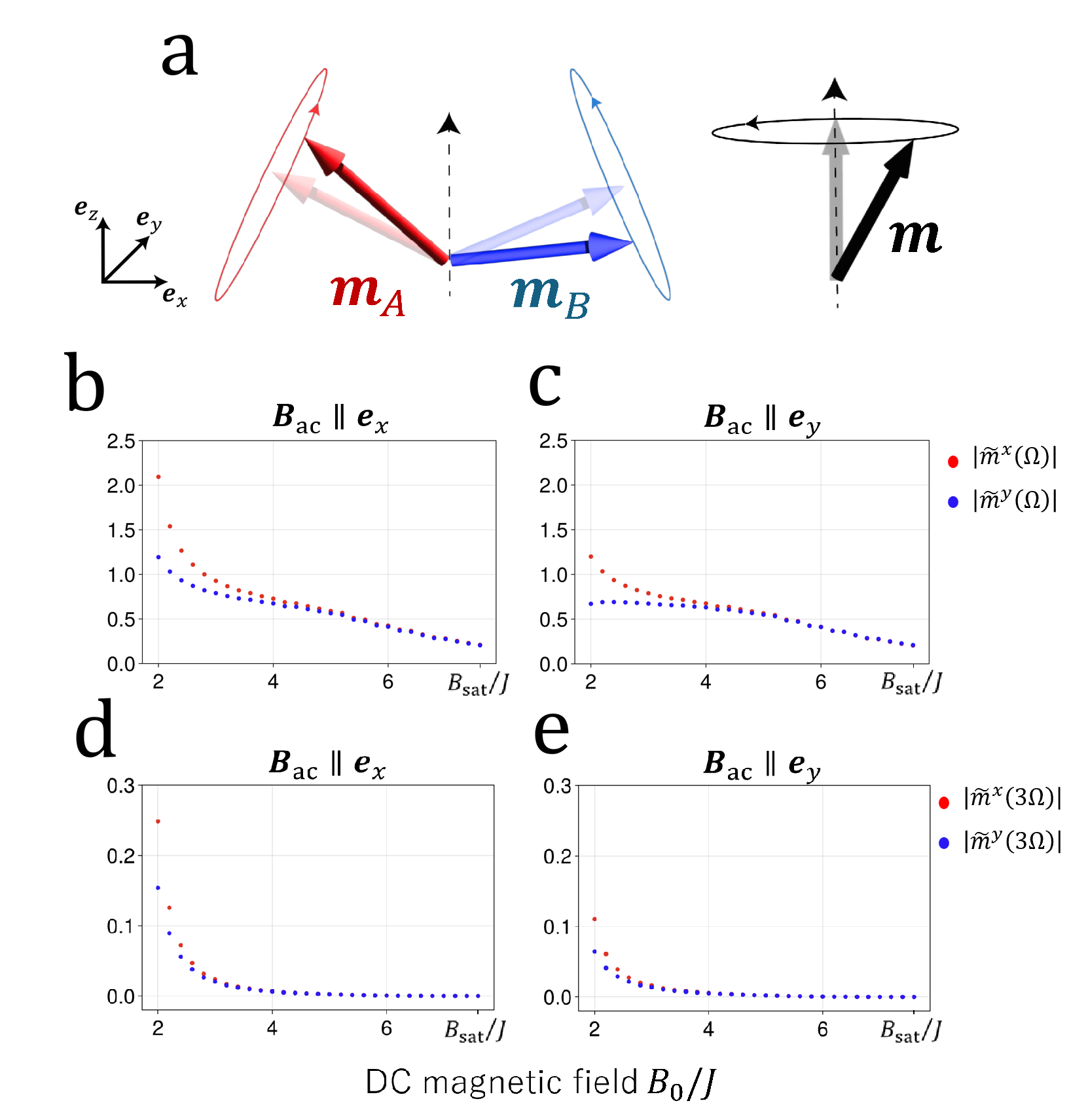}
  \caption{
    (a) Schematic image of precession motion in $\alpha$-mode magnetic resonance of canted phase. Symbols $\bm m_A$ and $\bm m_B$ denote the magnetic moments of A and B sublattices, respectively, while $\bm m=(\bm m_A + \bm m_B)/2$ is the uniform magnetization. Light-colored arrows mean the directions of magnetic moments in the equilibrium state. The sublattice moments draw an elliptical orbit in the coordinate $(S_{\bm r}^\xi,~S_{\bm r}^\eta,~S_{\bm r}^\zeta)$, in which the axis $S_{\bm r}^\xi$ is the major axis and the axis $S_{\bm r}^\eta$ is the minor axis. The uniform magnetization $\bm m$ also has an elliptical orbit whose major and minor axes are respectively the $S^x$ and $S^y$ axes. Panels (b) and (c) show how the first harmonic generation peaks $|\tilde m^a(\Omega)|$ ($a=x,~y$) depend on the static (dc) magnetic field $B_0$ when (b) $\bm B_{\rm ac} \parallel \bm e_x$ or (c) $\bm B_{\rm ac} \parallel \bm e_y$.
    Panels (d) and (e) show how the third harmonic generation peaks $|\tilde m^a(3\Omega)|$ ($a=x,~y$) depend on the static magnetic field $B_0$ when (d) $\bm B_{\rm ac} \parallel \bm e_x$ or (e) $\bm B_{\rm ac} \parallel \bm e_y$. We set $B_{\rm ac}/J=0.01$ for upper panels (b) and (c) and  $B_{\rm ac}/J=0.05$ for lower panels.
  }
  \label{fig:cantdaen_num}
\end{figure}

\section{Dependence on laser-field direction and static-field strength}

In Appendix~\ref{app:resonance}, we see that for the magnetic resonance of the $\alpha$ mode [see Fig.~\ref{fig:model} (f)], the trajectory of uniform magnetization precession is an ellipse (not circle) in the canted and WF phases, especially in a region of low static magnetic fields [see Fig.~\ref{fig:cantdaen_num} (a)]. From this fact, the intensities of harmonic generation spectra are also expected to depend on the direction of the ac magnetic field.
This section numerically shows how first and third harmonic generations depend on the ac-field direction in the canted phase.

Figures~\ref{fig:cantdaen_num} (b)-(e) show the $B_0$ dependence of heights of the first and third harmonic generation peaks.
Let us compare the upper panels (b) and (c), where $\bm B_{\rm ac}$ is parallel to (b) $\bm e_x$ or (c) $\bm e_y$.
Note that the spin flop transition occurs at $B_0=B_{\rm sf}$ with $B_{\rm sf} = 1.74~J$ as we increase $B_0/J$.
By looking at the panels (b) and (c), we find two characteristics.
First, $|\tilde m^x(\Omega)| > |\tilde m^y(\Omega)|$ holds in both panels for small $B_0/J$ and the difference $|\tilde m^x(\Omega)|- |\tilde m^y(\Omega)|$ increases as we decrease $B_0/J$ so that it approaches $B_{\rm sf}$.
Besides, the peak height $|\tilde m^x(\Omega)|$ of the first harmonic generation for $\bm B_{\rm ac} \parallel \bm e_x$ is higher than that for $\bm B_{\rm ac} \parallel \bm e_y$.
In other words, the stronger magnetic resonance occurs in the former case than the latter.
These two characteristics vanish as we increase $B_0/J$ and completely disappear at $B_0=B_{\rm sat}$ when the system exhibits the phase transition to the saturated phase, also known as the forced ferromagnetic phase.
One can find qualitatively the same features in the third harmonic generation peak heights [lower panels (d) and (e) of Fig.~\ref{fig:cantdaen_num}].

The parameter dependence discussed in this section is consistent with the classical picture of the magnetic resonance in Appendix~\ref{app:resonance}.
The spatial anisotropy of the ellipse trajectory explains why the harmonic generation peaks show the dependence on the laser-field direction.

\begin{figure}[t!]
  \centering
  \includegraphics[bb = 0 0 1081 780, width = 0.9\linewidth]{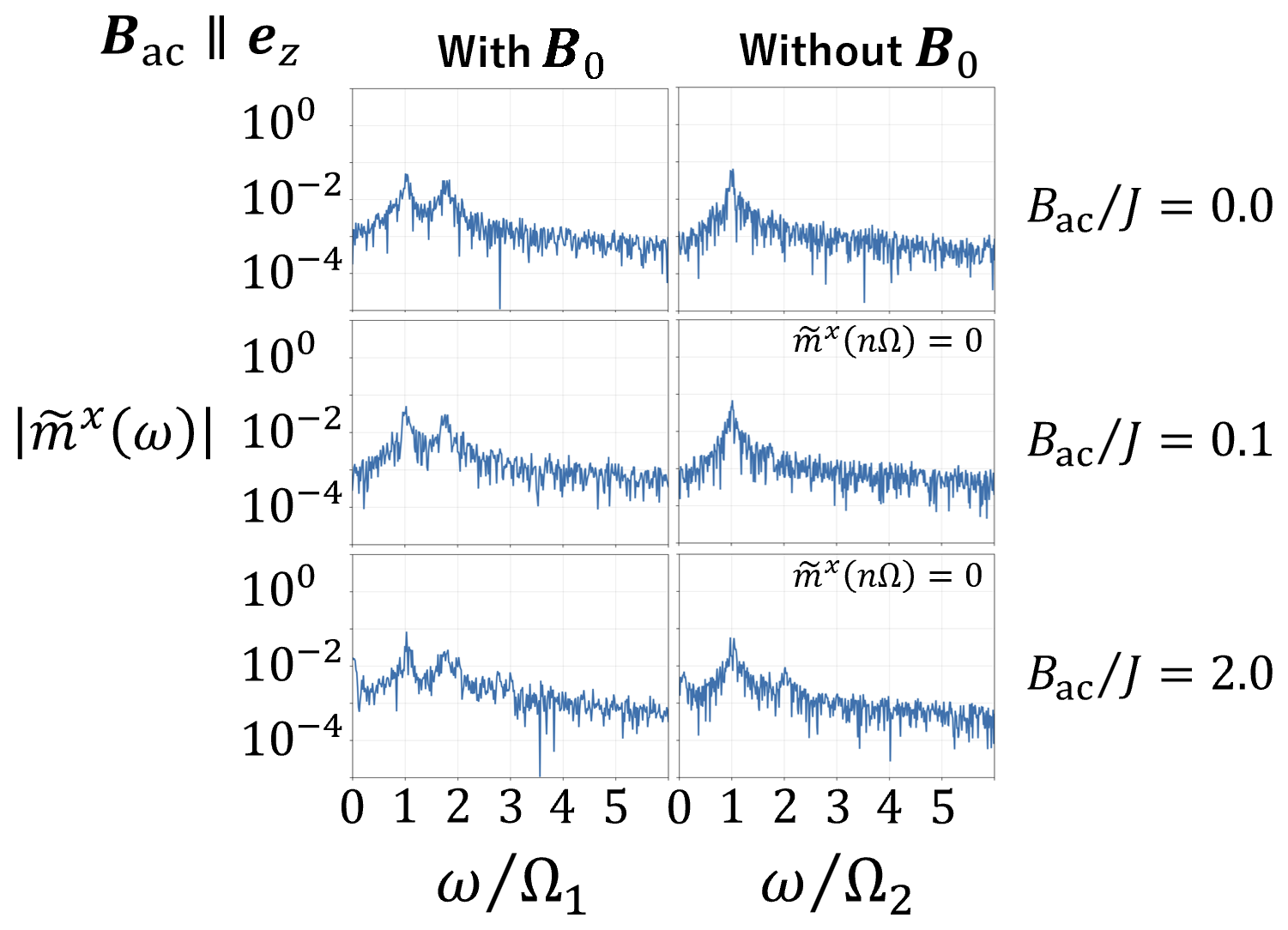}
  \caption{
    Spectra $|\tilde m^x(\omega)|$ in N\'eel phase when $\bm B_{\rm ac}\parallel \bm e_z$ with or without static magnetic field $B_0$.
    We compare the cases with $B_{\rm ac}/J= 0.0, 0.1, 2.0$.
    We set $B_0/J=0.5$ in the left column and $B_0/J=0$ in the right column.
    In the presence of $B_0$, no dynamical symmetry exists and every harmonic generation can potentially appear.
    By contrast, in the absence of $B_0$, dynamical symmetry forbids every harmonic generation.
    We set the laser-pulse frequency $\Omega$ so that it equals the magnon gap of the $\beta$ mode, $\epsilon_{\bm{0}}^{\text{N\'eel}, \beta}$: $\hbar\Omega_1/J=1.33$ with $B_0$ and $\hbar\Omega_2/J=1.83$ without $B_0$.
  }
  \label{fig:neelzstr}
\end{figure}


\section{Finite-temperature fluctuation in N\'eel phase for laser field parallel to $z$ axis}
\label{app:neelzstr}

As shown in Fig.~\ref{fig:neelhhg} (a) and (b), there are no sharp peaks in the harmonic generation of the N\'eel phase when the ac magnetic field is parallel to the $z$ axis. Therefore, in this section, we consider the temperature effect and the ac-field strength dependence of the harmonic generation spectrum in the N\'eel phase for ${\bm B}_{\rm ac}\parallel {\bm e}_z$.
First we look at $|\tilde m^a(\omega)|$ for $\bm B_{\rm ac} \parallel \bm e_z$ in the presence of the static magnetic field $B_0$.
As Fig.~\ref{fig:neelhhg}~(a) shows, there is neither dynamical symmetry nor nontrivial selection rule on the harmonic generation.
The left column of Fig.~\ref{fig:neelzstr} shows the harmonic generation spectra $|\tilde m^x(\omega)|$ for $B_{\rm ac}/J = 0.0$, $0.1$, and $2.0$.
We see qualitative agreements among these three cases.
In these figures, we have set the laser frequency to the magnetic resonance of the $\beta$ mode: $\hbar\Omega_1=\epsilon_{\bm 0}^{\text{N\'eel},\beta}=1.33J$.
This agreement of three cases implies that the laser field along the $z$ axis (with realistic strength) excites no magnons.
The broad peaks in these spectra are due to thermally excited magnons.
The three left panels in Fig.~\ref{fig:neelzstr} seem to have a second harmonic generation peak in addition to the first harmonic generation one. However, we note that the broad peak around $\hbar\omega\sim 2\hbar \Omega_1$ is the linear response of the $\alpha$ mode.
In fact, the peak position deviates from $\hbar\omega=2\hbar \Omega_1$.

Let us move on to the spectra in the absence of the static magnetic field $B_0$.
As we already pointed out in the main text, the dynamical symmetry forbids every harmonic generation peak i.e., $\tilde m^x(n\Omega) = 0$ for $n=1,2,3,\cdots$.
Nevertheless, we observe the broad first harmonic generation for $B_{\rm ac}/J = 0.0$ and $0.1$.
We can conclude that the emergence of the first harmonic generation peak would be attributed to the thermally excited magnon, by comparing the panels of $B_{\rm ac}/J = 0.0$ and $0.1$.
When $B_{\rm ac}/J=2.0$, we also observe a small second harmonic generation as a combination of the finite-temperature effect and the strong ac field.
We can infer that the dynamical symmetry is slightly violated at finite temperatures because thermal excitation of magnons cannot be ruled out.

\begin{figure}[t!]
  \centering
  \includegraphics[bb = 0 0 1135 1152, width = 0.9\linewidth]{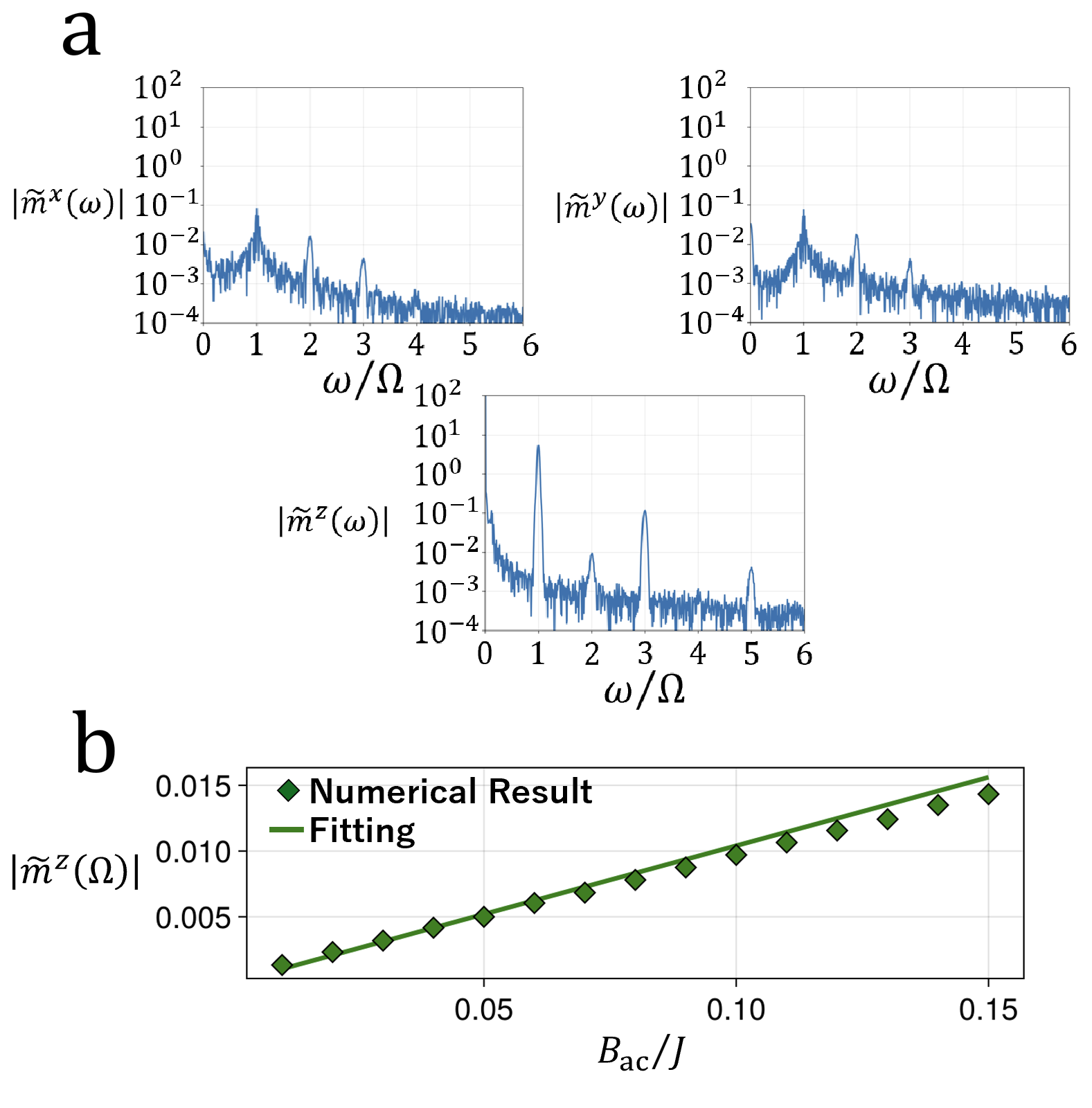}
  \caption{
    (a) Harmonic generation spectra $|\tilde m^a(\omega)|$ for $a=x,y,z$ in canted phase when laser field is parallel to $z$ axis and extremely strong.
    We set $B_{\rm ac}/J=3.0$.
    The laser frequency $\Omega$ is chosen to be the magnon gap of the $\alpha$ mode: $\hbar\Omega=\epsilon_{\bm{0}}^{\text{Cant}, \alpha} = 2.7J$.
    (b) Linear $B_{\rm ac}/J$ dependence of peak height $|\tilde m^z(\Omega)|$ of first harmonic generation in spectrum $|\tilde m^z(\omega)|$.
  }
  \label{fig:cantzstr}
\end{figure}


\section{
  Harmonic generation spectra in canted phase for laser field parallel to $z$ axis
}
\label{app:canted}

Similarly to the N\'eel phase, Fig.~\ref{fig:Canthhg} shows that only small first harmonic generation peaks appear when the ac field ${\bm B}_{\rm ac}$ is parallel to $\bm e_z$ in the canted phase.
Let us take a close look at the ac-field strength dependence of the harmonic generation spectrum $|\tilde m^a(\omega)|$ in the canted phase for $\bm B_{\rm ac} \parallel \bm e_z$.
Figure~\ref{fig:cantzstr}~(a) shows the spectra $|\tilde m^a(\omega)|$ for $a=x,y$, and $z$ under the extremely strong laser field, $B_{\rm ac}/J = 3.0$ when $\bm B_{\rm ac} \parallel \bm e_z$.
The harmonic generation peaks are attributed to the $\alpha$ mode of magnons.
The $\beta$ mode is gapless longitudinal mode and thus irrelevant to the spectrum.
Let us recall that Fig.~\ref{fig:Canthhg}~(a) shows the first harmonic generation only when $\bm B_{\rm ac} \parallel \bm e_z$.
The absence of second and higher harmonic generations results from the mismatch between the laser-field direction and the precession direction.
The precession around the $z$ axis is hardly induced by the laser field oscillating in the $z$ axis.
Nevertheless, Fig.~\ref{fig:cantzstr}~(a) shows second and higher harmonic generations thanks to the extremely strong laser field strength.
We observe harmonic generations up to third order in $|\tilde m^x(\omega)|$ and $|\tilde m^y(\omega)|$ and observe those up to fifth order in $|\tilde m^z(\omega)|$.
We note that it is difficult to generate a THz laser with intensity $B_{\rm ac}>J$ within the current experimental technique (see also Table~\ref{tab:laser}).

The peak height of the first harmonic generation in the spectrum $|\tilde m^z(\omega)|$ shows a linear dependence on $B_{\rm ac}/J$ [Fig.~\ref{fig:cantzstr}~(b)].
This linearity indicates that the first harmonic generation is the linear response of the $\alpha$ mode to the laser field $\bm B_{\rm ac}$.

\section{Three dimensionality}
\label{app:3d}
In this section, we briefly discuss the effects of three dimensionality on the harmonic generation spectrum in the N\'eel phase. In this study, we have concentrated on 2D antiferromagnets. This is because when we apply a laser field to 3D materials, the field is usually absorbed by atoms near the surface, and it does not affect atoms far from the surface. However, it is crucial to confirm that the harmonic generation spectrum of a 2D antiferromagnet exhibits a lineshape almost identical to that of a 3D or layered antiferromagnet possessing the same magnetic order. To this end, we numerically compute the harmonic generation spectra of quasi-2D $N_z$-layer antiferromagnet with a N\'eel order.
The Hamiltonian for the $N_z$-layer antiferromagnet is given by
\begin{align}
  \hat{\mathcal H}_{\text{layer}}=
  & \sum_{j=1}^{N_z}\Bigg[J\sum_{\braket{\bm r,\bm r'}} \hat{\bm S}_{\bm r,j} \cdot \hat{\bm S}_{\bm r',j}-K\sum_{\bm r} (\hat S_{\bm r,j}^z)^2 \nonumber\\
  &- B_0 \sum_{\bm r}\hat S_{\bm r,j}^z\Bigg] 
  +J_\parallel \sum_{j=1}^{N_z-1} \sum_{\bm r} \hat{\bm S}_{\bm r,j} \cdot \hat{\bm S}_{\bm r,j+1}\nonumber\\
  =& \sum_{j=1}^{N_z} \hat{\mathcal H}_{\text{N-C}}(j)+J_\parallel \sum_{j=1}^{N_z-1} \sum_{\bm r} \hat{\bm S}_{\bm r,j} \cdot \hat{\bm S}_{\bm r,j+1}.
  \label{H_Neel-Canted_def_Nz}
\end{align}
Here, $\hat{\bm S}_{\bm r,j}$ is the $j$-th layer spin operator on site $\bm r$,  $\hat{\mathcal H}_{\text{N-C}}(j)$ is the $j$-th layer Hamiltonian and $J_\parallel$ is the inter-layer exchange interaction. For simplicity, we set $J_\parallel=J$, namely, all the exchange couplings are antiferromagnetic.
We regard the spins $\hat{\bm S}_{\bm r,j}$ as a classical vector $m_{\bm r,j}^\alpha (t)$ and then we can compute the harmonic generation spectrum of the $N_z$-layer model by numerically solving the LLG equation.
In this model, the uniform (total) magnetization per one site is defined as
\begin{align}
  m^\alpha(t)=\frac{1}{N N_z}\sum_{\bm r,j}m_{\bm r,j}^\alpha (t),
\end{align}
instead of Eq.~\eqref{ma_tot_def}.
\begin{figure}
  \centering
  \includegraphics[width=0.9\linewidth]{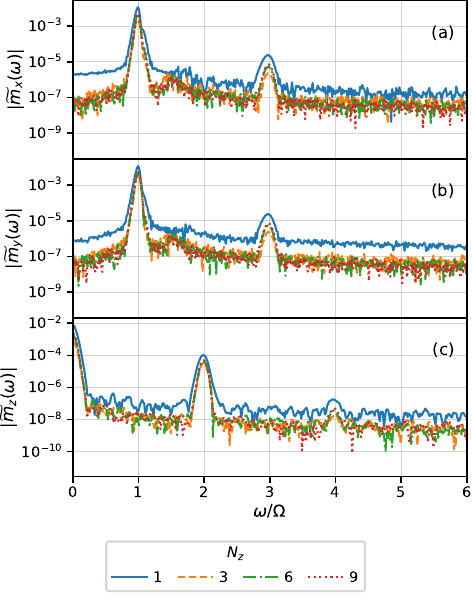}
  \caption{Layer-number ($N_z$) dependence of harmonic generation spectra in $N_z$-layer N\'eel phase of the model~\eqref{H_Neel-Canted_def_Nz}.
    The applied THz field $\bm B_{\rm ac}$ is polarized along the $x$ direction and we impose the open boundary condition along the interlayer ($j$) direction.
    We set $k_{\rm B}T/J=10^{-3}$, $B_{\rm ac}/J=0.1$, and $B_0/J=0.5$. The THz laser frequency $\Omega$ is set to the energy $\epsilon_{\bm{0}}^{\text{N\'eel}, \beta}$ of the $\beta$-mode magnon. Since the $\beta$-mode excitation energy depends on the layer number $N_z$, we used the $\beta$-mode energy of the 2D model for $N_z=1$ to determine $\Omega$, while the $\beta$-mode energy of the 3D model was used for $N_z=3$, $6$ and $9$.
    Blue, orange, green, and red curves are respectively the results of $N_z=1$, 3, 6 and 9. The result of $N_z=1$ corresponds to the first column of Fig.~\ref{fig:neelhhg} (a). To improve numerical convergence, we have used a shorter time step $J\Delta t/\hbar=0.00625$ for all calculations involving the 3D model here.
  }
  \label{fig:Neel-two-dim-vs-three-dim}
\end{figure}
Figure~\ref{fig:Neel-two-dim-vs-three-dim} shows the numerically estimated layer-number ($N_z$) dependence of the harmonic generation spectra in the $N_z$-layer N\'eel phases.
One easily sees that the lineshapes (especially the peaks and their positions) of the harmonic generation spectra show little dependence on the layer number $N_z$. Therefore, we can state that our numerical results for ideal 2D antiferromagnets may be viewed as those of 3D antiferromagnets with the same magnetic order at least at a qualitative level.

%

\end{document}